\documentclass[aps,10pt,twocolumn,superscriptaddress,nofootinbib,preprintnumbers]{revtex4-2} 

\usepackage{natbib}
\usepackage{multirow}
\usepackage{graphicx,rotating,amssymb}
\usepackage{enumerate}
\usepackage[percent]{overpic}

\usepackage{color}
\definecolor{rosso}{cmyk}{0,1,1,0.4}
\definecolor{rossos}{cmyk}{0,1,1,0.55}
\definecolor{rossoc}{cmyk}{0,1,1,0.2}
\definecolor{blu}{cmyk}{1,1,0,0.3}
\definecolor{blus}{cmyk}{1,1,0,0.6}
\definecolor{bluc}{cmyk}{1,1,0,0.1}
\definecolor{verde}{cmyk}{0.92,0,0.59,0.25}
\definecolor{verdec}{cmyk}{0.92,0,0.59,0.15}
\definecolor{verdes}{cmyk}{0.92,0,0.59,0.4}

\ifx\pdfoutput\undefined
\usepackage[dvips,bookmarks]{hyperref}  
\else
\usepackage{hyperref} 
\fi
\hypersetup{colorlinks,bookmarksopen,bookmarksnumbered,citecolor=rossoc,
linkcolor=rossoc,pdfstartview=FitH,urlcolor=blue}

\newcommand{\usine}{{\sc usine~v3.5}}
\newcommand{\minuit}{{\sc minuit}}

\newcommand{\BIG}{{\sc BIG}}
\newcommand{\SLIM}{{\sc SLIM}}
\newcommand{\QUAINT}{{\sc QUAINT}}

\newcommand{\beq}{\begin{equation}}
\newcommand{\eeq}{\end{equation}}
\newcommand{\ben}{\begin{eqnarray}}
\newcommand{\een}{\end{eqnarray}}
\newcommand{\bi}{\begin{itemize}}
\newcommand{\ei}{\end{itemize}}

\def\aap{A\&A}%
\def\apj{ApJ}%
\def\apjl{ApJ}%
\def\araa{ARA\&A}%
\def\crp{CR~Phy.}%
\def\jcap{J. Cosmology Astropart. Phys.}%
\def\nat{Nature}%
\def\physrep{Phys.~Rep.}%
\def\prd{Phys.~Rev.~D}%
\def\prl{Phys.~Rev.~Lett.}%

\newcommand{\pbar}{\ensuremath{\bar{p}}}

\begin{document}


\title{AMS-02 antiprotons' consistency with a secondary astrophysical origin}
\author{Mathieu Boudaud}
\email{Deceased}
\affiliation{LPTHE, Sorbonne Universit\'e, CNRS, 4 Place Jussieu,  F-75005 Paris, France}

\author{Yoann G\'enolini}
\email{yoann.genolini@ulb.ac.be}
\affiliation{Service de Physique Th\'eorique, Universit\'e Libre de Bruxelles, Boulevard du Triomphe, CP225, 1050 Brussels, Belgium}

\author{\\Laurent Derome}
\affiliation{LPSC, Univ. Grenoble Alpes, CNRS, 53 avenue des Martyrs, F-38000 Grenoble, France}

\author{Julien Lavalle}
\affiliation{LUPM, Universit\'e de Montpellier, CNRS, Place Eug\`ene Bataillon, F-34000 Montpellier, France}

\author{David Maurin}
\affiliation{LPSC, Univ. Grenoble Alpes, CNRS, 53 avenue des Martyrs, F-38000 Grenoble, France}

\author{Pierre Salati}
\affiliation{{LAPTh}, Univ. Grenoble Alpes, USMB, CNRS, F-74000 Annecy, France}

\author{Pasquale D. Serpico}
\affiliation{{LAPTh}, Univ. Grenoble Alpes, USMB, CNRS, F-74000 Annecy, France}

\preprint{
LAPTH-017/19, ULB-TH/19-04, LUPM:19-056
}
\begin{abstract}

    \begin{center}
    {\it {$\cal I$}n memory of Mathieu, our dear friend and colleague, who passed away while this work was under review.}
    \end{center}~\\
  The AMS-02 experiment has ushered cosmic-ray physics into precision era. In a companion paper, we designed an improved method to calibrate propagation models on B/C data. Here we provide a robust prediction of the \pbar{} flux, accounting for several sources of uncertainties and their correlations. Combined with a correlation matrix for the \pbar{} data, we show that the latter are consistent with a secondary origin. This paper presents key elements relevant to the dark matter search in this channel, notably by pointing out the inherent difficulties in achieving predictions at the percent-level precision.
\end{abstract}

\maketitle

\paragraph*{Introduction ---}
The spectrum of Galactic cosmic-ray (CR) antiprotons (\pbar{}'s) originating from interactions of CR protons on the interstellar (IS) gas was first calculated fifty years ago \cite{1967PhRv..158.1227R,1968PhRv..171.1344S}. CR \pbar{}'s were discovered a few years later in balloon-borne experiments \cite{1981ICRC....9..146B,1981ApJ...248.1179B}, but in excess compared to this expected {\em secondary} background; this was soon interpreted as the presence of an extra contribution, possibly from primordial black holes \cite{1981Natur.293..120K} or dark matter (DM) annihilation \cite{1984PhRvL..53..624S,1985PhRvL..55.2622S}. Over the past thirty years, a lot of progress has been made both theoretically and experimentally.

From the 90' to the 00's, balloon flights (MASS, IMAX, CAPRICE, HEAT-pbar, and BESS \cite{1996ApJ...467L..33H,1996PhRvL..76.3057M,1997ApJ...487..415B,2001ApJ...561..787B,2001PhRvL..87A1101B,1997ApJ...474..479M,1998PhRvL..81.4052M,2000PhRvL..84.1078O,2001APh....16..121M,2002PhRvL..88e1101A}) and AMS-01 onboard the shuttle \cite{2002PhR...366..331A} collected hundreds of \pbar{}'s, up to $\sim 10$ GeV. Improvements in the predictions, accounting for tertiary production \cite{1983JPhG....9..227T}, more accurate cross sections \cite{1992ApJ...394..174G}, and reacceleration \cite{1998ApJ...499..250S} paved the road to obtain compatibility of the data with modern diffusion models \cite{2001ApJ...563..172D,2002ApJ...565..280M}, the transport parameters of which were fitted on B/C \cite{2001ApJ...555..585M}.

In the following decade, the next generation of instruments---BESS-TeV/Polar \cite{2005ICRC....3...13H,2008PhLB..670..103B,2012PhRvL.108e1102A} and the PAMELA satellite \cite{2009PhRvL.102e1101A,2010PhRvL.105l1101A,2013JETPL..96..621A}---measured several thousands of \pbar{}'s up to a few hundreds of GeV. Updates of the model parameters and corresponding predictions, using better measured $p$ and He CR spectra, yielded a secondary flux in agreement with the data \cite[e.g.][]{2009PhRvL.102g1301D}.

The state-of-the-art AMS-02 experiment has been operating on the ISS since 2011. It has already recorded several tens of thousands of \pbar{}'s \cite{2016PhRvL.117i1103A}, up to TeV energies. For the first time, the measurement is dominated by systematic uncertainties (at a few percent level), implying new challenges for their interpretation. Here we address this issue, going beyond previous analyses of preliminary AMS-02 \pbar{}'s data \cite{2015JCAP...09..023G,2015JCAP...10..034K,2015JCAP...12..039E}.

We underline that CR \pbar{}'s are one of the most sensitive astroparticle probes of annihilating/decaying DM in the GeV-TeV range \cite{2002astro.ph.12111M,2012CRPhy..13..740L,Boudaud:2014qra,2017NatPh..13..224C,2018PhRvD..97b3003C}, and any constraint on DM candidates depends on how well the astrophysical background is controlled. This is especially important as claims for \pbar{} excesses attributed to DM are being debated \cite{2017PhRvL.118s1101C,2017PhRvL.118s1102C,Cholis:2019ejx,Cuoco:2019kuu,2019arXiv190309545L}.


\medskip
\paragraph*{Methodology ---}

The flux of CR \pbar{}'s at Earth depends on (i) the cross sections entering their production, scattering, and annihilation, (ii) the CR propagation model, (iii) the IS spectrum of the (most abundant) CR nuclei, and (iv) modulation of fluxes in the Solar cavity.

Uncertainties on relevant nuclear cross sections are among the dominant ones for the \pbar{} flux calculation \cite{2001ApJ...563..172D,2015JCAP...09..023G}. 
Recently, new data have been taken \cite{2010EPJC...65....9A,2017EPJC...77..671A,Aaij:2018svt}, leading to improvements in the cross section parametrisations \cite{2014PhRvD..90h5017D,2017PhRvD..96l3010L,2017PhRvD..96d3007D,2017JCAP...02..048W,2018PhRvD..97j3019K}. 
A crucial ingredient needed for the calculation is the Lorentz invariant cross section of prompt \pbar{}'s produced in $pp$ interactions.
We make use of the parametrisation proposed in Ref.~\cite{2017JCAP...02..048W} and improved in Ref.~\cite{2018PhRvD..97j3019K} (Param II).
For nucleon-nucleon interactions, we use the scaling relation B proposed in Ref.~\cite{2018PhRvD..97j3019K} which provides the best agreement with LHCb data. 
We use the covariance matrices of errors on the parameters to propagate the uncertainties to the \pbar{} flux calculation. 
Antineutrons ($\bar{n}$'s) and antihyperons are produced in hadronic interactions and decay into \pbar{}'s; their contribution has to be included in the total cross section.
The yield of \pbar{}'s produced via $\bar{n}$'s is larger than in the prompt \pbar{} channel, an energy-dependent effect studied and parameterised in Ref.~\cite{2017JCAP...02..048W}, in agreement with the very scarce experimental data. We introduce an exponentially decreasing energy-dependent parametrisation in order to reproduce the 1$\sigma$ CL interval of Fig. 8 in Ref.~\cite{2017JCAP...02..048W}.
Regarding the anti-hyperon contribution, we make use of the parametrisation determined in Ref.~\cite{2017JCAP...02..048W} and the covariance matrix of the corresponding parameters to propagate the uncertainties on the \pbar{} flux calculation.  More details on the nonprompt channels and on the way we propagate the related uncertainties are given in the App.~\ref{sec:antihyperons} and App.~\ref{sec:antineutrons}, respectively. 
Inelastic annihilating and non-annihilating interactions of \pbar{}'s  with the IS medium (ISM) are treated following the procedure described in~\cite{Boudaud:2014qra}.

We use the 1D diffusion model implemented in \usine{} \cite{2018arXiv180702968M} which assumes a thick diffusion halo size $L$, and a thin disc containing the sources and the gas. Computationally more expensive scenarios with extended geometries \cite{2016PhRvD..94l3019K,2016PhRvD..94l3007F,2017PhRvD..96l3010L} and time dependence \cite{2015PhRvL.115r1103K} have been considered in the literature. Yet, should simple 1D models prove sufficient to account for the data in a secondary production scenario, more complex scenarios would {\it a fortiori} work as well. We use the most generic transport model  defined in~\cite{Genolini:2019ewc} (called \BIG{}) where the transport parameters are fitted on B/C following the methodology described in \cite{Derome:2019jfs}, i.e. a model of the covariance matrix of B/C AMS-02 errors has also been incorporated in the fit. Using the \minuit{} \cite{1975CoPhC..10..343J} package to fit the B/C, we compute the best-fit values of the free parameters (transport and nuisance parameters) and their covariance matrix. 
We fully propagate the transport uncertainty to the \pbar{} flux from the covariance matrix. We have compared the distribution of the best fit parameters from the covariance matrix with the distribution obtained by resampling the B/C data to check that the Gaussian treatment of the uncertainties is valid in this context.

Important inputs for the \pbar{} calculation are the IS fluxes for all progenitors (mostly H, He, C and O, but also all nuclei up to Fe, see App.~\ref{SM:fraccontribs} for their detailed contributions; we also account for heavy elements in the ISM, see App.~\ref{SM:impact_heavy}). There are two options to propagate the associated uncertainties: a standard approach in semianalytical models \cite{2001ApJ...563..172D,2009PhRvL.102g1301D,2015JCAP...09..023G,2017JCAP...02..048W} is to demodulate top-of-atmosphere (TOA) data, in order to extract the parameters describing the IS fluxes and their uncertainties, which allows us to yield the \pbar{} flux with a fast procedure and to easily propagate the progenitors uncertainties. The preferred approach in fully numerical propagation codes is to directly use the IS fluxes calculated at the propagation stage, ensuring that the calculated fluxes fit the data \cite{2002ApJ...565..280M,2010APh....34..274D,2015JCAP...12..039E,2016PhRvD..94l3019K}. In this global fitting, the transport and source uncertainties are determined simultaneously. Although using a semianalytical model, we follow this second approach but do not rely on a global fitting of the transport and source parameters. Instead, we follow and extend the two-step procedure detailed in \cite{Genolini:2019ewc} (see Sec.~III therein): We start from the best-fit parameters obtained from the B/C analysis (\BIG{}) and then perform a simultaneous fit of H \cite{2015PhRvL.114q1103A}, He, C, and O \cite{2017PhRvL.119y1101A} AMS-02 data to determine the source parameters, i.e. four normalisations ($^1$H, $^4$He, $^{12}$C, and $^{16}$O) and three slopes ($\alpha_{\rm H}$, $\alpha_{\rm He}$, and a universal source slope $\alpha_{\rm Z>2}$ for all other species).
Source isotopic fractions are fixed to Solar System values \cite{2003ApJ...591.1220L}; the abundances of non-fitted elements up to $Z=30$ are fixed such as to match the HEAO-3 data \cite{1990A&A...233...96E} at 10.6 GeV/n.
We fit at the same time the high-rigidity diffusion break parameters which are better constrained by elemental fluxes than by the B/C~\cite{Genolini:2019ewc}.
We include in the fit the covariance matrices of errors on H, He, C, and O data, see App.~\ref{SM:covmat}. The outputs of this fitting procedure are discussed in App.~\ref{SM:scores}. Note that we fix the Fisk potential~\footnote{The force-field approximation \cite{1975ApJ...202..265G,2004ApJ...612..262C,2006AdSpR..37.1727O,2007APh....28..154S} has a single free parameter, and it was found to describe well modulated fluxes \cite{2017A&A...605C...2G}.} $\phi_{\rm FF}$ for H, He, C, and O data to the value yielded by the B/C data fit \cite{Genolini:2019ewc}. While published AMS-02 data are from the same time period for B/C, He, C, and O (05/2011 to 05/2016), they originate from a shorter period for H (05/2011 to 11/2013), so that the associated modulation level should be slightly different. However, as the fit is restricted to data above the \pbar{} production threshold ($E_{k/n}> 6\,m_p$ \cite{2001ApJ...563..172D}), the impact is feeble, and in any case negligible compared to other uncertainties.

State-of-the-art modulation models are 2D or 3D charge-dependent diffusion and drift in the Solar cavity~\cite{2013LRSP...10....3P}. They go beyond the force-field approximation \cite{1987A&A...184..119P} and at variance with the latter, they predict different modulations for positively- and negatively-charged particles. 
Charge-sign dependent effects on secondary antiprotons have been estimated by Ref.~\cite{Vittino:2017fuh} to be smaller than $\sim 10\%$ below 5 GeV and negligible above.
However, these models have a large number of parameters and are still under study. Moreover, the impact of the drift effect is strongly dependent on the data-collection period. The AMS-02 \pbar{} data \cite{2016PhRvL.117i1103A} were analysed from a time period of 4 years (May 2011 to May 2015) longer than that of H and shorter than that of He, C, and O data. As illustrated in \cite{2014JCAP...12..045C}, drift models can be used to derive effective and different Fisk potentials for nuclei and \pbar{}'s,  with $\phi_{\rm FF}^{\pbar{}}$ which is not clearly smaller or larger than $\phi_{\rm FF}^{\rm nuc}$ on a sufficiently long period. Assuming the same modulation for \pbar{} and H, He, C, and O actually already gives a satisfactory description at low rigidity. For this reason, we did not include further uncertainty on $\phi_{\rm FF}^{\pbar{}}$. If anything, this would slightly enlarge the uncertainty of the prediction at low rigidity, improving further the consistency between our calculated \pbar{} flux and the data.

Uncertainties in the parameters discussed above should be propagated to the \pbar{} flux calculation. To proceed, for each source of uncertainty (production cross section, transport in the Galaxy, fit to parents fluxes) we draw $\approx 10000$ realizations of the  parameters from their covariance matrix and we compute the corresponding \pbar{} fluxes. We then calculate the 1$\sigma$ confidence-level (CL) envelope for each source of uncertainty as well as the covariance matrix of the model $\mathcal{C}^{\rm model}$ (see App.~\ref{SM:correlations}). This enables us to soundly assess the compatibility of the model with the data.

\smallskip
In the context of the B/C analysis, we stressed the importance of using a realistic covariance matrix of the data errors to avoid misleading conclusions~\cite{Derome:2019jfs}.  We anticipate that the same is true for \pbar{}'s. However, since this matrix is not directly provided by the AMS-02 collaboration, we build it from the published systematic errors and associated description of their physics origin, in the same spirit as in \cite{Derome:2019jfs}.

\begin{figure}
\includegraphics[width=\columnwidth]{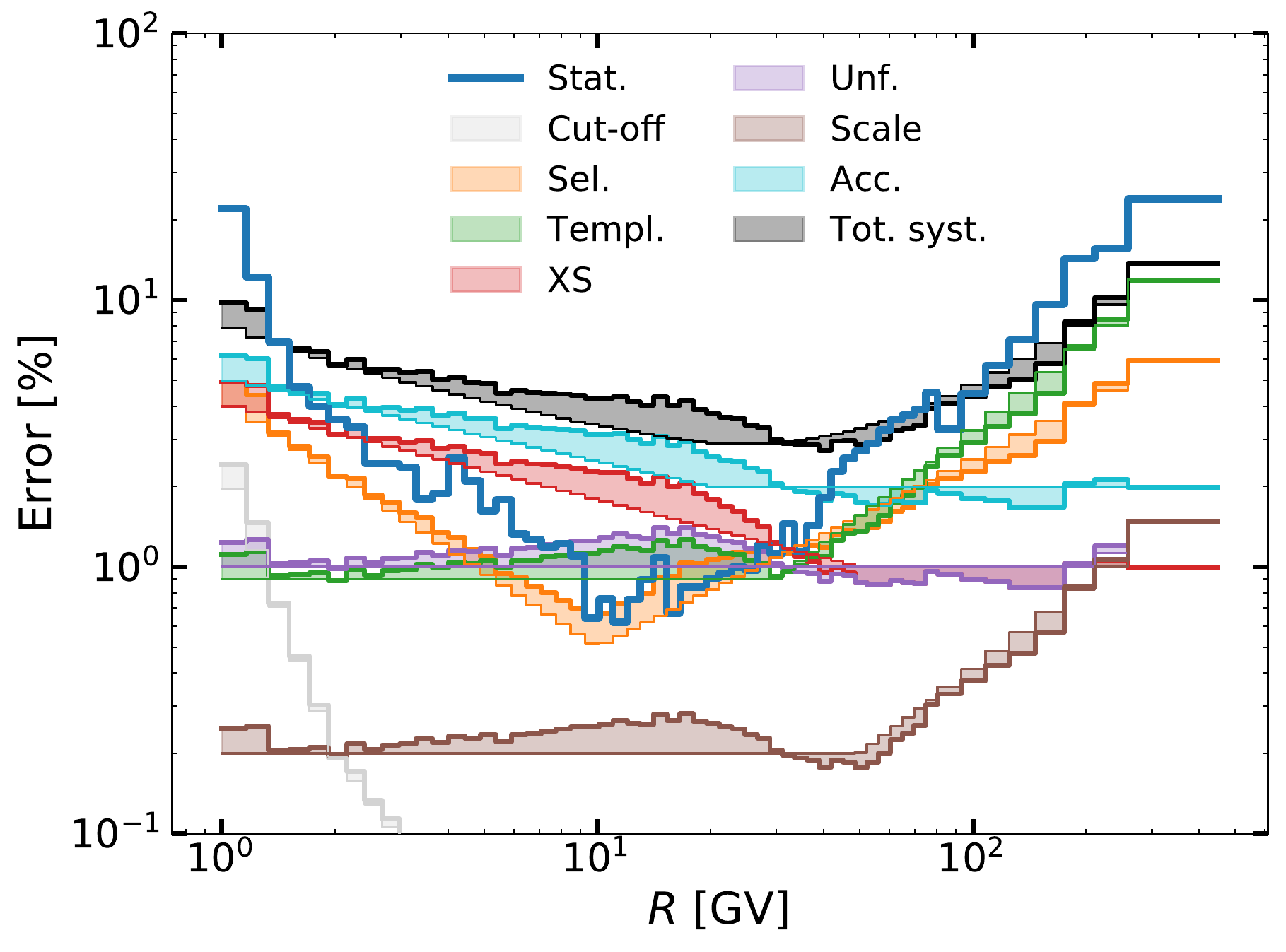}
\caption{AMS-02 errors for \pbar{} data. Statistical (Stat.) and Total Systematic (Tot. syst.) lines correspond to the errors provided in \cite{2018PhRvL.120b1101A}. Individual contributions in the systematic errors namely rigidity cut-off (Cut-off), selection (Sel.), template fitting (Templ.), cross sections (XS), unfolding (Unf.), rigidity scale (Scale) and acceptance (Acc.), built from information provided in \cite{2016PhRvL.117i1103A} are shown (coloured lines) before (thin) and after (thick) the rescaling applied to match the total systematic error.}
\label{fig:AMS_errors}
\end{figure}

The various contributions to the AMS-02 systematics are broadly described in \cite{2016PhRvL.117i1103A}. For instance, the text \textit{``This [selection] uncertainty amounts to 4\% at 1GV, 0.5\% at 10GV, and rises to 6\% at 450 GV''} is interpreted as a piecewise power-law behaviour, and is shown as an orange thin line in Fig.~\ref{fig:AMS_errors}. Thus we build the seven sources of systematics quoted by the AMS-02 collaboration (coloured thin lines), where the quadratic sum of all contributions leads to the black thin line. In order for the sum
to match the total systematic errors provided in \cite{2016PhRvL.117i1103A} (black thick line), we rescale for each rigidity point our separate contributions by the ratio of the thick to the thin black lines. This leads to our model for the AMS-02 \pbar{} systematics (coloured thick lines). The covariance matrix associated with these systematics is then built based on a choice of their correlation length, $\ell$. More details on this procedure are given in App.~\ref{SM:correlations} and, for  the B/C analysis, in \cite{Derome:2019jfs}.  For \pbar{}'s, we take as educated guesses for the correlations lengths (in unit of energy decade) $\ell_{\rm Acc.}=0.1$ (acceptance), $\ell_{\rm Cut}=1.0$ (rigidity cut-off), $\ell_{\rm Scale}=4.0$ (rigidity scale), $\ell_{\rm Templ.}=0.5$ (template fitting), $\ell_{\rm XS}=1.0$ (cross sections), $\ell_{\rm Unf.}=1.0$ (unfolding), and $\ell_{\rm Sel.}=0.5$ (selection).
%
%

\begin{figure}
\begin{overpic}[width=0.45\textwidth]{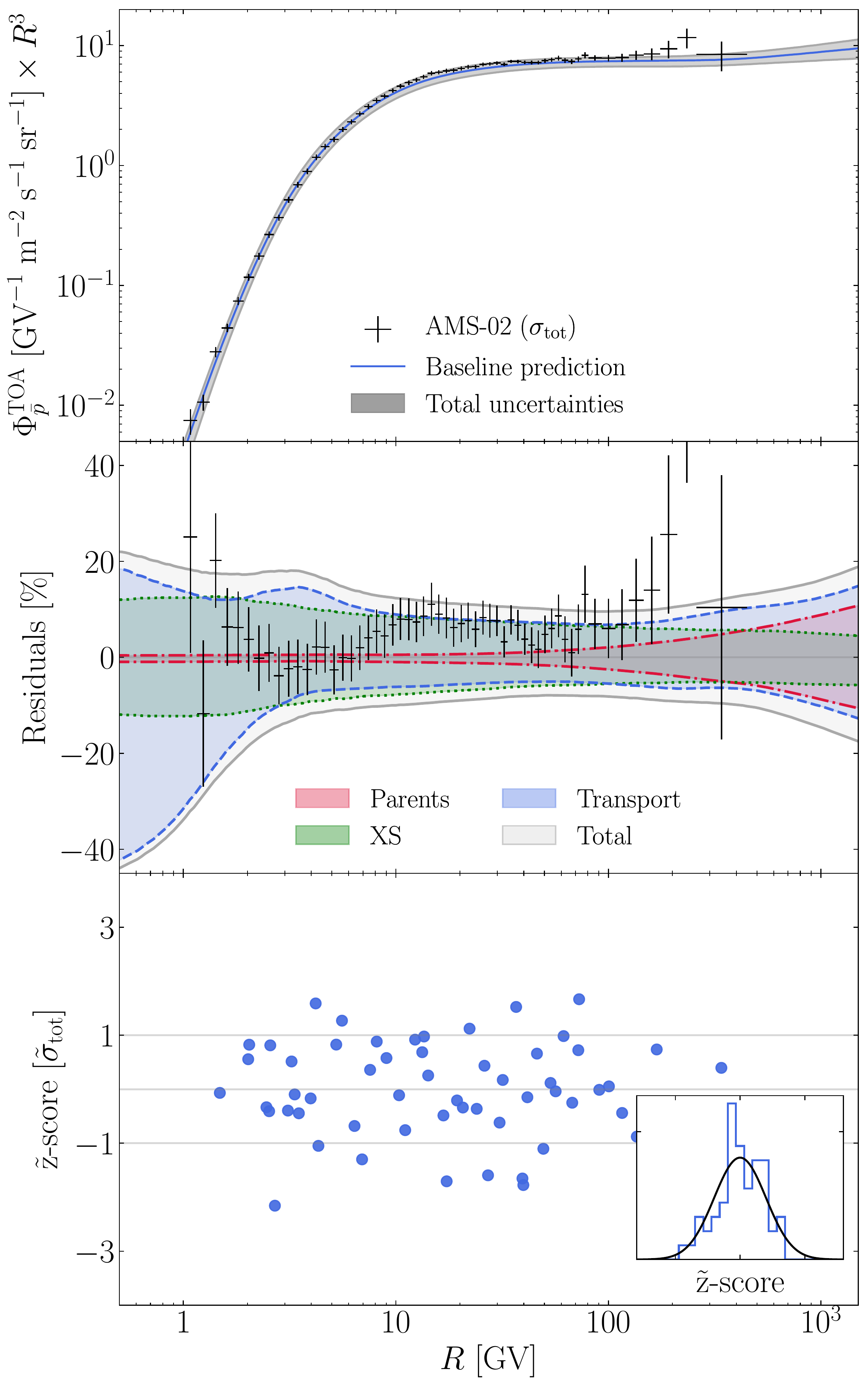}%
  \put (9.5,96.5) {(a)}  \put (9.5,65) {(b)} \put (9.5,34) {(c)}
\end{overpic}
\caption{Comparison of \pbar{} model and data (a), along with residuals and 68\% total confidence interval for the model (grey) together with the transport (blue), the parents (red) and the cross sections (green) contributions (b). 
The residuals of the eigen vectors of the total covariance matrix as a function of the pseudo-rigidity $\tilde{R}$, as well as their distribution are shown in (c) and in the inset.}
\label{fig:pbar}
\end{figure}

\medskip
\paragraph*{Results ---}
The top panel of Fig.~\ref{fig:pbar} shows our baseline \pbar{} flux prediction (not a fit) obtained from the best-fit values for the \pbar{} production cross sections, the transport (\BIG{}) and the associated parents fluxes, compared with AMS-02 data with errors taken as the quadratic sum of systematic and statistical errors (black crosses). The `standard' residuals with respect to the baseline model are displayed in the middle panel. Note that the points do not include the model uncertainties, nor correlations in the data uncertainties. We also show on the same plot the 68\% total confidence band for the model (grey band). This band could release the tension with the data, even before accounting for the information on the correlations in rigidity bins. The respective contributions of parents, cross sections and transport are also plotted. At tens of GV, the errors from transport and cross sections are almost constant and close to 10\%. At larger rigidities, the errors from transport and parents increase because of the increasing experimental uncertainty in the B/C ratio and parent fluxes, respectively. At low rigidity, the error from transport grows for the same reasons and encompasses the uncertainty in the prediction of the low-rigidity behaviour (see App.~\ref{SM:pbar_vs_model}). However, we remind the reader that a visual comparison can be deceiving: The presence of non-diagonal values in the covariance matrices is responsible for a better agreement between the model and data than perceived in the residuals (see Table~\ref{tab:results_chi2}).

To test the actual compatibility of our prediction with the \pbar{} data, we present two statistical tests which boil down to probabilistic statements in terms of $p$-value (see App.~\ref{SM:likelihood_test}). First, we propose a $\chi^2$ test, with the help of a covariance matrix of errors on both data and model:
\begin{equation}
  \chi^2 \!\!=\!\!({\rm data}\!-\!{\rm model})^{\rm T} ({\cal C}^{\rm model}\!+{\cal C}^{\rm data})^{-1} ({\rm data}\!-\!{\rm model})\,.
\label{eq:cov}
\end{equation}
The covariance matrices of the data ${\cal C}^{\rm data}$ and the model ${\cal C}^{\rm model}$ are given by the sum of the different contributions previously detailed (see SM, Sec.~II-A and III). We find $\chi^2\approx 44$, and identifying the number of degrees of freedom (dof) with the number of \pbar{} data points (57), we infer a corresponding $p$-value of 0.9 which is reported in the last line of Table.~\ref{tab:results_chi2}.  

Such a test does not directly assess a possible overestimate of the errors, and also relies on the notion of number of dof (which may be a shaky concept in some circumstances, see e.g. the discussion in~\cite{2010arXiv1012.3754A}). Thus, we also perform a Kolmogorov-Smirnov (KS) test, which obviates the above limitations. We compute the distribution of the `eigen residuals' ($\tilde{z}$-score) corresponding to the residuals of the eigen vectors (data-model) of the total covariance matrix. In the bottom panel of Fig.~\ref{fig:pbar} we show these `eigen residuals' as a function of `rigidity' (actually, the one rotated in the eigen basis, see App.~\ref{SM:likelihood_test}), and in the inset, we compare the corresponding histogram with a Gaussian. The KS test leads to a $p$-value of 0.27 which is also very good and is clearly consistent with the hypothesis that \pbar{}'s are of secondary origin. For completeness, we also report in Table.~\ref{tab:results_chi2} the $p$-values when considering different combinations of errors:
(i) If there was no uncertainty in our baseline model, the covariance matrix of data errors alone (${\cal C}^{\rm data}$) would already give enough freedom to allow for a very good agreement between the data and the secondary flux prediction;
(ii) Considering only the statistical uncertainties in the data and the uncertainties in the model ($\sigma_{\rm stat}$ and ${\cal C}^{\rm model}$), this prediction is marginally consistent with the data at the 2$\sigma$ level, with the KS test leading to an even better $p$-value. Also note the relevance of the KS test (as opposed to the $\chi^2$ test) to spot error overestimates, in the case of $\sigma_{\rm tot}$ and ${\cal C}^{\rm model}$;
(iii) In the most realistic case considering both ${\cal C}^{\rm data}$ and ${\cal C}^{\rm model}$, $p$-values are very good for both the $\chi^2$ and KS test.
Thus, not only is a secondary origin for the locally measured \pbar{}'s statistically consistent with the data, but, as shown by these considerations, it is also robust with respect to error mismodelling in either model or data errors.

%
\begin{table}[!htb]
\begin{center}
\caption{Respective $p$-values for different sources of errors. We take dof$=57$, i.e. the number of \pbar{} data. Total errors on data are defined to be $\sigma_{\rm tot}=\sqrt{\sigma_{\rm stat}^2+\sigma_{\rm syst}^2}$.}
\label{tab:results_chi2}
\begin{tabular}{c c c c}
\hline\hline
Error considered    & ~~$\chi^2/{\rm dof}$~~ & ~p-value $({\chi^2})$~ & ~p-value (KS)~ \\
\hline\hline
$\sigma_{\rm stat}$                              &  23    &        0           &  0   \\
$\sigma_{\rm tot}$                               & 1.69   & $8.3\times10^{-4}$ &  0   \\
${\cal C}^{\rm data}$                            & 0.85   &       0.79         & 0.97 \\
$\sigma_{\rm stat}$ and ${\cal C}^{\rm model}$   & 1.32   &       0.05         & 0.99 \\
$\sigma_{\rm tot}$ and ${\cal C}^{\rm model}$    & 0.37   &       1.0          & 0.01 \\
${\cal C}^{\rm data}$ and ${\cal C}^{\rm model}$ & 0.77   &       0.90         & 0.86 \\
\hline
\end{tabular}
\\
\end{center}
\end{table}
%
%
%


\medskip
\paragraph*{Conclusions ---}
Percent-level details in the model predictions now matter, as do more subtle aspects of the data error treatment. 
In this paper we have presented a major upgrade of the \pbar{} flux prediction and analysis by: (i) using the latest constraints on transport parameters from AMS-02 B/C data, (ii) propagating all uncertainties (with their correlations) on the predicted \pbar{} flux, and (iii) accounting for correlated errors in \pbar{} data. The multi-component nature of the systematic error, with different $R$-dependencies and correlation lengths, has a crucial impact on the analysis, and was not captured in more simplified treatments as in Ref.~\cite{Cuoco:2019kuu}. With these novelties,  we unambiguously show that the {\it AMS-02 data are consistent with a pure secondary astrophysical origin.} 
We stress that {\it  this conclusion is not based on a fit to the AMS-02 \pbar{} data, but on a prediction of the \pbar{} flux computed from external data.}
Our results should hold for any steady-stade propagation model of similar complexity, as they all amount to the same ``effective grammage'' crossed to produce boron nuclei (on which the analysis is calibrated), with roughly the same grammage entering the secondary \pbar{}'s.
We have checked that this conclusion is robust with respect to a variation by a factor of a few of the correlation lengths of the AMS-02 systematic uncertainties. Also, recent analyses of Fermi-LAT data are suggestive of a spatial dependent diffusion coefficient, notably different in the inner Galaxy~\cite{Gaggero:2014xla}. Moving to more complex scenarios containing the 1D framework considered here as limiting case would broaden theory space, but would not alter our conclusions on the viability of secondary production to explain antiproton data.
On the technical aspects, more computationally expensive methods could allow one to go beyond the quadratic assumption (i.e. assuming multi-Gaussian error distributions) embedded in the covariance matrix of errors. For more advanced applications, sampling techniques like Markov chain Monte Carlo could be used (e.g., \cite{2009A&A...497..991P}). 
However, a significant improvement in our perspectives for DM searches in the \pbar{} flux can only be achieved by simultaneously reducing the systematics in the data and the errors of the modelling. 
On the data side, a covariance matrix of errors directly provided by the AMS-02 collaboration would definitively be an important improvement to fully benefit from the precision achieved by AMS-02.
On the modelling side, the next step would be to combine more secondary-to-primary ratios (Li/C, Be/C, and B/C) to further decrease the propagation uncertainties. Of course, better data and modelling on \pbar{} and $\bar{n}$ production cross sections is also required, and the sub-leading error due to primary source parameters could be reduced by combining AMS-02 data with higher energy data from CREAM, TRACER and CALET \cite{2019PhRvL.122r1102A}.
In the current state of our analysis, we can anticipate that, from the frequentist point of view, a clear statistical preference for an additional feature in the data is unlikely. However, this conclusion must rely on a quantitative analysis that we postpone for a forthcoming paper.
%

\smallskip
{\it \bf Acknowledgements ---}
MB is grateful to Michael Korsmeier and Martin Winkler for very useful discussions. We are grateful to all the members of the Cosmic Rays Alpine Collaboration. This work has been supported by the ``Investissements d'avenir, Labex ENIGMASS'', by Univ. de Savoie AAP ``DISE", and by the French ANR, Project DMAstro-LHC, ANR-12-BS05-0006. The work of Y.G. is supported by the IISN, the FNRS-FRS and a ULB ARC. We also acknowledge a partial support from the Agence Nationale pour la Recherche (ANR) Project No. ANR-18-CE31-0006, the Origines, Constituants, et EVolution de l'Univers (OCEVU) Labex (No. ANR-11-LABX-0060), the CNRS IN2P3-Theory/INSU-PNHE-PNCG project ``Galactic Dark Matter'', and the European Union's Horizon 2020 research and innovation program under the Marie Sklodowska-Curie Grant Agreements No. 690575 and No. 674896.

\appendix

\section{Antiproton production cross sections}
\label{sec:pbar_XS}

Most of the \pbar{}'s are produced via decays of antihyperons and $\bar{n}$'s. These contributions are not included in the prompt \pbar{} cross section measured by the collider experiments used in Ref.~\cite{2018PhRvD..97j3019K}.
Ref~\cite{Winkler2017} proposed energy-dependent parametrisations based on the most up-to-date collider data. The total \pbar{} production cross section is 
\beq
\sigma_{\rm inv}^{\rm tot} = \sigma_{\rm inv}(2 + \Delta_{\rm IS} + 2\Delta_{\Lambda}),
\label{eq:total_to_prompt}
\eeq
where $\sigma_{\rm inv}$ is the Lorentz invariant cross section of promptly produced \pbar{}'s, $\Delta_{\rm IS}$ and $\Delta_{\Lambda}$ the isospin asymmetry and anti-hyperon corrections, respectively.

\subsection{Antihyperons}
\label{sec:antihyperons}

A sizeable fraction of \pbar{}'s is produced via the decay of antihyperons $\bar{\Lambda}$ and $\bar{\Sigma}$. The ratio of hyperon-induced to promptly produced \pbar{}'s is $\Delta_{\Lambda} = (0.81 \pm 0.04) (\bar{\Lambda}/\bar{p})$, where $\bar{\Lambda}/\bar{p}$ is the ratio of \pbar{}'s produced via the decay of $\bar{\Lambda}$ to the total yield. Here, it is assumed that antihyperons decay equally into \pbar{}'s and $\bar{n}$'s. The energy dependence of $\bar{\Lambda}/\bar{p}$ is given by
\beq
\bar{\Lambda}/\bar{p} = c_1 + \frac{c_2}{1 + (c_3/s)^{c_4}},
\eeq
where $\sqrt{s}$ is the centre-of-mass energy, $c_1 = 0.31$, $c_2 = 0.30$, $c_3 = (146 \, \rm GeV)^2$, and $c_4 = 0.9$.
We determine the uncertainty by randomly drawing 10000 realisations of the parameters from the covariance matrix. This method enables us to reproduce the median and the 1$\sigma$ CL obtained in Ref.\cite{Winkler2017}. They are shown in the top panel of Fig.~\ref{fig:isospin_hyperons}.
%

\subsection{Antineutrons}
\label{sec:antineutrons}

\begin{figure}[!t]
\includegraphics[width=0.8\columnwidth]{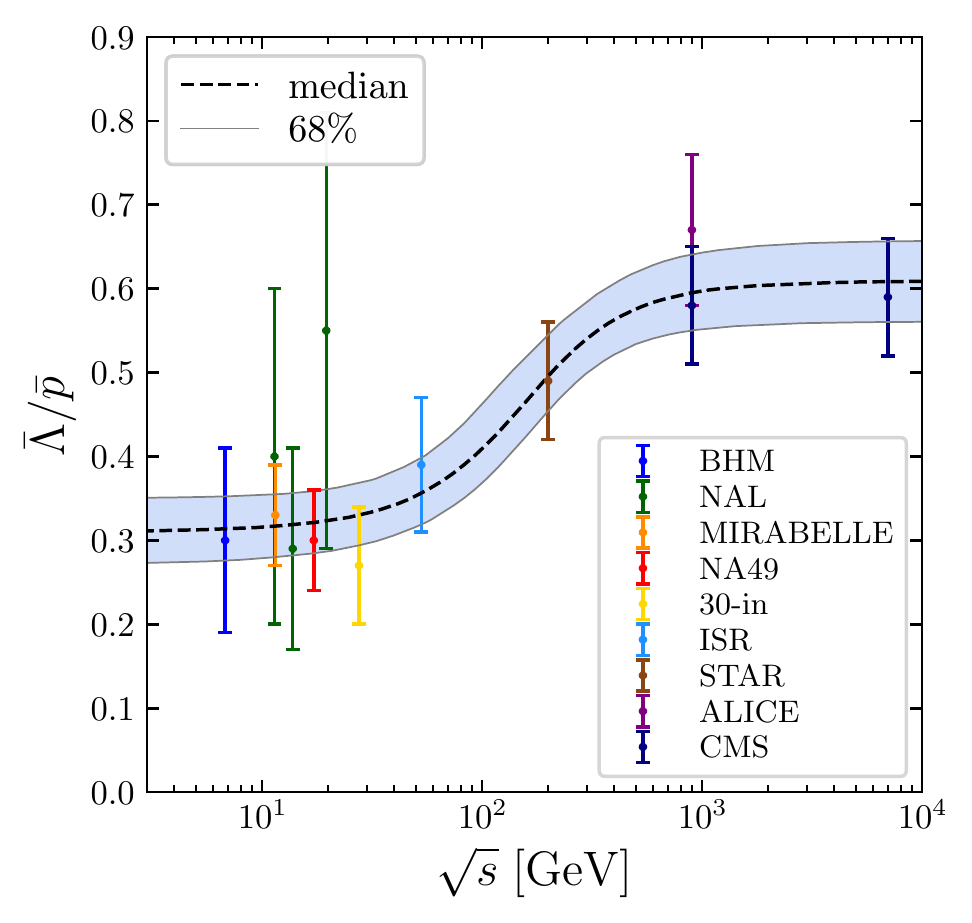}
\includegraphics[width=0.8\columnwidth]{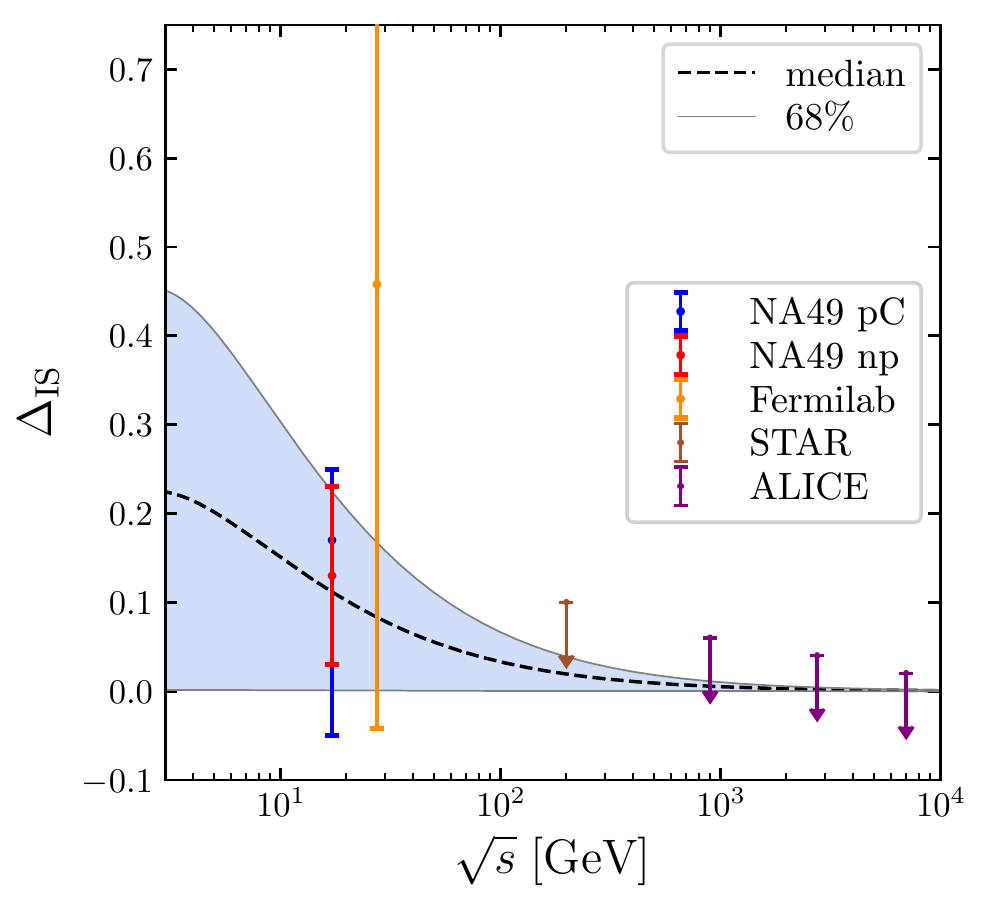}
\includegraphics[width=0.8\columnwidth]{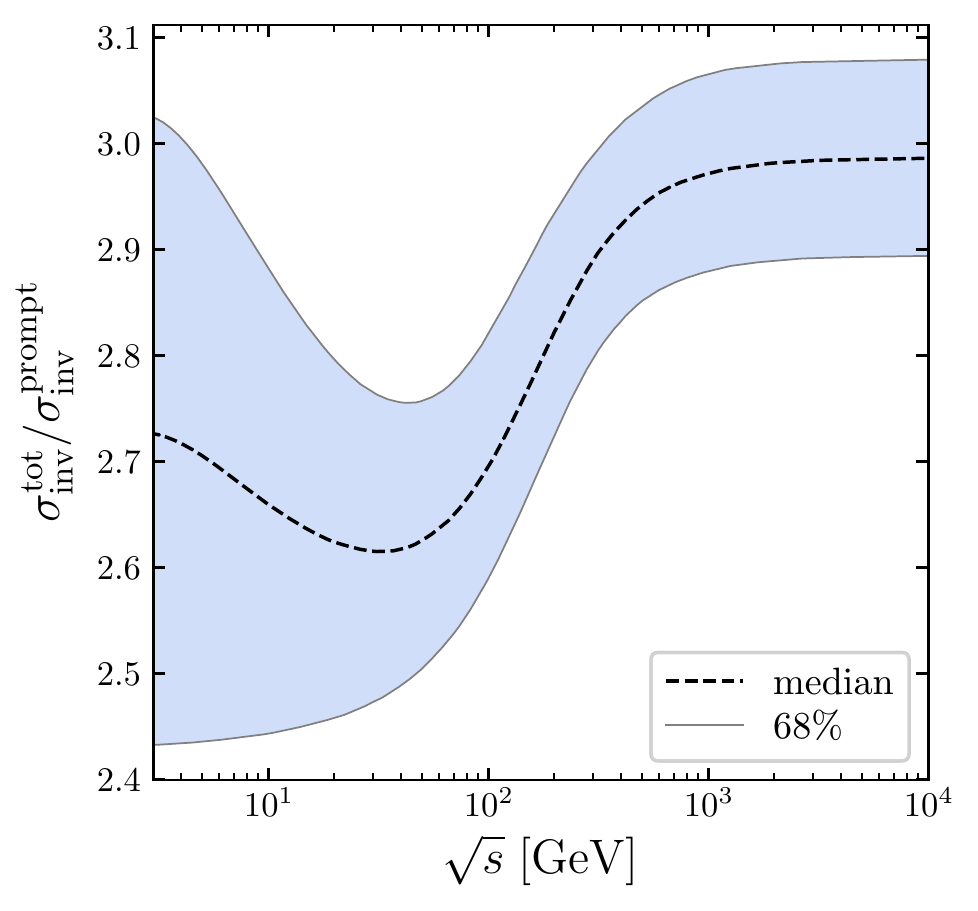}
\caption{Hyperon correction (top panel) and isospin asymmetry (middle panel) as a function of the center-of-mass energy. We also show the ratio between total production cross section and promptly produced \pbar{} cross section (bottom panel).
\label{fig:isospin_hyperons}}
\end{figure}

Antiprotons are preferentially produced via the decay of $\bar{n}$'s. The energy dependence of the isospin asymmetry proposed in Ref.~\cite{Winkler2017} is motivated by the fact that a pion has to be produced when \pbar{}'s are promptly produced. We found that taking the parametrisation and the covariance matrix of parameters from Ref.\cite{Winkler2017} does not allow to reproduce the median and the 1$\sigma$ CL interval the authors show in Fig.~8. Instead, we introduce the parametrisation
\beq
\Delta_{\rm IS} = c_0  (x + c_2)^{c_3} \exp(-x/c_1)
\eeq
where $x = \log(\sqrt s)$, $c_0 = 0.33$, $c_1  = 6.42$, $c_2  = 0.5$ and $c_3  = 1.6$. We randomly draw 10000 realisations of $c_0$ using a Gaussian distribution with a standard deviation of $\sigma_{c_0} = 0.36$ and with the lower bound $c_0 > 0.04$. 
This way, we recover the median and the 1$\sigma$ CL interval of Fig.~8 in Ref.~\cite{Winkler2017}. The results are shown in the middle panel of Fig.~\ref{fig:isospin_hyperons}.
Since we adopt the nucleon-nucleon scaling relation determined in Ref.~\cite{2018PhRvD..97j3019K}, which depends (though slightly) on a different parametrisation of $\Delta_{\rm IS}$, we rescale the parameter $D_2$ entering Eq.~(17) of Ref.~\cite{2018PhRvD..97j3019K} in order to obtain the same value of $f^{A_1A_2}$ at the energy of NA49.

The ratio between $\sigma_{\rm inv}^{\rm tot}$ as expressed in Eq.~(\ref{eq:total_to_prompt}) and $\sigma_{\rm inv}$ is shown in the bottom panel of Fig.~\ref{fig:isospin_hyperons}. In the case of vanishing nonprompt contribution and isospin violation, one would expect a constant value  equal to 2. Hence, not only the \pbar{} flux receives an upwards correction of $\sim20\%$ to $50\%$, but also acquires peculiar energy-dependent features, further affected by relatively large uncertainties.
%
%

\section{Antiproton parents}

\subsection{AMS-02 data errors}
\label{SM:covmat}

In H, He, C, and O fluxes published by the AMS-02 collaboration~\cite{2015PhRvL.114q1103A,2017PhRvL.119y1101A}, uncertainties are broken-down into several contributions: statistical (Stat.), unfolding (Unf.), rigidity scale (Scale), acceptance (Acc.), and even trigger (Trig.) for protons data. As discussed in \cite{Derome:2019jfs}, the information provided in the Supplemental Material of AMS-02 publications allows us to build a first estimate of the covariance matrix of uncertainties $\mathcal{C}^\alpha$. For each uncertainty $\alpha$ (Stat., Unf., Scale, Acc., and Trig.), the $i\!j$-th element of the matrix is taken to be
\begin{equation}
\mathcal{C}^\alpha_{ij} = \sigma^\alpha_i \sigma^\alpha_j \exp\left[-\frac{1}{2}
\left(\frac{\log_{10} (R_i/R_j)}{\ell_\alpha} \right)^2\right]\,,
\label{eq:covmat}
\end{equation}
where $\sigma^\alpha_i$ is the systematic uncertainty of type $\alpha$ taken at rigidity bin $R_i$, and $\ell_\alpha$ is the correlation length indicating the range over which rigidity bins are correlated.

As the data analysis of AMS-02 nuclei follows many similar steps, the correlation lengths derived in \cite{Derome:2019jfs} for B/C data \cite{2018PhRvL.120b1101A} should not be too different from the ones for the proton \cite{2015PhRvL.114q1103A} and He, C, and O data \cite{2017PhRvL.119y1101A}. Consistently with \cite{Derome:2019jfs}, we take:
\begin{itemize}
   \item $\ell_{\rm Stat.} = 0$ (statistical uncertainties are uncorrelated);
   \item $\ell_{\rm Scale} = \infty$ (rigidity scale impacts all rigidities);
   \item $\ell_{\rm Unf.} = 0.5$ (unfolding acts on intermediate scales) though the exact value is not critical \cite{Derome:2019jfs};
   \item $\ell_{\rm Trig.} = 1$ (intermediate error, close to a normalisation error, for the H data only); 
   \item $\ell_{\rm Acc.}$ is ill-defined as it comes from a mixture of uncertainties with different correlation lengths (data/Monte Carlo corrections, cross section normalisation). Following \cite{Derome:2019jfs}, we split it into
   \begin{itemize}
      \item $\ell_{\rm Acc.~norm.}\sim 1.0$ (normalisation error);
      \item $\ell_{\rm Acc.\,LE} \sim 0.3$ (low-energy rise);
      \item $\ell_{\rm Acc.~res.}= 0.1$ (residual acceptance).
   \end{itemize}
\end{itemize}
Note that $\ell_{\rm Acc.~norm}$ is not well constrained, while it was found to be critical for the B/C analysis \cite{Derome:2019jfs}. Its value relies on indirect consistency arguments (analysis of transport parameters), and we refer the reader to \cite{Derome:2019jfs} for the discussion and caveats about this choice.
%

%
\begin{figure}[!t]
\includegraphics[width=0.49\textwidth]{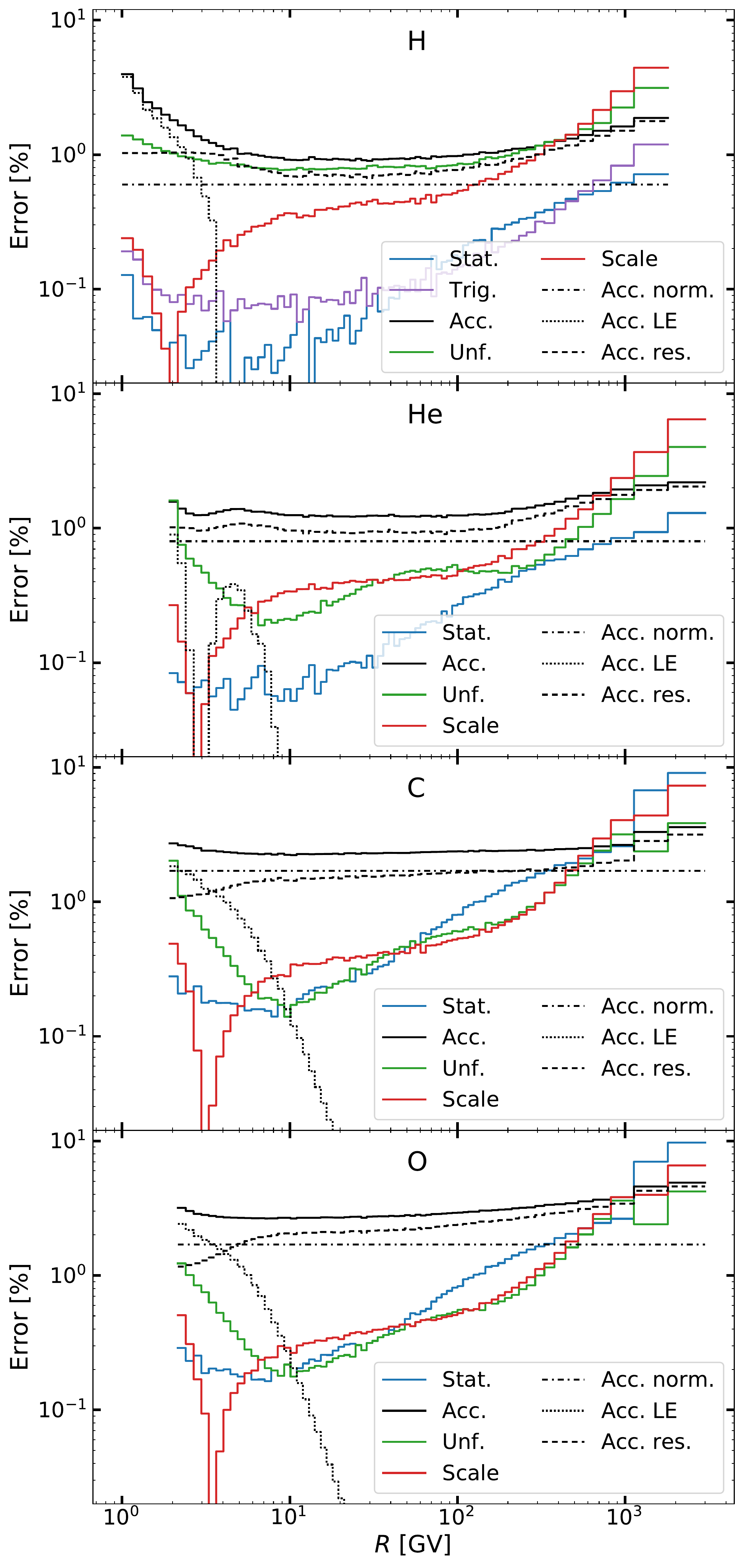}
\caption{
AMS-02 statistical and systematic uncertainties for the H, He, C, and O nuclei data. Solid lines correspond to the uncertainties provided by the AMS-02 collaboration \cite{2015PhRvL.114q1103A,2017PhRvL.119y1101A}, namely those coming from the statistical treatment, the acceptance, the absolute energy scale, and the unfolding procedure (and also from the trigger for protons). The black lines correspond to a further splitting in the acceptance errors---normalisation (norm.), low energy (LE), and residual (res.)---as discussed in the text. The step-like evolution is due to rounding errors from the published tables.
\label{fig:SM_AMS_errors}
}
\end{figure}
%
The above numbers and various systematics (shown in Fig.~\ref{fig:SM_AMS_errors}) are plugged in Eq.~(\ref{eq:covmat}) to generate the covariance matrices for H, He, C, and O. These matrices are used for the determination of source parameters.

\subsection{Primary CR source parameters from the AMS-02 H, He, C, and O data}
\label{SM:scores}

As discussed in App.~\ref{SM:correlations}, for each propagation configuration (\BIG{}, \SLIM{}, and \QUAINT{}, see \cite{Genolini:2019ewc} and also Sec.~\ref{SM:pbar_vs_model}), we fit the source parameters and the high-rigidity break of the diffusion coefficient on H, He, C, and O AMS-02 data, relying on their covariance matrix of errors introduced in Sec.~\ref{SM:covmat}. As we are only interested in the region where these species produce \pbar{}'s, we limit the fit to data above the threshold production $E_k/n=6m_p$. As the H, He, C, and O nuclei are the dominant \pbar{} progenitors (see Sec.~\ref{SM:fraccontribs}), we check here that good fits are achieved in order to ensure the reliability of our \pbar{} prediction.

Table~\ref{tab:SM_fitsHHeCO} gathers the outputs of the fits, i.e. the $\chi^2/{\rm dof}$ values, best-fit values for the sources parameters (4 normalisations and 3 slopes) and high-rigidity diffusion break parameters (break slope, position, and sharpness). First, the values obtained here for the diffusion break parameters and source slope $\alpha_{Z>2}$ are consistent with those of the analysis carried out in \cite{Genolini:2019ewc} based on the C and O nuclei only (and only using total errors instead of the covariance matrix of errors in the fit). The additional information brought from this analysis is a comparison of the H, He, and $Z>2$ source spectral indices\footnote{He, C, and O data were analysed at a later stage by the AMS-02 collaboration, by which time improvements on the alignment and rigidity scale were made. This could slightly change the best-fit break values and source parameter for this quantity.}. While this is not the goal of this study, it is worth mentioning that all models prefer a slightly harder spectral index for He than for heavier species, $\sim 0.02$ harder (uncertainties, not shown in the Table, are $\sim 0.007$). We also confirm a different source slope, $\sim 0.06$, between H and He~\cite{2011Sci...332...69A,2015PhRvL.114q1103A}, except in the \QUAINT{} propagation model configuration. However, the latter give a bad fit to the primary elemental data\footnote{The bad fit of H data in \QUAINT{} is understood as this model has high levels of reacceleration and convection, which affects more strongly $A/Z=1$ species (i.e. H) than $A/Z\approx2$ species (He, C, O).} with $\chi^2/{\rm dof}=1.93$.

%
\begin{table}
\begin{center}
\caption{Best-fit parameter values and $\chi^2/{\rm dof}$ on H, He, C, and O AMS-02 data for the three benchmark transport models \BIG{}, \SLIM{}, and \QUAINT{}~\cite{Genolini:2019ewc}. The first line recalls the modulation parameter value fixed by the analysis of AMS-02 B/C data \cite{Genolini:2019ewc}, which is also taken for H, He, C, and O AMS-02 data. See text for discussion.\label{tab:SM_fitsHHeCO}}
\begin{tabular}{l c c c}
\hline\hline
   & \hspace{0.5cm}\BIG\hspace{0.5cm} & \hspace{0.5cm}\SLIM\hspace{0.5cm} & \hspace{0.5cm}\QUAINT\hspace{0.5cm} \\
\hline
   $\phi_{\rm FF}^\star$ [GV]          & 0.731   & 0.734   & 0.725   \\
   $\chi^2_{\rm min}/{\rm dof}^\dagger$& 0.90    & 0.71    & 1.93    \\[2mm]
   \hline
   \multicolumn{4}{c}{Source parameters}\\
   \hline
   $\alpha_{\rm H}$                    & 2.37    & 2.41    & 2.34    \\
   $\alpha_{\rm He}$                   & 2.31    & 2.34    & 2.30    \\
   $\alpha_{Z>2}$                      & 2.33    & 2.36    & 2.33    \\
   $\log_{10}(q_{\rm H})^\ddagger$     & -3.98   & -3.92   & -4.05   \\
   $\log_{10}(q_{\rm He})^\ddagger$    & -4.44   & -4.39   & -4.46   \\
   $\log_{10}(q_{\rm C})^\ddagger$     & -5.86   & -5.82   & -5.86   \\
   $\log_{10}(q_{\rm O})^\ddagger$     & -5.75   & -5.71   & -5.75   \\[2mm]
   \hline
   \multicolumn{4}{c}{High-rigidity break parameters}\\
   \hline
   $\Delta_{\rm h}$                    & 0.17    & 0.26    & 0.13    \\
   $R_{\rm h} $ [GV]                   & 358.    & 303.    & 436.    \\
   $s_{\rm h}$                         & 0.04    & 0.15    & 0.002   \\
\hline\hline
\end{tabular}
\\
$^\star$ Best-fit modulation level from B/C analysis \cite{Genolini:2019ewc}.\\
$^\dagger$ We adopt \#dof= 209  (219 data, 10 free parameters).\\
$^\ddagger$ Source abundances $q$ are in unit of [GeV/n~m$^3$~Myr]$^{-1}$.
\end{center}
\end{table}
%

\begin{figure*}[t]
\includegraphics[width=\textwidth]{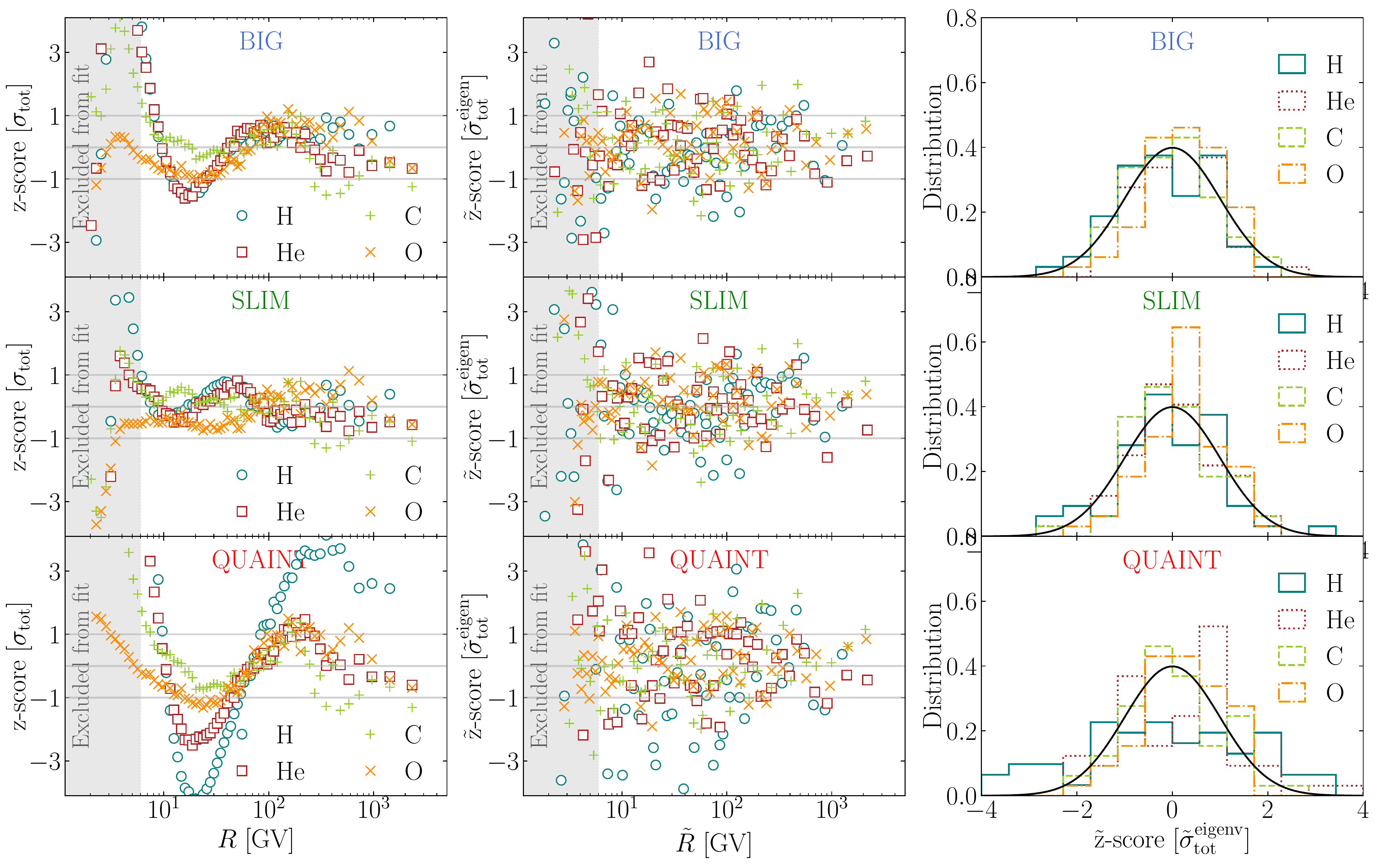}
\caption{Z-score in the original basis (left) and in the eigenvector basis (centre), the distribution of the latter being shown as histograms (right). The rows correspond to the different propagation configurations introduced in \cite{Genolini:2019ewc}, namely \BIG{}, \SLIM{}, and \QUAINT{} (from top to bottom). In each panel, the element H, He, C, and O fit to AMS-02 data are symbol- and colour-coded. See text for discussion and definitions.
\label{fig:SM_HHeCO_scores}
}
\end{figure*}

Figure~\ref{fig:SM_HHeCO_scores} shows further details with a graphical view of the goodness-of-fit for the H (blue), He (red), C (green), and O (orange) nuclei data in and outside the fit regions (grey shaded area on the left and centre panels). The left panels show $z$-scores, i.e. the residuals measured in unit of $\sigma_{\rm tot}$, (model-data)/$\sigma_{\rm tot}$, for the \BIG{} (top), the \SLIM{} (middle), and the \QUAINT{} (bottom) propagation model configurations: a good fit implies that 68\% of the points should be distributed within $1\sigma_{\rm tot}$. However, this representation does not account for correlations in the systematic errors. As discussed in Sec.~\ref{SM:likelihood_test}, it is more interesting to show the same quantity in a rotated basis in which the covariance matrix of errors is diagonal (called $\tilde{z}$-scores)\footnote{The price to pay is that now the matrix of rigidity positions is no longer diagonal, and the value we pick to represent in the middle panels of Fig.~\ref{fig:SM_HHeCO_scores} is the {\it rotated} $z$-score (dubbed $\tilde{z}$-score) defined in Sec.~\ref{SM:likelihood_test}.}. The $\tilde{z}$-scores shown in the second column of Fig.~\ref{fig:SM_HHeCO_scores} no longer display the spectral features seen in the left panels. This illustrates the importance of properly visualising scores with the covariance accounted for, as we may otherwise be visually biased by these structures. The right panels of Fig.~\ref{fig:SM_HHeCO_scores} are the histograms of the $\tilde z$-score points displayed in the middle panels. They indicate that whenever a good $\chi^2/{\rm dof}$ value is obtained, the normalised residuals follow a $1\sigma$-width Gaussian distribution (solid black line). These plots provide further details on the separated contributions from H, He, C, and O data: in all cases, the H data are always less well fitted (broader distribution), especially in the \QUAINT{} model configuration, which explains the bad $\chi^2/{\rm dof}$.

In the light of this study, we correctly reproduce the H, He, C, and O AMS-02 data (except for the \QUAINT{} model configuration) which gives good confidence in the self-consistency of the \pbar{} flux calculation.

%
%
\subsection{Ranking parent contributions}
\label{SM:fraccontribs}
%
\begin{figure}[t]
\includegraphics[width=\columnwidth]{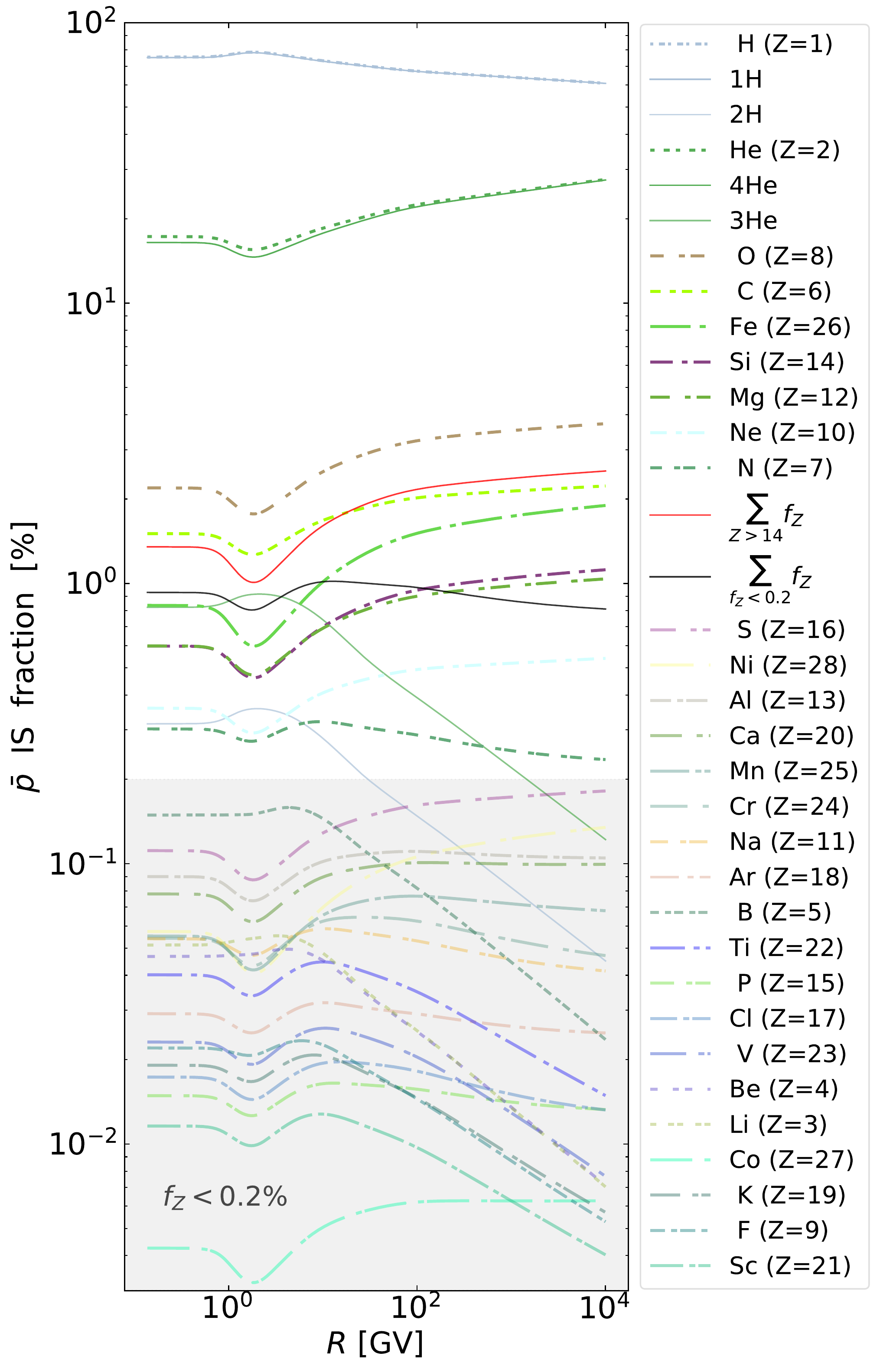}
\caption{Relative contributing fraction $f_Z$ (in percent) of \pbar{} (thick dash-dotted lines) for propagation model \BIG{} used in this study \cite{Genolini:2019ewc}. Specific contributions highlighted in thin solid lines: (i) broken-down isotopic contribution of $^{1}$H and $^{2}$H (blue), and $^{3}$He and $^{4}$He (green);  (ii) contribution from all elements $Z>14$ (red); (iii) contribution from all elements whose fraction is $f_Z<0.2\%$ (black). The grey area delimits the region within which contributions are below $0.2\%$.\label{fig:SM_pbarfrac}}
\end{figure}
%
From the \BIG{} model configuration, we show in Fig.~\ref{fig:SM_pbarfrac} the fractional IS \pbar{} contribution per parent element, properly accounting for each CR isotopes (and summed over all ISM components, here H and He). 
We wish to highlight a few features in this plot: the most significant channels are those involving the most abundant CR fluxes, i.e. H ($\sim80\%$), He ($\sim20\%$), then O, C, Fe, Si, and Mg ($\sim 1-3\%$), followed by Ne and N ($\sim 0.5\%$). All the remaining ones give sub-percent contributions, but we stress that all the contributions below 0.2\% (grey shaded area) add up to a total of $\sim 1\%$ (solid black line). Also, if we only account the contributions of elements heavier than Si, they add up to a total of $\sim 2\%$ (solid red line). 

The contributions of the two most important elements (H and He) are broken down into their isotopic content (blue and green solid lines for H and He isotopes respectively), highlighting the contributions from primary species ($^1$H and $^4$He) w.r.t. sub-dominant secondary ones ($^2$H and $^3$He, lighter-coloured lines).

Focusing on the rigidity dependence, the feature at a few GV is related to secondary species: the peaks/dips seen at these rigidities directly reflect the peaks in secondary-to-primary ratios (e.g., B/C, $^3$He/$^4$He, etc.). Also the spectra of secondary species are one $\delta$ (slope of the diffusion coefficient) steeper than those of primary species, so that their contributions fall off with rigidity. Among the primary species, a difference is observed between H and all heavier elements: H has a steeper source spectrum than the other species (see Table~\ref{tab:SM_fitsHHeCO}), so that its relative contribution w.r.t. these elements decreases with rigidity.

%
\begin{figure}[t]
\includegraphics[width=\columnwidth]{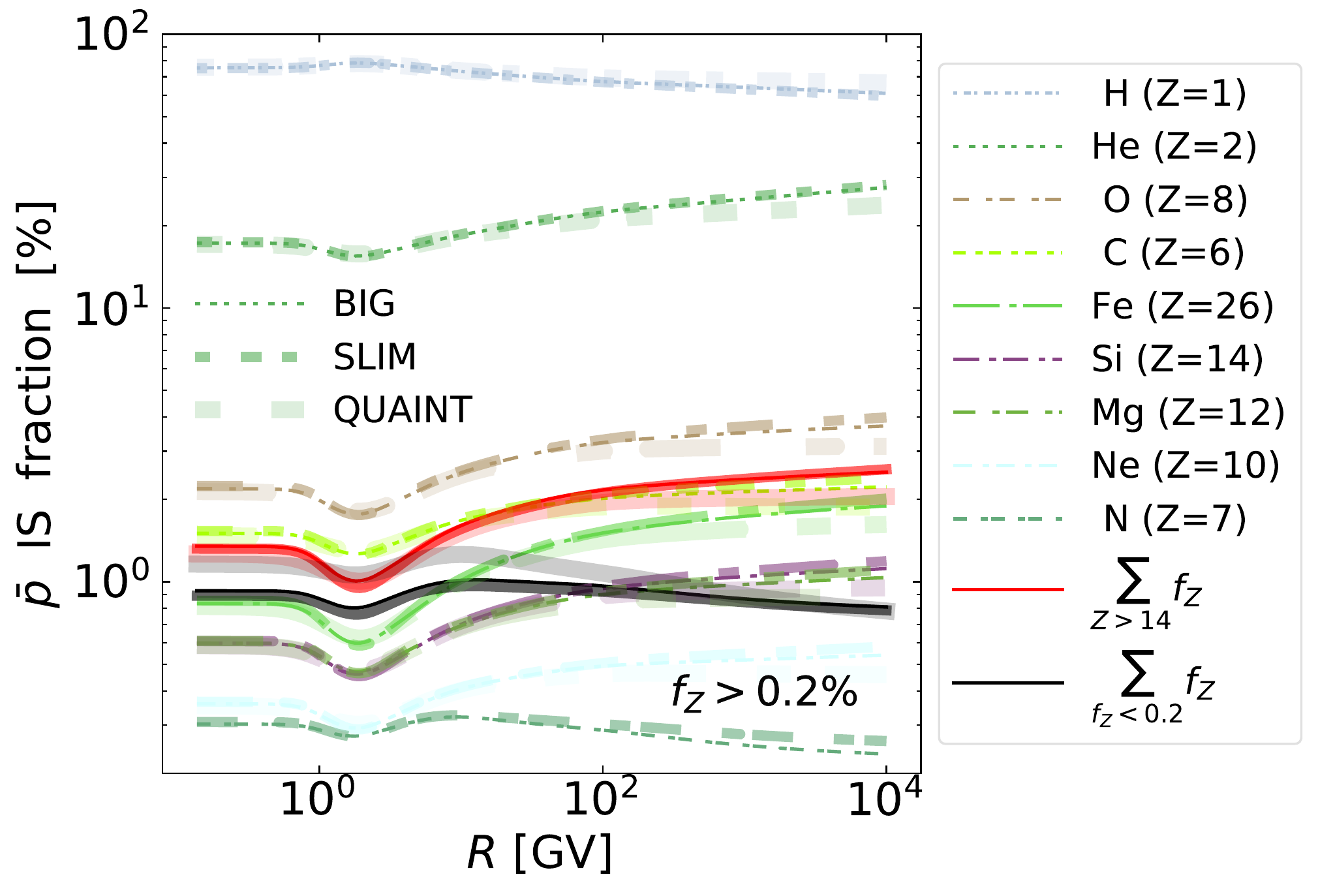}
\caption{Same as Fig.~\ref{fig:SM_pbarfrac} for contributions from \BIG{}, \SLIM{}, and \QUAINT{} (thinner to thicker lines) transport configurations for $f_Z>0.2\%$ only. See text for discussion.\label{fig:SM_pbarfrac2}}
\end{figure}
%

For the sake of illustration, we show in Fig.~\ref{fig:SM_pbarfrac2} that these fractional contributions are not very sensitive to the transport configuration considered (\BIG{}, \SLIM{}, and \QUAINT{} setups, see Sec.~\ref{SM:pbar_vs_model}). This is not surprising as all three are tuned to match the B/C, H, He, C, and O data. The main difference is observed at high rigidity in the \QUAINT{} configuration, because in this model, contrary to the \BIG{} and \SLIM{} ones, H and other species have a similar source spectral indices (see Table~\ref{tab:SM_fitsHHeCO} in Sec.~\ref{SM:scores}).

\subsection{Impact of heavy species in the ISM}
\label{SM:impact_heavy}

To be able to perform tens of thousands fits of the H, He, C, and O  data together with the associated $\bar p$ flux predictions, almost all our calculations were carried out restricting ourselves to CR parents up to $^{30}$Si. Moreover, a standard practice in the field is to assume that the IS gas is only made of H and He. However, in the context of high-precision \pbar{} data, it is important to reassess the impact of heavier species in both CRs and in the ISM, to control the accuracy of the \pbar{} flux calculation. For instance, the fraction of CNO in the ISM is estimated to be $\sim 0.5-1\%$ \cite{1998ApJ...499..250S,2005PhRvD..71h3013D,2018PhRvD..97j3019K}, whereas that of heavy species in CRs (e.g. Fe) is estimated to be $\lesssim 1\%$ \cite{2018PhRvD..97j3019K} (see Fig.~\ref{fig:SM_pbarfrac}).

Before coming to our prescription to account for these contributions, a few remarks are in order, since `heavy' CRs and ISM enter the calculation at several steps: to extract the transport parameters from the B/C analysis, in the calculation and fit of H, He, C, and O fluxes, and then in the \pbar{} flux calculation. Our reference calculation here is based on the transport configuration \BIG{} (see Sec.~\ref{fig:SM_pbar_vs_configs} of \cite{Genolini:2019ewc} for more details), and the best-fit transport parameters proposed in \cite{Genolini:2019ewc} only consider CR parents up to $^{30}$Si with H and He only in the ISM. Hence, for a self-consistent calculation, we must refit the B/C (using the same procedure as in \cite{Genolini:2019ewc}), to extract the transport parameters for the same CR and ISM content as for the \pbar{} propagation. The last missing piece is the detail of the implementation of the ISM composition and cross sections for our calculations. The IS gas composition and species abundances are taken up to Fe from \cite{2009ARA&A..47..481A}, the production cross sections for \pbar{}'s were already discussed in the main text and in Sec.~\ref{sec:pbar_XS}, and for the nuclear inelastic and production cross sections, we rely on a $A_{\rm target}^{0.31}$ scaling relation proposed in \cite{1995ICRC....2..622D} for $^3$He production.

\begin{figure}[t]
\includegraphics[width=\columnwidth]{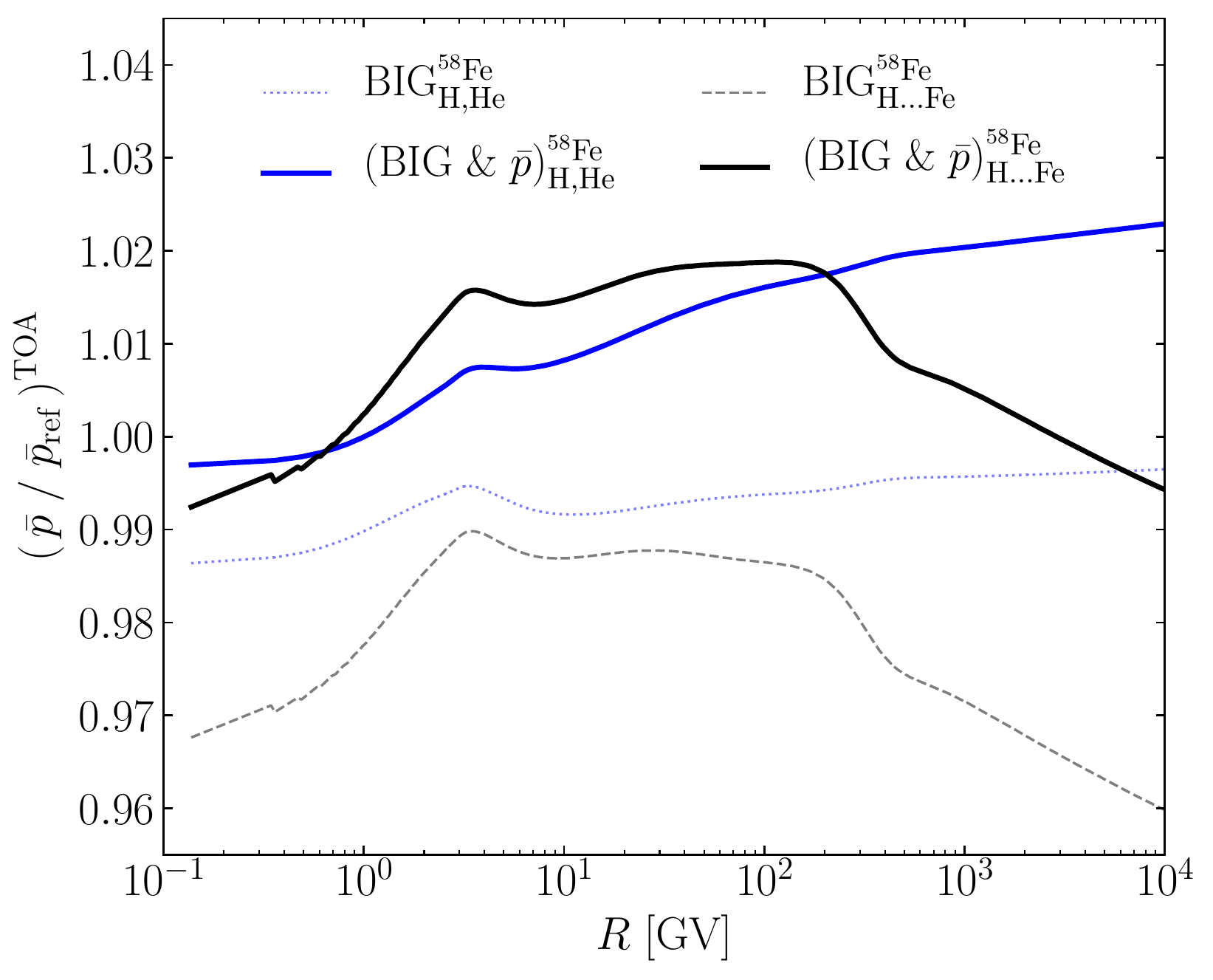}
\caption{Ratio of calculated \pbar{} TOA flux to reference, when considering CR projectiles up to $^{58}$Fe (reference is up to $^{30}$Si) and ISM targets H to Fe (reference is H and He only). The thin lines show---keeping \pbar{} calculated from parents up to $^{30}$Si on H+He only---how the heavy component in CRs (blue dotted line) or both in CRs and ISM (black dashed line) impact the best-fit transport parameters and directly reflect on the ensuing \pbar{}. The solid lines show the ratio between \pbar{} with the heavy component to the one without (transport parameters and \pbar{} calculation consistently done). See text for more details.\label{fig:SM_pbar_vs_heavy}}
\end{figure}

We show in Fig.~\ref{fig:SM_pbar_vs_heavy} the ratio of a given TOA calculation to our reference \pbar{} flux. To better understand each effect, we show separately the impact from the `propagation stage' and from the fully consistent calculation: in the former case, we keep the same CRs and ISM as in the reference for the \pbar{} flux calculation, but use the best-fit transport parameters with `heavy' CRs and/or ISM species, whereas for the latter case, `heavy' CRs and ISM species are self-consistently used at all stages of the calculation.

\begin{itemize}

\item `Propagation stage only' calculation (dashed and dotted thin lines): the thin dotted blue line shows that accounting for CRs up to Fe\footnote{Heavier species are negligible: Ni contributes less than 0.05\% to \pbar{} (see Fig.~\ref{fig:SM_pbarfrac}) and the abundances of $Z\geq30$ elements drop further by several orders of magnitude.} impacts on the transport coefficients---and thus \pbar{}---at the percent level; adding an extra abundance of heavier ISM species (black dashed line) further increases it to $\lesssim4\%$ effect, but with a very different spectral shape. Intuitively, these changes go in the right direction, as opening more channels for boron production means less grammage to get the same B/C data, and thus less \pbar{}'s. The fine details are difficult to track down as they involve a mixture of increased production (CRs and ISM species) balanced by an increased destruction on the heavier ISM species (more efficient both at lower rigidities and for heavier species).

\item `Fully consistent' calculation (thick solid lines): accounting for heavy CRs and heavy ISM species in the transport parameters and in the \pbar{} flux calculation leads to an overall impact of $\sim 1-2\%$, i.e. smaller than in the `propagation stage only' case discussed above. This is not so surprising since as we open more channels for boron production, we also open more channels for \pbar{} production, the two effects mostly cancelling out\footnote{The rigidity-dependence of the extra contributions for nuclei and \pbar{}'s are not exactly the same, leading to different spectral features. For instance, considering heavier CRs (solid blue line) leads to a \pbar{} flux increasing with rigidity up to $2\%$ at 1 TV, likely to be related to the increased fraction of Fe w.r.t. lighter CRs: lighter elements fragment less than heavy ones at low rigidity.}. Finally, accounting for heavier species in the ISM leads to a plateau between 3 and 200 GV ($\sim 2\%$ increase), whose shape is similar to that obtained from the `propagation stage only' calculation (dashed thin black line).

\end{itemize}

There are three important messages to take away from this analysis.
\begin{enumerate}
   \item Reaching a few \% accuracy in the \pbar{} flux calculation, even within the framework of a single propagation model, requires to carefully account for heavy CRs up to Fe, and also to account for heavy species in the ISM, notwithstanding the very large uncertainties on the specific cross sections to consider to do so.

   \item These contributions do not amount to a simple normalisation effect, but have subtle spectral features. If heavy species are not consistently taken into account, they may lead to spectral features at the few percent level.

   \item Because we want to have a calculation as accurate as possible, all our \pbar{} fluxes calculated from `CRs up to $^{30}$Si' and `ISM = H, He' are rescaled by the solid black curve of Fig.~\ref{fig:SM_pbar_vs_heavy}; the analysis presented in the main paper thus accounts for the contribution of all relevant heavy species.
\end{enumerate}

\section{Correlations of data and model uncertainties}
\label{SM:correlations}

\begin{figure}[t]
\includegraphics[width=\columnwidth]{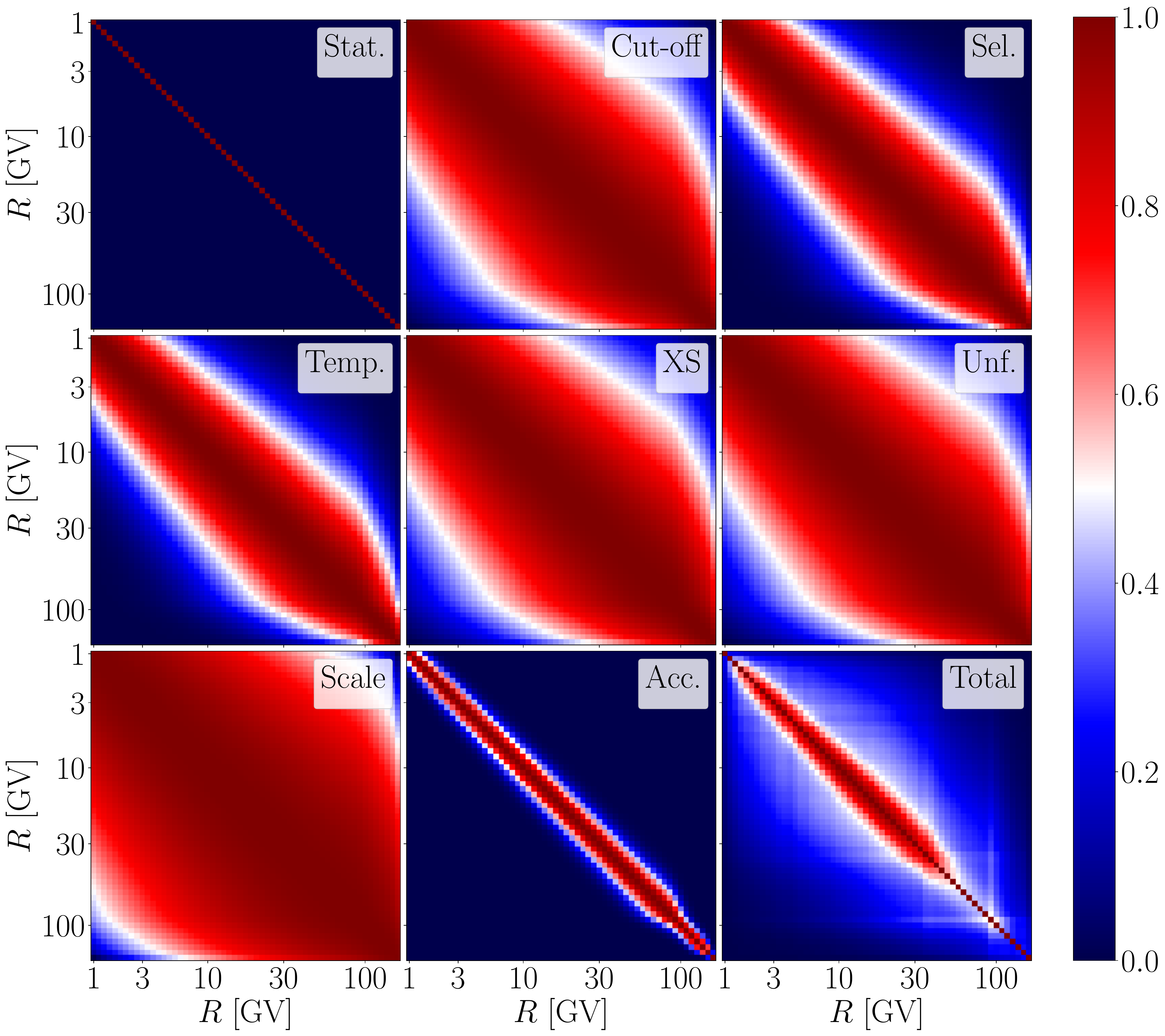}
\caption{
Correlation matrices  between the various rigidity bins for the sources of experimental errors featured in Fig.~1 of the main text. Each component contributes to the covariance matrix ${\cal C}^{\rm data}$. The warmer the color, the larger the correlation. Statistical errors are not correlated. For the next seven panels, the width of the red bands directly reflects the choice of the correlation lengths discussed in the main paper. The last panel (bottom right) shows the data overall correlations (Total).
\label{fig:SM_C_data}}
\end{figure}

To compute the likelihood of the AMS-02 \pbar{} flux, we first assume the data to follow a Gaussian distribution, with covariance matrix ${\cal C}^{\rm data}$, around the theoretical prediction. The various sources of experimental errors, each contributing to ${\cal C}^{\rm data}$, are discussed in the main text. Here, for each of these, we plot in Fig.~\ref{fig:SM_C_data} the corresponding {\it correlation} matrix defined as
\begin{equation}
{\rm c}_{ij}^{\alpha} = {\displaystyle \frac{{\cal C}_{ij}^{\alpha}}{\sqrt{{\cal C}_{ii}^{\alpha}} \sqrt{{\cal C}_{jj}^{\alpha}}}} \,,
\label{eq:definition_correlation}
\end{equation}
where $i$ and $j$ stand for rigidity and ${\cal C}$ is the covariance matrix. We notice that ${\rm c}_{ij}^{\alpha}$ lies in the range from 0 to 1.
The first eight panels show the rigidity correlation of the individual uncertainty contributions. The bottom right panel shows the total correlation matrix for the data, which combines the individual correlation matrices weighted by their associated uncertainties (Fig.~\ref{fig:AMS_errors} of the main text). This latter panel shows that at low- and high-rigidities, the bins are weakly correlated (dominated by statistical uncertainties). At intermediate rigidities, the dominant uncertainties is from the Acc., hence a correlation on short rigidity ranges (bottom centre panel), but contributions from other systematics add significant larger-range correlations (white and light blue shade in bottom right panel).

The predicted fluxes also suffer from uncertainties related to model parameters. The likelihood is built assuming these fluxes also follow a Gaussian distribution, with covariance matrix
\begin{equation}
{\cal C}^{\rm model} = {\displaystyle \sum_{\rm a}} \; {\cal C}^{\rm a} = {\cal C}^{\rm xs}+{\cal C}^{\rm transport}+{\cal C}^{\rm parents} \,,
\label{eq:C_model}
\end{equation}
each of the above-mentioned sources contributing independently to the covariance matrix ${\cal C}^{\rm a}$.
Strictly speaking, predictions would follow a Gaussian distribution, should they vary linearly with respect to Gaussian-distributed model parameters. As errors are small, this idealized case is not far from reality. But cuts on input parameters, such as the Alfv\'enic speed or the convective wind, and the presence of insuperable although small non-linearities, have motivated us to go a step further. We still stick to the Gaussian assumption, but we directly explore the \pbar{} flux behaviour at each rigidity bin $i$ while deriving how different rigidities are correlated.

\begin{figure}[t]
\includegraphics[width=\columnwidth]{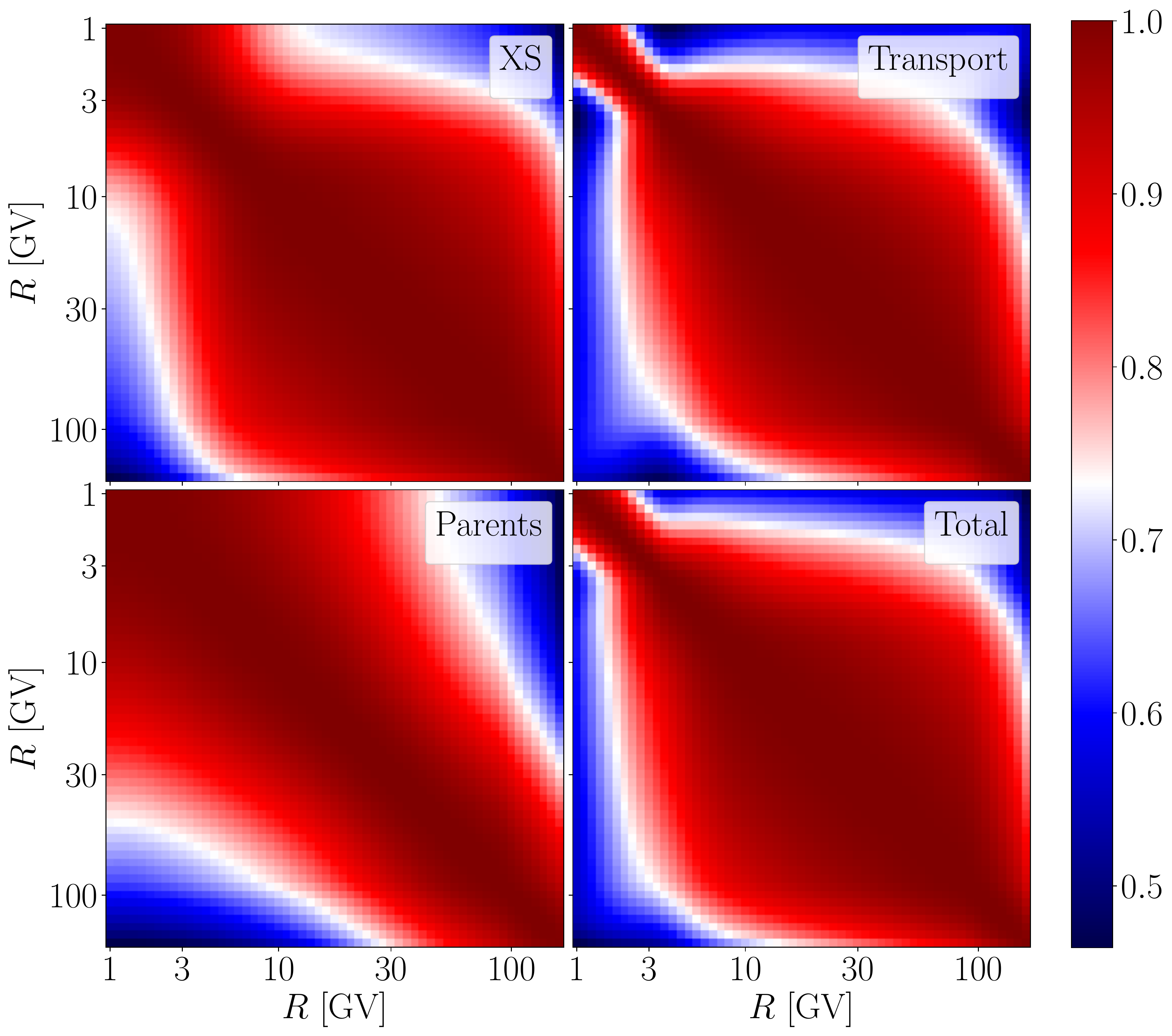}
\caption{
The {\pbar} flux correlation matrices of the theoretical uncertainties arising from {\pbar} production cross section, CR transport and parents are showed in the first three panels. The overall correlation matrix (Total), associated to the covariance matrix ${\cal C}^{\rm model}$ of Eq.~(\ref{eq:C_model}), corresponds to the lower-right panel. The warmer the color, the larger the correlation.
\label{fig:SM_C_model}}
\end{figure}

To do so, for each source  ${\rm a}$ in Eq.~(\ref{eq:C_model}), we first draw randomly $N \sim 10^{4}$ realizations of the input parameters, assuming they are Gaussian-distributed with a covariance matrix which we borrow from previous analyses. In the case of CR transport, for instance, we use the best-fit values and covariance matrix of the best-fit low- and intermediate-rigidity parameters as published in~\cite{Genolini:2019ewc}. For each set $n$ of input parameters, we calculate the \pbar{} flux $\Phi_{i,n}^{\rm a}$ at AMS-02 \pbar{} rigidity $i$. From the resulting distribution, we compute the $1\sigma$ CL flux envelope. We also infer the average prediction $\mu_{i}^{\rm a}$
and covariance matrix
\begin{equation}
{\cal C}^{\rm a}_{ij} = {\displaystyle \frac{1}{N}} \, {\displaystyle \sum_{n=1}^{N}}
\left( \Phi_{i,n}^{\rm a} - \mu_{i}^{\rm a} \right) \left( \Phi_{j,n}^{\rm a} - \mu_{j}^{\rm a} \right) .
\label{eq:def_pbar_C_th}
\end{equation}
The associated correlation matrix ${\rm c}_{ij}^{\rm a}$ is defined as in Eq.~(\ref{eq:definition_correlation}). It is plotted in Fig.~\ref{fig:SM_C_model} for the three sources of model uncertainties, i.e. \pbar{} production cross section, CR transport and parents. Notice how strongly the different rigidity bins are correlated. This plainly justifies the use of covariance matrices, which turn out to be of paramount importance.
The three sources of uncertainties show three different behaviours for their correlation matrices: (i) `Parent' shows a strong correlation between all bins; (ii) `Transport' shows a weaker correlation between low- ($\lesssim 4$~GV) and high-rigidities, probably related to the low-energy break in the transport model which decouples them; (iii) `XS' is in-between, reflecting the various cross section contributions; (iv) `Total', as for the data case, combines the individual correlation matrices weighted by their associated uncertainties (middle panel in Fig.~\ref{fig:AMS_errors} of the main text). It is dominated by `Transport' at low rigidities, it is a mixture of `Transport' and `XS' at intermediate rigidities, and a mixture of `Transport' and 'Parents' at high rigidities. We stress that the total correlation matrix for the model is very different to that of the data (Fig.~\ref{fig:SM_C_data}).

%
%
%
%

\section{The likelihood tests}
\label{SM:likelihood_test}

Our analysis aims at establishing whether or not the \pbar{} flux measured by AMS-02 is compatible with a pure secondary origin. We do not perform any fit of the data. We just gauge how likely they are, once our model for CR transport and secondary \pbar{} production is given. Although our approach is frequentist, priors on the theoretical parameters are incorporated in the definition of the likelihood. The sources of model uncertainties are to be found in \pbar{} production cross section, CR transport and parent fluxes.

The likelihood tests are performed by constructing a $\chi^2$ that accounts for both experimental and theoretical uncertainties, with their correlations. The global covariance matrix is defined as the sum
\begin{equation}
{\cal C}={\cal C}^{\rm data}+{\cal C}^{\rm model} \,,
\end{equation}
while the $\chi^2$ is taken to be
\begin{equation}
\chi^2 = {\displaystyle \sum_{i,j}} \; x_{i} \left( {\cal C}^{-1} \right)_{ij} x_{j} \equiv
{\mathbf x}^{\rm T} {\cal C}^{-1} {\mathbf x} \,,
\end{equation}
where $i$ and $j$ are rigidity indices. The i-th element $x_{i}$ of vector ${\mathbf x}$ is the residuals at rigidity $i$, i.e. the difference between the AMS-02 measurement
and the baseline theoretical prediction
(inferred with the best-fit CR parameters of model {\BIG} of \cite{Genolini:2019ewc} and using full covariance errors for parent fluxes).
%
Once the $\chi^2$ is determined, the $p-$value gauges how good is the agreement between data and baseline model.

A more direct yet qualititive test is provided by the visual inspection of the ${z}$-score which, at rigidity $i$, is defined as $z_{i} = {x_{i}}/{\sqrt{{\cal C}_{ii}}}$. However, in the case of non-vanishing correlations between different rigidity bins, the ${z}$-score could be dangerously misleading. Residuals $x_{i}$ need actually to be rotated in a new basis where the covariance matrix becomes diagonal. Such a rotation is performed with the orthogonal matrix $U$ and yields
\begin{equation}
\tilde{x}_{i} = U_{ij\,} x_{j}
\;\;\text{while}\;\;
\tilde{\cal C} = U \, {\cal C} \, U^{\rm T} \,.
\end{equation}
By construction, the covariance matrix $\tilde{\cal C}$ is diagonal, with elements $\tilde{\cal C}_{ii} = \tilde{\sigma}_{i}^{2}$. The rotated  ${z}$-score, dubbed $\tilde{z}$-score, is defined as $\tilde{z}_{i} = {\tilde{x}_{i}}/{\tilde{\sigma}}_{i}$ and is directly related to the $\chi^2$ through
\begin{equation}
\chi^2 = {\displaystyle \sum_{i}} \; \tilde{z}_{i}^{2} \,.
\label{eq:chi2_z_tilde_relation}
\end{equation}
The index $i$ labels now the pseudo (or rotated) rigidity which we define as
\begin{equation}
\tilde{R}_{i} = {\displaystyle \sum_{j}} \; U_{ij}^{2} \, R_{j} \,,
\end{equation}
with $R_{j}$ the physical rigidity. In the basis of the eigenvectors of the covariance matrix $\tilde{\cal C}$, the rigidity is no longer diagonal. In our case, the rotation is small, with $U$ close to unity. The pseudo rigidity $\tilde{R}_{i}$ is not very different from the physical value $R_{i}$.

Finally, by virtue of Eq.~(\ref{eq:chi2_z_tilde_relation}), the $\tilde{z}$-score can be interpreted as the correct distance between the data and the baseline model. With our assumptions (see above), each value $\tilde{z}_{i}$ is expected to be distributed normally, with mean 0 and variance 1. This motivates us to perform a Kolmogorov-Smirnov (KS) test on the AMS-02 data, to supplement the $\chi^2$ analysis.

\section{Uncertainties from the SLIM/BIG/QUAINT transport configurations}
\label{SM:pbar_vs_model}

In \cite{Genolini:2019ewc}, we introduced three propagation configurations: \SLIM{} (pure diffusion with low- and high-rigidity breaks), \QUAINT{} (`standard' convection/reacceleration transport with high-rigidity break), both limiting cases of the more general configuration \BIG{} (low- and high-rigidity breaks, convection, reacceleration). From a statistical point of view, \BIG{} is slightly preferred by the B/C analysis, but the other two also provide satisfactory $\chi^2/{\rm dof}$ \cite{Genolini:2019ewc}.
Repeating our procedure for the \pbar{} flux calculation (take the best-fit transport parameters, fit high-rigidity break and source parameters on H, He, C, and O data, and calculate \pbar{}), we investigate here how sensitive the results are to the propagation configuration.

%
\begin{figure}[t]
\includegraphics[width=\columnwidth]{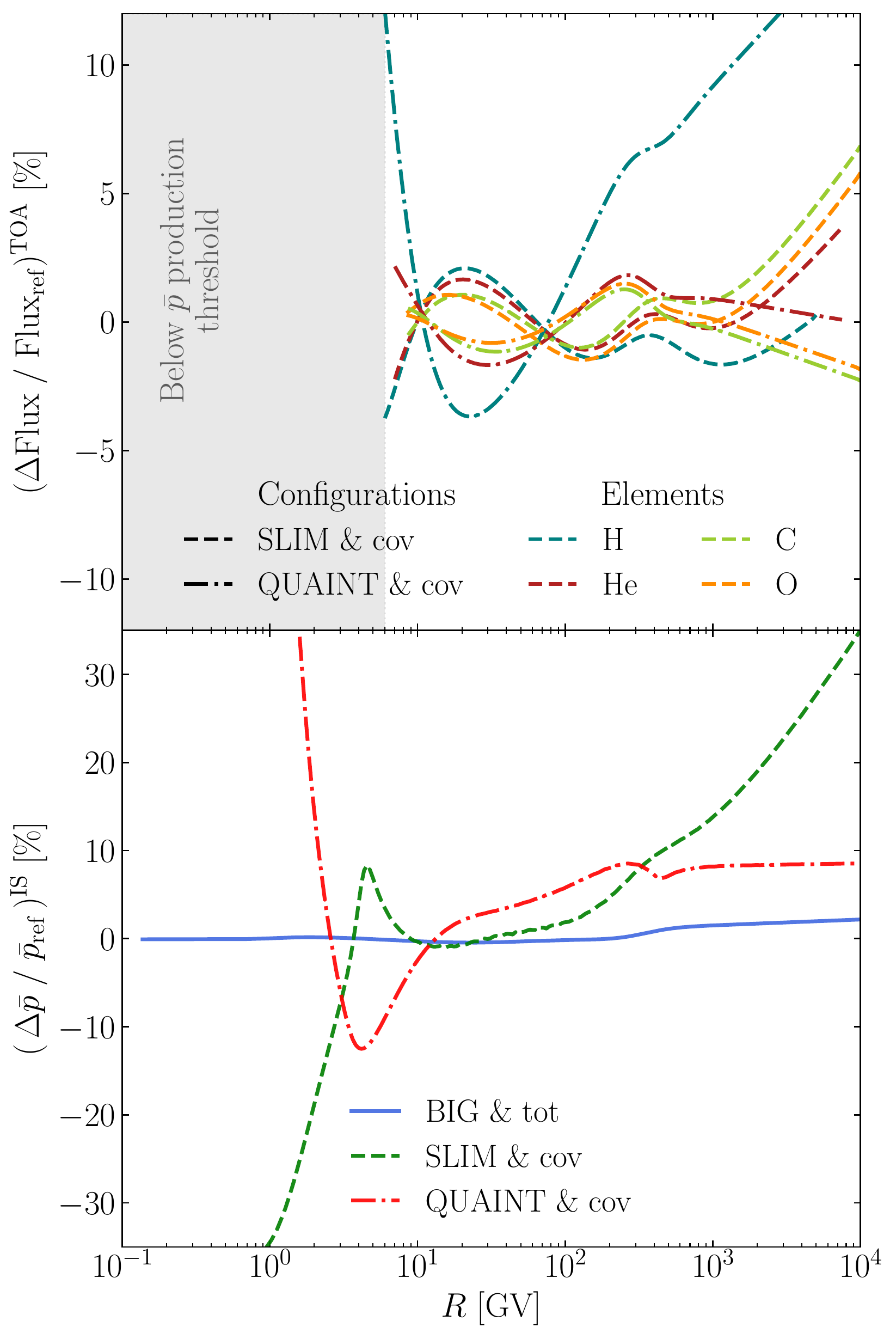}
\caption{Relative variations of fluxes in \SLIM{}, and \QUAINT{} propagation configurations w.r.t. \BIG{}. {\bf Top panel:} best-fit H, He, C, and O TOA fluxes (fit on AMS-02 data). {\bf Bottom panel:} calculated \pbar{} IS flux. See text for discussion.\label{fig:SM_pbar_vs_configs}}
\end{figure}
%

\subsection{\pbar{} from the BIG/SLIM/QUAINT setups}

Figure~\ref{fig:SM_pbar_vs_configs} shows the relative difference w.r.t. \BIG{}, for calculations based on the \SLIM{} (dashed) and \QUAINT{} (dash-dotted) best-fit parameters. The top panel shows the relative difference between the best-fit H (blue), He (red), C (green), and O (orange) TOA fluxes, while the bottom panel shows the relative difference for the corresponding IS \pbar{} flux prediction\footnote{We recall that the different configurations have different best-fit values for $\phi_{\rm FF}$, see Table~\ref{tab:SM_fitsHHeCO}. The  H, He, C, and O fits are performed assuming the same modulation level as obtained in the B/C analysis. See discussion in the main paper.}.
The grey shaded area in the top panel highlights rigidities below the kinetic threshold of \pbar{} production\footnote{The production threshold is $E_k=6 m_p$, which translates into slightly different rigidities for H ($A/Z=1$) and other elements ($A/Z\sim2$), but for simplicity we use a single region for all elements in the plot.}. Both panels share the same rigidity axis, but to interpret these graphs, we recall that \pbar{}'s at rigidity $R$ GV are created from parents whose rigidities are ten to thousand times larger. Indeed, the low-rigidity variation of the \pbar{} flux between configurations can be attributed to the configurations themselves, because the few percent variation on the fit parents above the threshold (top panel) cannot provide the $\gtrsim 10\%$ \pbar{} variation below a few GV. At high rigidity, the $5-10\%$ increase in \pbar{} in the \QUAINT{} setup seems to be related to the rise of the H `bad' fit (discussed in Sec.~\ref{SM:scores}). However, most of the difference observed for the $\bar p$ flux in the \SLIM{} configuration may be related to the different high-rigidity break $\Delta_h$ (0.26 vs 0.17 in \BIG{}, see Table~\ref{tab:SM_fitsHHeCO}). The latter break in the diffusion coefficient appears only once in the formula of primary fluxes, whereas it appears twice for secondary fluxes: for given fluxes of parents (top panel), the high-rigidity slope of the \pbar{} predictions (bottom panel) will differ by $\Delta_h$, asymptotically.

We have two last comments on Fig.~\ref{fig:SM_pbar_vs_configs}. First, all the fits to the H, He, C, and O data make use of the covariance matrices of errors introduced in Sec.~\ref{SM:covmat}, except for the blue solid line shown in the bottom panel. The latter was calculated for the same configuration as for the reference case (\BIG{}), but using for the H, He, C, and O data uncertainties the quadratic sum of statistical and systematic errors. This shows that a better description of the data uncertainties for the \pbar{} parents amounts to a $\sim1-2\%$ impact on the calculated \pbar{} flux. Second, in the range where AMS-02 measured the TOA \pbar{} flux ($1-300$~GV), the calculations differ by $\lesssim 20\%$, meaning that the inter-model variation is already larger or similar to the uncertainties on the \pbar{} flux data. We also stress that the difference is not a pure normalisation one but also a spectral one; this is important to keep in mind as potential percent excesses in \pbar{} data (w.r.t. to the astrophysical prediction) may vanish using another propagation model.

\subsection{Intra- vs inter-model comparison and $z$-scores}

\begin{figure}[!t]
\includegraphics[width=\columnwidth]{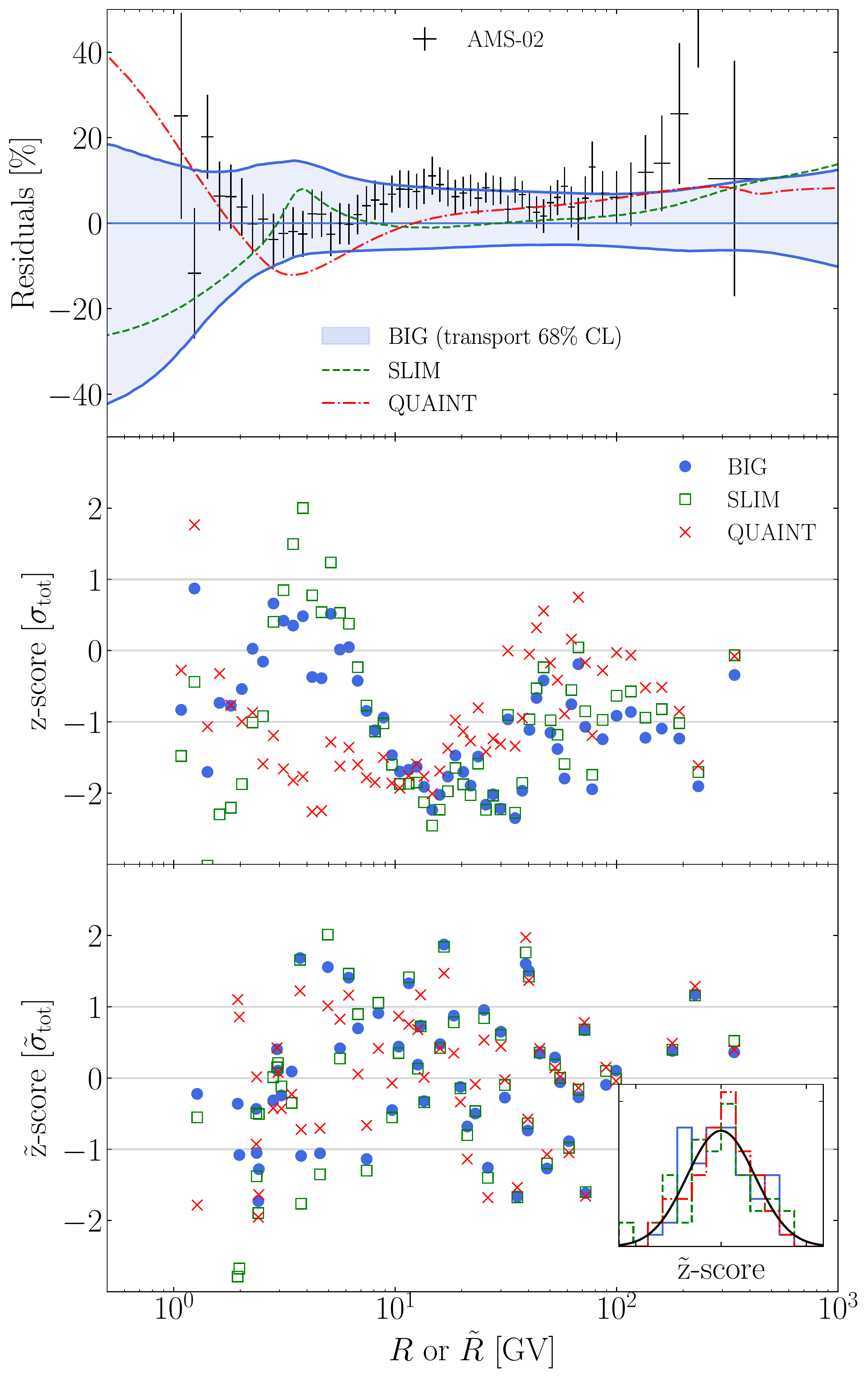}
\caption{Compatibility of TOA \pbar{} data with model calculations. {\bf Top panel:} residuals w.r.t. the \BIG{} configuration for: AMS-02 data (symbols), $68\%$~CL on \BIG{} transport parameters (blue solid lines), the \SLIM{} and the \QUAINT{} configurations (dashed green and dashed-dotted red lines respectively). {\bf Middle panel:} $z$-score of \BIG{}, \SLIM{}, and \QUAINT{} best-fit configurations (blue circles, green squares, and red crosses respectively) against AMS-02 \pbar{} data {\bf Bottom panel:} same but for `rotated' score in the basis where AMS-02 \pbar{} systematics are diagonal. The insert shows the same $\tilde{z}$-scores projected in an histogram, with a $1\sigma$-width Gaussian distribution on top (solid black line). See text for discussion.\label{fig:SM_intervsintra}}
\end{figure}

The top panel of Fig.~\ref{fig:SM_intervsintra} shows the relative difference of various quantities, {\em (config-ref)/ref}, with {\em ref} taken to be our baseline configuration with the transport model \BIG{} and where the covariance matrix of errors of AMS-02 H, He C and O have been used. The symbols show the residuals of AMS-02 data, which have spectral features. However, other propagation configurations (\SLIM{} and \QUAINT{}) have different spectral residuals, highlighting once more that bumps and dips in the residuals should not be taken too seriously. The solid blue lines bracket the $68\%$~CL on the transport uncertainties for \BIG{}. It is worth noting that the intra-model transport uncertainty is at the same level as the inter-model uncertainty. This is not surprising since \SLIM{} and \QUAINT{} are actually subcases of the \BIG{} model (see \cite{Genolini:2019ewc}).

The second and third panels of Fig.~\ref{fig:SM_intervsintra} show the $z$-score in the standard and `rotated' basis respectively (see Sec.~\ref{SM:likelihood_test}). Indeed, the middle panel shows some spectral features (bumps) that the eyes would be eager to interpret as potentially significant DM induced excesses. However, the middle panel does not account properly for correlations in the data between different rigidity bins, whereas the bottom panel does. There is then no significant correlation left, and the inset shows that the $\tilde{z}$-scores distribution is close to the expected normal distribution (solid black line). This shows that the three transport configurations proposed in \cite{Genolini:2019ewc} are all consistent with the \pbar{} data. For a quantitative analysis (using the \BIG{} transport configuration), see the main paper.
%
%
%

\bibliographystyle{apsrev4-1}
\bibliography{pbar}

\begin{thebibliography}{84}%
\makeatletter
\providecommand \@ifxundefined [1]{%
 \@ifx{#1\undefined}
}%
\providecommand \@ifnum [1]{%
 \ifnum #1\expandafter \@firstoftwo
 \else \expandafter \@secondoftwo
 \fi
}%
\providecommand \@ifx [1]{%
 \ifx #1\expandafter \@firstoftwo
 \else \expandafter \@secondoftwo
 \fi
}%
\providecommand \natexlab [1]{#1}%
\providecommand \enquote  [1]{``#1''}%
\providecommand \bibnamefont  [1]{#1}%
\providecommand \bibfnamefont [1]{#1}%
\providecommand \citenamefont [1]{#1}%
\providecommand \href@noop [0]{\@secondoftwo}%
\providecommand \href [0]{\begingroup \@sanitize@url \@href}%
\providecommand \@href[1]{\@@startlink{#1}\@@href}%
\providecommand \@@href[1]{\endgroup#1\@@endlink}%
\providecommand \@sanitize@url [0]{\catcode `\\12\catcode `\$12\catcode
  `\&12\catcode `\#12\catcode `\^12\catcode `\_12\catcode `\%12\relax}%
\providecommand \@@startlink[1]{}%
\providecommand \@@endlink[0]{}%
\providecommand \url  [0]{\begingroup\@sanitize@url \@url }%
\providecommand \@url [1]{\endgroup\@href {#1}{\urlprefix }}%
\providecommand \urlprefix  [0]{URL }%
\providecommand \Eprint [0]{\href }%
\providecommand \doibase [0]{http://dx.doi.org/}%
\providecommand \selectlanguage [0]{\@gobble}%
\providecommand \bibinfo  [0]{\@secondoftwo}%
\providecommand \bibfield  [0]{\@secondoftwo}%
\providecommand \translation [1]{[#1]}%
\providecommand \BibitemOpen [0]{}%
\providecommand \bibitemStop [0]{}%
\providecommand \bibitemNoStop [0]{.\EOS\space}%
\providecommand \EOS [0]{\spacefactor3000\relax}%
\providecommand \BibitemShut  [1]{\csname bibitem#1\endcsname}%
\let\auto@bib@innerbib\@empty
\bibitem [{\citenamefont {{Rosen}}(1967)}]{1967PhRv..158.1227R}%
  \BibitemOpen
  \bibfield  {author} {\bibinfo {author} {\bibfnamefont {S.}~\bibnamefont
  {{Rosen}}},\ }\href {\doibase 10.1103/PhysRev.158.1227} {\bibfield  {journal}
  {\bibinfo  {journal} {Physical Review}\ }\textbf {\bibinfo {volume} {158}},\
  \bibinfo {pages} {1227} (\bibinfo {year} {1967})}\BibitemShut {NoStop}%
\bibitem [{\citenamefont {{Shen}}\ and\ \citenamefont
  {{Berkey}}(1968)}]{1968PhRv..171.1344S}%
  \BibitemOpen
  \bibfield  {author} {\bibinfo {author} {\bibfnamefont {C.~S.}\ \bibnamefont
  {{Shen}}}\ and\ \bibinfo {author} {\bibfnamefont {G.~B.}\ \bibnamefont
  {{Berkey}}},\ }\href {\doibase 10.1103/PhysRev.171.1344} {\bibfield
  {journal} {\bibinfo  {journal} {Physical Review}\ }\textbf {\bibinfo {volume}
  {171}},\ \bibinfo {pages} {1344} (\bibinfo {year} {1968})}\BibitemShut
  {NoStop}%
\bibitem [{\citenamefont {{Bogomolov}}\ \emph {et~al.}(1981)\citenamefont
  {{Bogomolov}}, \citenamefont {{Lubianaia}}, \citenamefont {{Romanov}},
  \citenamefont {{Stepanov}},\ and\ \citenamefont
  {{Shulakova}}}]{1981ICRC....9..146B}%
  \BibitemOpen
  \bibfield  {author} {\bibinfo {author} {\bibfnamefont {E.~A.}\ \bibnamefont
  {{Bogomolov}}}, \bibinfo {author} {\bibfnamefont {N.~D.}\ \bibnamefont
  {{Lubianaia}}}, \bibinfo {author} {\bibfnamefont {V.~A.}\ \bibnamefont
  {{Romanov}}}, \bibinfo {author} {\bibfnamefont {S.~V.}\ \bibnamefont
  {{Stepanov}}}, \ and\ \bibinfo {author} {\bibfnamefont {M.~S.}\ \bibnamefont
  {{Shulakova}}},\ }\href@noop {} {\bibfield  {journal} {\bibinfo  {journal}
  {International Cosmic Ray Conference}\ }\textbf {\bibinfo {volume} {9}},\
  \bibinfo {pages} {146} (\bibinfo {year} {1981})}\BibitemShut {NoStop}%
\bibitem [{\citenamefont {{Buffington}}\ \emph {et~al.}(1981)\citenamefont
  {{Buffington}}, \citenamefont {{Schindler}},\ and\ \citenamefont
  {{Pennypacker}}}]{1981ApJ...248.1179B}%
  \BibitemOpen
  \bibfield  {author} {\bibinfo {author} {\bibfnamefont {A.}~\bibnamefont
  {{Buffington}}}, \bibinfo {author} {\bibfnamefont {S.~M.}\ \bibnamefont
  {{Schindler}}}, \ and\ \bibinfo {author} {\bibfnamefont {C.~R.}\ \bibnamefont
  {{Pennypacker}}},\ }\href {\doibase 10.1086/159247} {\bibfield  {journal}
  {\bibinfo  {journal} {\apj}\ }\textbf {\bibinfo {volume} {248}},\ \bibinfo
  {pages} {1179} (\bibinfo {year} {1981})}\BibitemShut {NoStop}%
\bibitem [{\citenamefont {{Kiraly}}\ \emph {et~al.}(1981)\citenamefont
  {{Kiraly}}, \citenamefont {{Szabelski}}, \citenamefont {{Wdowczyk}},\ and\
  \citenamefont {{Wolfendale}}}]{1981Natur.293..120K}%
  \BibitemOpen
  \bibfield  {author} {\bibinfo {author} {\bibfnamefont {P.}~\bibnamefont
  {{Kiraly}}}, \bibinfo {author} {\bibfnamefont {J.}~\bibnamefont
  {{Szabelski}}}, \bibinfo {author} {\bibfnamefont {J.}~\bibnamefont
  {{Wdowczyk}}}, \ and\ \bibinfo {author} {\bibfnamefont {A.~W.}\ \bibnamefont
  {{Wolfendale}}},\ }\href {\doibase 10.1038/293120a0} {\bibfield  {journal}
  {\bibinfo  {journal} {\nat}\ }\textbf {\bibinfo {volume} {293}},\ \bibinfo
  {pages} {120} (\bibinfo {year} {1981})}\BibitemShut {NoStop}%
\bibitem [{\citenamefont {{Silk}}\ and\ \citenamefont
  {{Srednicki}}(1984)}]{1984PhRvL..53..624S}%
  \BibitemOpen
  \bibfield  {author} {\bibinfo {author} {\bibfnamefont {J.}~\bibnamefont
  {{Silk}}}\ and\ \bibinfo {author} {\bibfnamefont {M.}~\bibnamefont
  {{Srednicki}}},\ }\href {\doibase 10.1103/PhysRevLett.53.624} {\bibfield
  {journal} {\bibinfo  {journal} {Physical Review Letters}\ }\textbf {\bibinfo
  {volume} {53}},\ \bibinfo {pages} {624} (\bibinfo {year} {1984})}\BibitemShut
  {NoStop}%
\bibitem [{\citenamefont {{Stecker}}\ \emph {et~al.}(1985)\citenamefont
  {{Stecker}}, \citenamefont {{Rudaz}},\ and\ \citenamefont
  {{Walsh}}}]{1985PhRvL..55.2622S}%
  \BibitemOpen
  \bibfield  {author} {\bibinfo {author} {\bibfnamefont {F.~W.}\ \bibnamefont
  {{Stecker}}}, \bibinfo {author} {\bibfnamefont {S.}~\bibnamefont {{Rudaz}}},
  \ and\ \bibinfo {author} {\bibfnamefont {T.~F.}\ \bibnamefont {{Walsh}}},\
  }\href {\doibase 10.1103/PhysRevLett.55.2622} {\bibfield  {journal} {\bibinfo
   {journal} {Physical Review Letters}\ }\textbf {\bibinfo {volume} {55}},\
  \bibinfo {pages} {2622} (\bibinfo {year} {1985})}\BibitemShut {NoStop}%
\bibitem [{\citenamefont {{Hof}}\ \emph {et~al.}(1996)\citenamefont {{Hof}},
  \citenamefont {{Menn}}, \citenamefont {{Pfeifer}}, \citenamefont {{Simon}},
  \citenamefont {{Golden}}, \citenamefont {{Stochaj}}, \citenamefont
  {{Stephens}}, \citenamefont {{Basini}}, \citenamefont {{Ricci}},
  \citenamefont {{Brancaccio}} \emph {et~al.}}]{1996ApJ...467L..33H}%
  \BibitemOpen
  \bibfield  {author} {\bibinfo {author} {\bibfnamefont {M.}~\bibnamefont
  {{Hof}}}, \bibinfo {author} {\bibfnamefont {W.}~\bibnamefont {{Menn}}},
  \bibinfo {author} {\bibfnamefont {C.}~\bibnamefont {{Pfeifer}}}, \bibinfo
  {author} {\bibfnamefont {M.}~\bibnamefont {{Simon}}}, \bibinfo {author}
  {\bibfnamefont {R.~L.}\ \bibnamefont {{Golden}}}, \bibinfo {author}
  {\bibfnamefont {S.~J.}\ \bibnamefont {{Stochaj}}}, \bibinfo {author}
  {\bibfnamefont {S.~A.}\ \bibnamefont {{Stephens}}}, \bibinfo {author}
  {\bibfnamefont {G.}~\bibnamefont {{Basini}}}, \bibinfo {author}
  {\bibfnamefont {M.}~\bibnamefont {{Ricci}}}, \bibinfo {author} {\bibfnamefont
  {F.~M.}\ \bibnamefont {{Brancaccio}}},  \emph {et~al.},\ }\href {\doibase
  10.1086/310185} {\bibfield  {journal} {\bibinfo  {journal} {\apjl}\ }\textbf
  {\bibinfo {volume} {467}},\ \bibinfo {pages} {L33} (\bibinfo {year}
  {1996})}\BibitemShut {NoStop}%
\bibitem [{\citenamefont {{Mitchell}}\ \emph {et~al.}(1996)\citenamefont
  {{Mitchell}}, \citenamefont {{Barbier}}, \citenamefont {{Christian}},
  \citenamefont {{Krizmanic}}, \citenamefont {{Krombel}}, \citenamefont
  {{Ormes}}, \citenamefont {{Streitmatter}}, \citenamefont {{Labrador}},
  \citenamefont {{Davis}}, \citenamefont {{Mewaldt}} \emph
  {et~al.}}]{1996PhRvL..76.3057M}%
  \BibitemOpen
  \bibfield  {author} {\bibinfo {author} {\bibfnamefont {J.~W.}\ \bibnamefont
  {{Mitchell}}}, \bibinfo {author} {\bibfnamefont {L.~M.}\ \bibnamefont
  {{Barbier}}}, \bibinfo {author} {\bibfnamefont {E.~R.}\ \bibnamefont
  {{Christian}}}, \bibinfo {author} {\bibfnamefont {J.~F.}\ \bibnamefont
  {{Krizmanic}}}, \bibinfo {author} {\bibfnamefont {K.}~\bibnamefont
  {{Krombel}}}, \bibinfo {author} {\bibfnamefont {J.~F.}\ \bibnamefont
  {{Ormes}}}, \bibinfo {author} {\bibfnamefont {R.~E.}\ \bibnamefont
  {{Streitmatter}}}, \bibinfo {author} {\bibfnamefont {A.~W.}\ \bibnamefont
  {{Labrador}}}, \bibinfo {author} {\bibfnamefont {A.~J.}\ \bibnamefont
  {{Davis}}}, \bibinfo {author} {\bibfnamefont {R.~A.}\ \bibnamefont
  {{Mewaldt}}},  \emph {et~al.},\ }\href {\doibase 10.1103/PhysRevLett.76.3057}
  {\bibfield  {journal} {\bibinfo  {journal} {\prl}\ }\textbf {\bibinfo
  {volume} {76}},\ \bibinfo {pages} {3057} (\bibinfo {year}
  {1996})}\BibitemShut {NoStop}%
\bibitem [{\citenamefont {{Boezio}}\ \emph {et~al.}(1997)\citenamefont
  {{Boezio}}, \citenamefont {{Carlson}}, \citenamefont {{Francke}},
  \citenamefont {{Weber}}, \citenamefont {{Suffert}}, \citenamefont {{Hof}},
  \citenamefont {{Menn}}, \citenamefont {{Simon}}, \citenamefont {{Stephens}},
  \citenamefont {{Bellotti}} \emph {et~al.}}]{1997ApJ...487..415B}%
  \BibitemOpen
  \bibfield  {author} {\bibinfo {author} {\bibfnamefont {M.}~\bibnamefont
  {{Boezio}}}, \bibinfo {author} {\bibfnamefont {P.}~\bibnamefont {{Carlson}}},
  \bibinfo {author} {\bibfnamefont {T.}~\bibnamefont {{Francke}}}, \bibinfo
  {author} {\bibfnamefont {N.}~\bibnamefont {{Weber}}}, \bibinfo {author}
  {\bibfnamefont {M.}~\bibnamefont {{Suffert}}}, \bibinfo {author}
  {\bibfnamefont {M.}~\bibnamefont {{Hof}}}, \bibinfo {author} {\bibfnamefont
  {W.}~\bibnamefont {{Menn}}}, \bibinfo {author} {\bibfnamefont
  {M.}~\bibnamefont {{Simon}}}, \bibinfo {author} {\bibfnamefont {S.~A.}\
  \bibnamefont {{Stephens}}}, \bibinfo {author} {\bibfnamefont
  {R.}~\bibnamefont {{Bellotti}}},  \emph {et~al.},\ }\href {\doibase
  10.1086/304593} {\bibfield  {journal} {\bibinfo  {journal} {\apj}\ }\textbf
  {\bibinfo {volume} {487}},\ \bibinfo {pages} {415} (\bibinfo {year}
  {1997})}\BibitemShut {NoStop}%
\bibitem [{\citenamefont {{Boezio}}\ \emph {et~al.}(2001)\citenamefont
  {{Boezio}}, \citenamefont {{Bonvicini}}, \citenamefont {{Schiavon}},
  \citenamefont {{Vacchi}}, \citenamefont {{Zampa}}, \citenamefont
  {{Bergstr{\"o}m}}, \citenamefont {{Carlson}}, \citenamefont {{Francke}},
  \citenamefont {{Grinstein}}, \citenamefont {{Suffert}} \emph
  {et~al.}}]{2001ApJ...561..787B}%
  \BibitemOpen
  \bibfield  {author} {\bibinfo {author} {\bibfnamefont {M.}~\bibnamefont
  {{Boezio}}}, \bibinfo {author} {\bibfnamefont {V.}~\bibnamefont
  {{Bonvicini}}}, \bibinfo {author} {\bibfnamefont {P.}~\bibnamefont
  {{Schiavon}}}, \bibinfo {author} {\bibfnamefont {A.}~\bibnamefont
  {{Vacchi}}}, \bibinfo {author} {\bibfnamefont {N.}~\bibnamefont {{Zampa}}},
  \bibinfo {author} {\bibfnamefont {D.}~\bibnamefont {{Bergstr{\"o}m}}},
  \bibinfo {author} {\bibfnamefont {P.}~\bibnamefont {{Carlson}}}, \bibinfo
  {author} {\bibfnamefont {T.}~\bibnamefont {{Francke}}}, \bibinfo {author}
  {\bibfnamefont {S.}~\bibnamefont {{Grinstein}}}, \bibinfo {author}
  {\bibfnamefont {M.}~\bibnamefont {{Suffert}}},  \emph {et~al.},\ }\href
  {\doibase 10.1086/323366} {\bibfield  {journal} {\bibinfo  {journal} {\apj}\
  }\textbf {\bibinfo {volume} {561}},\ \bibinfo {pages} {787} (\bibinfo {year}
  {2001})},\ \Eprint {http://arxiv.org/abs/arXiv:astro-ph/0103513}
  {arXiv:astro-ph/0103513} \BibitemShut {NoStop}%
\bibitem [{\citenamefont {{Beach}}\ \emph {et~al.}(2001)\citenamefont
  {{Beach}}, \citenamefont {{Beatty}}, \citenamefont {{Bhattacharyya}},
  \citenamefont {{Bower}}, \citenamefont {{Coutu}}, \citenamefont
  {{Duvernois}}, \citenamefont {{Labrador}}, \citenamefont {{McKee}},
  \citenamefont {{Minnick}}, \citenamefont {{M{\"u}ller}}, \citenamefont
  {{Musser}}, \citenamefont {{Nutter}}, \citenamefont {{Schubnell}},
  \citenamefont {{Swordy}}, \citenamefont {{Tarl{\'e}}},\ and\ \citenamefont
  {{Tomasch}}}]{2001PhRvL..87A1101B}%
  \BibitemOpen
  \bibfield  {author} {\bibinfo {author} {\bibfnamefont {A.~S.}\ \bibnamefont
  {{Beach}}}, \bibinfo {author} {\bibfnamefont {J.~J.}\ \bibnamefont
  {{Beatty}}}, \bibinfo {author} {\bibfnamefont {A.}~\bibnamefont
  {{Bhattacharyya}}}, \bibinfo {author} {\bibfnamefont {C.}~\bibnamefont
  {{Bower}}}, \bibinfo {author} {\bibfnamefont {S.}~\bibnamefont {{Coutu}}},
  \bibinfo {author} {\bibfnamefont {M.~A.}\ \bibnamefont {{Duvernois}}},
  \bibinfo {author} {\bibfnamefont {A.~W.}\ \bibnamefont {{Labrador}}},
  \bibinfo {author} {\bibfnamefont {S.}~\bibnamefont {{McKee}}}, \bibinfo
  {author} {\bibfnamefont {S.~A.}\ \bibnamefont {{Minnick}}}, \bibinfo {author}
  {\bibfnamefont {D.}~\bibnamefont {{M{\"u}ller}}}, \bibinfo {author}
  {\bibfnamefont {J.}~\bibnamefont {{Musser}}}, \bibinfo {author}
  {\bibfnamefont {S.}~\bibnamefont {{Nutter}}}, \bibinfo {author}
  {\bibfnamefont {M.}~\bibnamefont {{Schubnell}}}, \bibinfo {author}
  {\bibfnamefont {S.}~\bibnamefont {{Swordy}}}, \bibinfo {author}
  {\bibfnamefont {G.}~\bibnamefont {{Tarl{\'e}}}}, \ and\ \bibinfo {author}
  {\bibfnamefont {A.}~\bibnamefont {{Tomasch}}},\ }\href {\doibase
  10.1103/PhysRevLett.87.271101} {\bibfield  {journal} {\bibinfo  {journal}
  {Physical Review Letters}\ }\textbf {\bibinfo {volume} {87}},\ \bibinfo {eid}
  {271101} (\bibinfo {year} {2001})},\ \Eprint
  {http://arxiv.org/abs/arXiv:astro-ph/0111094} {arXiv:astro-ph/0111094}
  \BibitemShut {NoStop}%
\bibitem [{\citenamefont {{Moiseev}}\ \emph {et~al.}(1997)\citenamefont
  {{Moiseev}}, \citenamefont {{Yoshimura}}, \citenamefont {{Ueda}},
  \citenamefont {{Anraku}}, \citenamefont {{Golden}}, \citenamefont {{Imori}},
  \citenamefont {{Inaba}}, \citenamefont {{Kimball}}, \citenamefont {{Kimura}},
  \citenamefont {{Makida}},\ and\ \citenamefont {{BESS
  Collaboration}}}]{1997ApJ...474..479M}%
  \BibitemOpen
  \bibfield  {author} {\bibinfo {author} {\bibfnamefont {A.}~\bibnamefont
  {{Moiseev}}}, \bibinfo {author} {\bibfnamefont {K.}~\bibnamefont
  {{Yoshimura}}}, \bibinfo {author} {\bibfnamefont {I.}~\bibnamefont {{Ueda}}},
  \bibinfo {author} {\bibfnamefont {K.}~\bibnamefont {{Anraku}}}, \bibinfo
  {author} {\bibfnamefont {R.}~\bibnamefont {{Golden}}}, \bibinfo {author}
  {\bibfnamefont {M.}~\bibnamefont {{Imori}}}, \bibinfo {author} {\bibfnamefont
  {S.}~\bibnamefont {{Inaba}}}, \bibinfo {author} {\bibfnamefont
  {B.}~\bibnamefont {{Kimball}}}, \bibinfo {author} {\bibfnamefont
  {N.}~\bibnamefont {{Kimura}}}, \bibinfo {author} {\bibfnamefont
  {Y.}~\bibnamefont {{Makida}}}, \ and\ \bibinfo {author} {\bibnamefont {{BESS
  Collaboration}}},\ }\href {\doibase 10.1086/303463} {\bibfield  {journal}
  {\bibinfo  {journal} {\apj}\ }\textbf {\bibinfo {volume} {474}},\ \bibinfo
  {pages} {479} (\bibinfo {year} {1997})}\BibitemShut {NoStop}%
\bibitem [{\citenamefont {{Matsunaga}}\ \emph {et~al.}(1998)\citenamefont
  {{Matsunaga}}, \citenamefont {{Orito}}, \citenamefont {{Matsumoto}},
  \citenamefont {{Yoshimura}}, \citenamefont {{Moiseev}}, \citenamefont
  {{Anraku}}, \citenamefont {{Golden}}, \citenamefont {{Imori}}, \citenamefont
  {{Makida}}, \citenamefont {{Mitchell}} \emph {et~al.}}]{1998PhRvL..81.4052M}%
  \BibitemOpen
  \bibfield  {author} {\bibinfo {author} {\bibfnamefont {H.}~\bibnamefont
  {{Matsunaga}}}, \bibinfo {author} {\bibfnamefont {S.}~\bibnamefont
  {{Orito}}}, \bibinfo {author} {\bibfnamefont {H.}~\bibnamefont
  {{Matsumoto}}}, \bibinfo {author} {\bibfnamefont {K.}~\bibnamefont
  {{Yoshimura}}}, \bibinfo {author} {\bibfnamefont {A.}~\bibnamefont
  {{Moiseev}}}, \bibinfo {author} {\bibfnamefont {K.}~\bibnamefont {{Anraku}}},
  \bibinfo {author} {\bibfnamefont {R.}~\bibnamefont {{Golden}}}, \bibinfo
  {author} {\bibfnamefont {M.}~\bibnamefont {{Imori}}}, \bibinfo {author}
  {\bibfnamefont {Y.}~\bibnamefont {{Makida}}}, \bibinfo {author}
  {\bibfnamefont {J.}~\bibnamefont {{Mitchell}}},  \emph {et~al.},\ }\href
  {\doibase 10.1103/PhysRevLett.81.4052} {\bibfield  {journal} {\bibinfo
  {journal} {\prl}\ }\textbf {\bibinfo {volume} {81}},\ \bibinfo {pages} {4052}
  (\bibinfo {year} {1998})},\ \Eprint
  {http://arxiv.org/abs/arXiv:astro-ph/9809326} {arXiv:astro-ph/9809326}
  \BibitemShut {NoStop}%
\bibitem [{\citenamefont {{Orito}}\ \emph {et~al.}(2000)\citenamefont
  {{Orito}}, \citenamefont {{Maeno}}, \citenamefont {{Matsunaga}},
  \citenamefont {{Abe}}, \citenamefont {{Anraku}}, \citenamefont {{Asaoka}},
  \citenamefont {{Fujikawa}}, \citenamefont {{Imori}}, \citenamefont
  {{Ishino}}, \citenamefont {{Makida}} \emph {et~al.}}]{2000PhRvL..84.1078O}%
  \BibitemOpen
  \bibfield  {author} {\bibinfo {author} {\bibfnamefont {S.}~\bibnamefont
  {{Orito}}}, \bibinfo {author} {\bibfnamefont {T.}~\bibnamefont {{Maeno}}},
  \bibinfo {author} {\bibfnamefont {H.}~\bibnamefont {{Matsunaga}}}, \bibinfo
  {author} {\bibfnamefont {K.}~\bibnamefont {{Abe}}}, \bibinfo {author}
  {\bibfnamefont {K.}~\bibnamefont {{Anraku}}}, \bibinfo {author}
  {\bibfnamefont {Y.}~\bibnamefont {{Asaoka}}}, \bibinfo {author}
  {\bibfnamefont {M.}~\bibnamefont {{Fujikawa}}}, \bibinfo {author}
  {\bibfnamefont {M.}~\bibnamefont {{Imori}}}, \bibinfo {author} {\bibfnamefont
  {M.}~\bibnamefont {{Ishino}}}, \bibinfo {author} {\bibfnamefont
  {Y.}~\bibnamefont {{Makida}}},  \emph {et~al.},\ }\href {\doibase
  10.1103/PhysRevLett.84.1078} {\bibfield  {journal} {\bibinfo  {journal}
  {\prl}\ }\textbf {\bibinfo {volume} {84}},\ \bibinfo {pages} {1078} (\bibinfo
  {year} {2000})},\ \Eprint {http://arxiv.org/abs/arXiv:astro-ph/9906426}
  {arXiv:astro-ph/9906426} \BibitemShut {NoStop}%
\bibitem [{\citenamefont {{Maeno}}\ \emph {et~al.}(2001)\citenamefont
  {{Maeno}}, \citenamefont {{Orito}}, \citenamefont {{Matsunaga}},
  \citenamefont {{Abe}}, \citenamefont {{Anraku}}, \citenamefont {{Asaoka}},
  \citenamefont {{Fujikawa}}, \citenamefont {{Imori}}, \citenamefont
  {{Makida}}, \citenamefont {{Matsui}} \emph {et~al.}}]{2001APh....16..121M}%
  \BibitemOpen
  \bibfield  {author} {\bibinfo {author} {\bibfnamefont {T.}~\bibnamefont
  {{Maeno}}}, \bibinfo {author} {\bibfnamefont {S.}~\bibnamefont {{Orito}}},
  \bibinfo {author} {\bibfnamefont {H.}~\bibnamefont {{Matsunaga}}}, \bibinfo
  {author} {\bibfnamefont {K.}~\bibnamefont {{Abe}}}, \bibinfo {author}
  {\bibfnamefont {K.}~\bibnamefont {{Anraku}}}, \bibinfo {author}
  {\bibfnamefont {Y.}~\bibnamefont {{Asaoka}}}, \bibinfo {author}
  {\bibfnamefont {M.}~\bibnamefont {{Fujikawa}}}, \bibinfo {author}
  {\bibfnamefont {M.}~\bibnamefont {{Imori}}}, \bibinfo {author} {\bibfnamefont
  {Y.}~\bibnamefont {{Makida}}}, \bibinfo {author} {\bibfnamefont
  {N.}~\bibnamefont {{Matsui}}},  \emph {et~al.},\ }\href {\doibase
  10.1016/S0927-6505(01)00107-4} {\bibfield  {journal} {\bibinfo  {journal}
  {Astroparticle Physics}\ }\textbf {\bibinfo {volume} {16}},\ \bibinfo {pages}
  {121} (\bibinfo {year} {2001})},\ \Eprint
  {http://arxiv.org/abs/arXiv:astro-ph/0010381} {arXiv:astro-ph/0010381}
  \BibitemShut {NoStop}%
\bibitem [{\citenamefont {{Asaoka}}\ \emph {et~al.}(2002)\citenamefont
  {{Asaoka}}, \citenamefont {{Shikaze}}, \citenamefont {{Abe}} \emph
  {et~al.}}]{2002PhRvL..88e1101A}%
  \BibitemOpen
  \bibfield  {author} {\bibinfo {author} {\bibfnamefont {Y.}~\bibnamefont
  {{Asaoka}}}, \bibinfo {author} {\bibfnamefont {Y.}~\bibnamefont {{Shikaze}}},
  \bibinfo {author} {\bibfnamefont {K.}~\bibnamefont {{Abe}}},  \emph
  {et~al.},\ }\href {\doibase 10.1103/PhysRevLett.88.051101} {\bibfield
  {journal} {\bibinfo  {journal} {\prl}\ }\textbf {\bibinfo {volume} {88}},\
  \bibinfo {eid} {051101} (\bibinfo {year} {2002})},\ \Eprint
  {http://arxiv.org/abs/arXiv:astro-ph/0109007} {arXiv:astro-ph/0109007}
  \BibitemShut {NoStop}%
\bibitem [{\citenamefont {{AMS Collaboration}}\ \emph
  {et~al.}(2002)\citenamefont {{AMS Collaboration}}, \citenamefont {{Aguilar}},
  \citenamefont {{Alcaraz}}, \citenamefont {{Allaby}}, \citenamefont {{Alpat}},
  \citenamefont {{Ambrosi}}, \citenamefont {{Anderhub}}, \citenamefont {{Ao}},
  \citenamefont {{Arefiev}}, \citenamefont {{Azzarello}},\ and\ \citenamefont
  {et~al.}}]{2002PhR...366..331A}%
  \BibitemOpen
  \bibfield  {author} {\bibinfo {author} {\bibnamefont {{AMS Collaboration}}},
  \bibinfo {author} {\bibfnamefont {M.}~\bibnamefont {{Aguilar}}}, \bibinfo
  {author} {\bibfnamefont {J.}~\bibnamefont {{Alcaraz}}}, \bibinfo {author}
  {\bibfnamefont {J.}~\bibnamefont {{Allaby}}}, \bibinfo {author}
  {\bibfnamefont {B.}~\bibnamefont {{Alpat}}}, \bibinfo {author} {\bibfnamefont
  {G.}~\bibnamefont {{Ambrosi}}}, \bibinfo {author} {\bibfnamefont
  {H.}~\bibnamefont {{Anderhub}}}, \bibinfo {author} {\bibfnamefont
  {L.}~\bibnamefont {{Ao}}}, \bibinfo {author} {\bibfnamefont {A.}~\bibnamefont
  {{Arefiev}}}, \bibinfo {author} {\bibfnamefont {P.}~\bibnamefont
  {{Azzarello}}}, \ and\ \bibinfo {author} {\bibnamefont {et~al.}},\ }\href
  {\doibase 10.1016/S0370-1573(02)00013-3} {\bibfield  {journal} {\bibinfo
  {journal} {\physrep}\ }\textbf {\bibinfo {volume} {366}},\ \bibinfo {pages}
  {331} (\bibinfo {year} {2002})}\BibitemShut {NoStop}%
\bibitem [{\citenamefont {{Tan}}\ and\ \citenamefont
  {{Ng}}(1983)}]{1983JPhG....9..227T}%
  \BibitemOpen
  \bibfield  {author} {\bibinfo {author} {\bibfnamefont {L.~C.}\ \bibnamefont
  {{Tan}}}\ and\ \bibinfo {author} {\bibfnamefont {L.~K.}\ \bibnamefont
  {{Ng}}},\ }\href {\doibase 10.1088/0305-4616/9/2/015} {\bibfield  {journal}
  {\bibinfo  {journal} {Journal of Physics G Nuclear Physics}\ }\textbf
  {\bibinfo {volume} {9}},\ \bibinfo {pages} {227} (\bibinfo {year}
  {1983})}\BibitemShut {NoStop}%
\bibitem [{\citenamefont {{Gaisser}}\ and\ \citenamefont
  {{Schaefer}}(1992)}]{1992ApJ...394..174G}%
  \BibitemOpen
  \bibfield  {author} {\bibinfo {author} {\bibfnamefont {T.~K.}\ \bibnamefont
  {{Gaisser}}}\ and\ \bibinfo {author} {\bibfnamefont {R.~K.}\ \bibnamefont
  {{Schaefer}}},\ }\href {\doibase 10.1086/171568} {\bibfield  {journal}
  {\bibinfo  {journal} {\apj}\ }\textbf {\bibinfo {volume} {394}},\ \bibinfo
  {pages} {174} (\bibinfo {year} {1992})}\BibitemShut {NoStop}%
\bibitem [{\citenamefont {{Simon}}\ \emph {et~al.}(1998)\citenamefont
  {{Simon}}, \citenamefont {{Molnar}},\ and\ \citenamefont
  {{Roesler}}}]{1998ApJ...499..250S}%
  \BibitemOpen
  \bibfield  {author} {\bibinfo {author} {\bibfnamefont {M.}~\bibnamefont
  {{Simon}}}, \bibinfo {author} {\bibfnamefont {A.}~\bibnamefont {{Molnar}}}, \
  and\ \bibinfo {author} {\bibfnamefont {S.}~\bibnamefont {{Roesler}}},\ }\href
  {\doibase 10.1086/305606} {\bibfield  {journal} {\bibinfo  {journal} {\apj}\
  }\textbf {\bibinfo {volume} {499}},\ \bibinfo {pages} {250} (\bibinfo {year}
  {1998})}\BibitemShut {NoStop}%
\bibitem [{\citenamefont {{Donato}}\ \emph {et~al.}(2001)\citenamefont
  {{Donato}}, \citenamefont {{Maurin}}, \citenamefont {{Salati}}, \citenamefont
  {{Barrau}}, \citenamefont {{Boudoul}},\ and\ \citenamefont
  {{Taillet}}}]{2001ApJ...563..172D}%
  \BibitemOpen
  \bibfield  {author} {\bibinfo {author} {\bibfnamefont {F.}~\bibnamefont
  {{Donato}}}, \bibinfo {author} {\bibfnamefont {D.}~\bibnamefont {{Maurin}}},
  \bibinfo {author} {\bibfnamefont {P.}~\bibnamefont {{Salati}}}, \bibinfo
  {author} {\bibfnamefont {A.}~\bibnamefont {{Barrau}}}, \bibinfo {author}
  {\bibfnamefont {G.}~\bibnamefont {{Boudoul}}}, \ and\ \bibinfo {author}
  {\bibfnamefont {R.}~\bibnamefont {{Taillet}}},\ }\href {\doibase
  10.1086/323684} {\bibfield  {journal} {\bibinfo  {journal} {\apj}\ }\textbf
  {\bibinfo {volume} {563}},\ \bibinfo {pages} {172} (\bibinfo {year}
  {2001})},\ \Eprint {http://arxiv.org/abs/astro-ph/0103150} {astro-ph/0103150}
  \BibitemShut {NoStop}%
\bibitem [{\citenamefont {{Moskalenko}}\ \emph {et~al.}(2002)\citenamefont
  {{Moskalenko}}, \citenamefont {{Strong}}, \citenamefont {{Ormes}},\ and\
  \citenamefont {{Potgieter}}}]{2002ApJ...565..280M}%
  \BibitemOpen
  \bibfield  {author} {\bibinfo {author} {\bibfnamefont {I.~V.}\ \bibnamefont
  {{Moskalenko}}}, \bibinfo {author} {\bibfnamefont {A.~W.}\ \bibnamefont
  {{Strong}}}, \bibinfo {author} {\bibfnamefont {J.~F.}\ \bibnamefont
  {{Ormes}}}, \ and\ \bibinfo {author} {\bibfnamefont {M.~S.}\ \bibnamefont
  {{Potgieter}}},\ }\href {\doibase 10.1086/324402} {\bibfield  {journal}
  {\bibinfo  {journal} {\apj}\ }\textbf {\bibinfo {volume} {565}},\ \bibinfo
  {pages} {280} (\bibinfo {year} {2002})},\ \Eprint
  {http://arxiv.org/abs/astro-ph/0106567} {astro-ph/0106567} \BibitemShut
  {NoStop}%
\bibitem [{\citenamefont {{Maurin}}\ \emph {et~al.}(2001)\citenamefont
  {{Maurin}}, \citenamefont {{Donato}}, \citenamefont {{Taillet}},\ and\
  \citenamefont {{Salati}}}]{2001ApJ...555..585M}%
  \BibitemOpen
  \bibfield  {author} {\bibinfo {author} {\bibfnamefont {D.}~\bibnamefont
  {{Maurin}}}, \bibinfo {author} {\bibfnamefont {F.}~\bibnamefont {{Donato}}},
  \bibinfo {author} {\bibfnamefont {R.}~\bibnamefont {{Taillet}}}, \ and\
  \bibinfo {author} {\bibfnamefont {P.}~\bibnamefont {{Salati}}},\ }\href
  {\doibase 10.1086/321496} {\bibfield  {journal} {\bibinfo  {journal} {\apj}\
  }\textbf {\bibinfo {volume} {555}},\ \bibinfo {pages} {585} (\bibinfo {year}
  {2001})},\ \Eprint {http://arxiv.org/abs/astro-ph/0101231} {astro-ph/0101231}
  \BibitemShut {NoStop}%
\bibitem [{\citenamefont {{Haino}}\ \emph {et~al.}(2005)\citenamefont
  {{Haino}}, \citenamefont {{Abe}}, \citenamefont {{Fuke}}, \citenamefont
  {{Maeno}}, \citenamefont {{Makida}}, \citenamefont {{Matsumoto}},
  \citenamefont {{Mitchell}}, \citenamefont {{Moiseev}}, \citenamefont
  {{Nishimura}}, \citenamefont {{Nozaki}} \emph
  {et~al.}}]{2005ICRC....3...13H}%
  \BibitemOpen
  \bibfield  {author} {\bibinfo {author} {\bibfnamefont {S.}~\bibnamefont
  {{Haino}}}, \bibinfo {author} {\bibfnamefont {K.}~\bibnamefont {{Abe}}},
  \bibinfo {author} {\bibfnamefont {H.}~\bibnamefont {{Fuke}}}, \bibinfo
  {author} {\bibfnamefont {T.}~\bibnamefont {{Maeno}}}, \bibinfo {author}
  {\bibfnamefont {Y.}~\bibnamefont {{Makida}}}, \bibinfo {author}
  {\bibfnamefont {H.}~\bibnamefont {{Matsumoto}}}, \bibinfo {author}
  {\bibfnamefont {J.~W.}\ \bibnamefont {{Mitchell}}}, \bibinfo {author}
  {\bibfnamefont {A.~A.}\ \bibnamefont {{Moiseev}}}, \bibinfo {author}
  {\bibfnamefont {J.}~\bibnamefont {{Nishimura}}}, \bibinfo {author}
  {\bibfnamefont {M.}~\bibnamefont {{Nozaki}}},  \emph {et~al.},\ }in\
  \href@noop {} {\emph {\bibinfo {booktitle} {International Cosmic Ray
  Conference}}},\ \bibinfo {series} {International Cosmic Ray Conference},
  Vol.~\bibinfo {volume} {3}\ (\bibinfo {year} {2005})\ p.~\bibinfo {pages}
  {13}\BibitemShut {NoStop}%
\bibitem [{\citenamefont {{BESS Collaboration}}\ \emph
  {et~al.}(2008)\citenamefont {{BESS Collaboration}}, \citenamefont {{Abe}},
  \citenamefont {{Fuke}}, \citenamefont {{Haino}}, \citenamefont {{Hams}},
  \citenamefont {{Itazaki}}, \citenamefont {{Kim}}, \citenamefont {{Kumazawa}},
  \citenamefont {{Lee}}, \citenamefont {{Makida}}, \citenamefont {{Matsuda}},
  \citenamefont {{Matsumoto}} \emph {et~al.}}]{2008PhLB..670..103B}%
  \BibitemOpen
  \bibfield  {author} {\bibinfo {author} {\bibnamefont {{BESS Collaboration}}},
  \bibinfo {author} {\bibfnamefont {K.}~\bibnamefont {{Abe}}}, \bibinfo
  {author} {\bibfnamefont {H.}~\bibnamefont {{Fuke}}}, \bibinfo {author}
  {\bibfnamefont {S.}~\bibnamefont {{Haino}}}, \bibinfo {author} {\bibfnamefont
  {T.}~\bibnamefont {{Hams}}}, \bibinfo {author} {\bibfnamefont
  {A.}~\bibnamefont {{Itazaki}}}, \bibinfo {author} {\bibfnamefont {K.~C.}\
  \bibnamefont {{Kim}}}, \bibinfo {author} {\bibfnamefont {T.}~\bibnamefont
  {{Kumazawa}}}, \bibinfo {author} {\bibfnamefont {M.~H.}\ \bibnamefont
  {{Lee}}}, \bibinfo {author} {\bibfnamefont {Y.}~\bibnamefont {{Makida}}},
  \bibinfo {author} {\bibfnamefont {S.}~\bibnamefont {{Matsuda}}}, \bibinfo
  {author} {\bibfnamefont {K.}~\bibnamefont {{Matsumoto}}},  \emph {et~al.},\
  }\href {\doibase 10.1016/j.physletb.2008.10.053} {\bibfield  {journal}
  {\bibinfo  {journal} {Physics Letters B}\ }\textbf {\bibinfo {volume}
  {670}},\ \bibinfo {pages} {103} (\bibinfo {year} {2008})},\ \Eprint
  {http://arxiv.org/abs/0805.1754} {arXiv:0805.1754} \BibitemShut {NoStop}%
\bibitem [{\citenamefont {{Abe}}\ \emph {et~al.}(2012)\citenamefont {{Abe}},
  \citenamefont {{Fuke}}, \citenamefont {{Haino}}, \citenamefont {{Hams}},
  \citenamefont {{Hasegawa}}, \citenamefont {{Horikoshi}}, \citenamefont
  {{Kim}}, \citenamefont {{Kusumoto}}, \citenamefont {{Lee}}, \citenamefont
  {{Makida}} \emph {et~al.}}]{2012PhRvL.108e1102A}%
  \BibitemOpen
  \bibfield  {author} {\bibinfo {author} {\bibfnamefont {K.}~\bibnamefont
  {{Abe}}}, \bibinfo {author} {\bibfnamefont {H.}~\bibnamefont {{Fuke}}},
  \bibinfo {author} {\bibfnamefont {S.}~\bibnamefont {{Haino}}}, \bibinfo
  {author} {\bibfnamefont {T.}~\bibnamefont {{Hams}}}, \bibinfo {author}
  {\bibfnamefont {M.}~\bibnamefont {{Hasegawa}}}, \bibinfo {author}
  {\bibfnamefont {A.}~\bibnamefont {{Horikoshi}}}, \bibinfo {author}
  {\bibfnamefont {K.~C.}\ \bibnamefont {{Kim}}}, \bibinfo {author}
  {\bibfnamefont {A.}~\bibnamefont {{Kusumoto}}}, \bibinfo {author}
  {\bibfnamefont {M.~H.}\ \bibnamefont {{Lee}}}, \bibinfo {author}
  {\bibfnamefont {Y.}~\bibnamefont {{Makida}}},  \emph {et~al.},\ }\href
  {\doibase 10.1103/PhysRevLett.108.051102} {\bibfield  {journal} {\bibinfo
  {journal} {Physical Review Letters}\ }\textbf {\bibinfo {volume} {108}},\
  \bibinfo {eid} {051102} (\bibinfo {year} {2012})},\ \Eprint
  {http://arxiv.org/abs/1107.6000} {arXiv:1107.6000 [astro-ph.HE]} \BibitemShut
  {NoStop}%
\bibitem [{\citenamefont {{Adriani}}\ \emph {et~al.}(2009)\citenamefont
  {{Adriani}}, \citenamefont {{Barbarino}}, \citenamefont {{Bazilevskaya}},
  \citenamefont {{Bellotti}}, \citenamefont {{Boezio}}, \citenamefont
  {{Bogomolov}}, \citenamefont {{Bonechi}}, \citenamefont {{Bongi}},
  \citenamefont {{Bonvicini}}, \citenamefont {{Bottai}} \emph
  {et~al.}}]{2009PhRvL.102e1101A}%
  \BibitemOpen
  \bibfield  {author} {\bibinfo {author} {\bibfnamefont {O.}~\bibnamefont
  {{Adriani}}}, \bibinfo {author} {\bibfnamefont {G.~C.}\ \bibnamefont
  {{Barbarino}}}, \bibinfo {author} {\bibfnamefont {G.~A.}\ \bibnamefont
  {{Bazilevskaya}}}, \bibinfo {author} {\bibfnamefont {R.}~\bibnamefont
  {{Bellotti}}}, \bibinfo {author} {\bibfnamefont {M.}~\bibnamefont
  {{Boezio}}}, \bibinfo {author} {\bibfnamefont {E.~A.}\ \bibnamefont
  {{Bogomolov}}}, \bibinfo {author} {\bibfnamefont {L.}~\bibnamefont
  {{Bonechi}}}, \bibinfo {author} {\bibfnamefont {M.}~\bibnamefont {{Bongi}}},
  \bibinfo {author} {\bibfnamefont {V.}~\bibnamefont {{Bonvicini}}}, \bibinfo
  {author} {\bibfnamefont {S.}~\bibnamefont {{Bottai}}},  \emph {et~al.},\
  }\href {\doibase 10.1103/PhysRevLett.102.051101} {\bibfield  {journal}
  {\bibinfo  {journal} {\prl}\ }\textbf {\bibinfo {volume} {102}},\ \bibinfo
  {eid} {051101} (\bibinfo {year} {2009})},\ \Eprint
  {http://arxiv.org/abs/0810.4994} {arXiv:0810.4994} \BibitemShut {NoStop}%
\bibitem [{\citenamefont {{Adriani}}\ \emph {et~al.}(2010)\citenamefont
  {{Adriani}}, \citenamefont {{Barbarino}}, \citenamefont {{Bazilevskaya}}
  \emph {et~al.}}]{2010PhRvL.105l1101A}%
  \BibitemOpen
  \bibfield  {author} {\bibinfo {author} {\bibfnamefont {O.}~\bibnamefont
  {{Adriani}}}, \bibinfo {author} {\bibfnamefont {G.~C.}\ \bibnamefont
  {{Barbarino}}}, \bibinfo {author} {\bibfnamefont {G.~A.}\ \bibnamefont
  {{Bazilevskaya}}},  \emph {et~al.},\ }\href {\doibase
  10.1103/PhysRevLett.105.121101} {\bibfield  {journal} {\bibinfo  {journal}
  {\prl}\ }\textbf {\bibinfo {volume} {105}},\ \bibinfo {eid} {121101}
  (\bibinfo {year} {2010})},\ \Eprint {http://arxiv.org/abs/1007.0821}
  {arXiv:1007.0821 [astro-ph.HE]} \BibitemShut {NoStop}%
\bibitem [{\citenamefont {{Adriani}}\ \emph {et~al.}(2013)\citenamefont
  {{Adriani}}, \citenamefont {{Bazilevskaya}}, \citenamefont {{Barbarino}},
  \citenamefont {{Bellotti}}, \citenamefont {{Boezio}}, \citenamefont
  {{Bogomolov}}, \citenamefont {{Bonvicini}}, \citenamefont {{Bongi}},
  \citenamefont {{Bonechi}} \emph {et~al.}}]{2013JETPL..96..621A}%
  \BibitemOpen
  \bibfield  {author} {\bibinfo {author} {\bibfnamefont {O.}~\bibnamefont
  {{Adriani}}}, \bibinfo {author} {\bibfnamefont {G.~A.}\ \bibnamefont
  {{Bazilevskaya}}}, \bibinfo {author} {\bibfnamefont {G.~C.}\ \bibnamefont
  {{Barbarino}}}, \bibinfo {author} {\bibfnamefont {R.}~\bibnamefont
  {{Bellotti}}}, \bibinfo {author} {\bibfnamefont {M.}~\bibnamefont
  {{Boezio}}}, \bibinfo {author} {\bibfnamefont {E.~A.}\ \bibnamefont
  {{Bogomolov}}}, \bibinfo {author} {\bibfnamefont {V.}~\bibnamefont
  {{Bonvicini}}}, \bibinfo {author} {\bibfnamefont {M.}~\bibnamefont
  {{Bongi}}}, \bibinfo {author} {\bibfnamefont {L.}~\bibnamefont {{Bonechi}}},
  \emph {et~al.},\ }\href {\doibase 10.1134/S002136401222002X} {\bibfield
  {journal} {\bibinfo  {journal} {Soviet Journal of Experimental and
  Theoretical Physics Letters}\ }\textbf {\bibinfo {volume} {96}},\ \bibinfo
  {pages} {621} (\bibinfo {year} {2013})}\BibitemShut {NoStop}%
\bibitem [{\citenamefont {{Donato}}\ \emph {et~al.}(2009)\citenamefont
  {{Donato}}, \citenamefont {{Maurin}}, \citenamefont {{Brun}}, \citenamefont
  {{Delahaye}},\ and\ \citenamefont {{Salati}}}]{2009PhRvL.102g1301D}%
  \BibitemOpen
  \bibfield  {author} {\bibinfo {author} {\bibfnamefont {F.}~\bibnamefont
  {{Donato}}}, \bibinfo {author} {\bibfnamefont {D.}~\bibnamefont {{Maurin}}},
  \bibinfo {author} {\bibfnamefont {P.}~\bibnamefont {{Brun}}}, \bibinfo
  {author} {\bibfnamefont {T.}~\bibnamefont {{Delahaye}}}, \ and\ \bibinfo
  {author} {\bibfnamefont {P.}~\bibnamefont {{Salati}}},\ }\href {\doibase
  10.1103/PhysRevLett.102.071301} {\bibfield  {journal} {\bibinfo  {journal}
  {Physical Review Letters}\ }\textbf {\bibinfo {volume} {102}},\ \bibinfo
  {eid} {071301} (\bibinfo {year} {2009})},\ \Eprint
  {http://arxiv.org/abs/0810.5292} {arXiv:0810.5292} \BibitemShut {NoStop}%
\bibitem [{\citenamefont {{Aguilar}}\ \emph {et~al.}(2016)\citenamefont
  {{Aguilar}}, \citenamefont {{Ali Cavasonza}}, \citenamefont {{Alpat}},
  \citenamefont {{Ambrosi}}, \citenamefont {{Arruda}}, \citenamefont {{Attig}},
  \citenamefont {{Aupetit}}, \citenamefont {{Azzarello}}, \citenamefont
  {{Bachlechner}}, \citenamefont {{Barao}},\ and\ \citenamefont
  {et~al.}}]{2016PhRvL.117i1103A}%
  \BibitemOpen
  \bibfield  {author} {\bibinfo {author} {\bibfnamefont {M.}~\bibnamefont
  {{Aguilar}}}, \bibinfo {author} {\bibfnamefont {L.}~\bibnamefont {{Ali
  Cavasonza}}}, \bibinfo {author} {\bibfnamefont {B.}~\bibnamefont {{Alpat}}},
  \bibinfo {author} {\bibfnamefont {G.}~\bibnamefont {{Ambrosi}}}, \bibinfo
  {author} {\bibfnamefont {L.}~\bibnamefont {{Arruda}}}, \bibinfo {author}
  {\bibfnamefont {N.}~\bibnamefont {{Attig}}}, \bibinfo {author} {\bibfnamefont
  {S.}~\bibnamefont {{Aupetit}}}, \bibinfo {author} {\bibfnamefont
  {P.}~\bibnamefont {{Azzarello}}}, \bibinfo {author} {\bibfnamefont
  {A.}~\bibnamefont {{Bachlechner}}}, \bibinfo {author} {\bibfnamefont
  {F.}~\bibnamefont {{Barao}}}, \ and\ \bibinfo {author} {\bibnamefont
  {et~al.}},\ }\href {\doibase 10.1103/PhysRevLett.117.091103} {\bibfield
  {journal} {\bibinfo  {journal} {Physical Review Letters}\ }\textbf {\bibinfo
  {volume} {117}},\ \bibinfo {eid} {091103} (\bibinfo {year}
  {2016})}\BibitemShut {NoStop}%
\bibitem [{\citenamefont {{Giesen}}\ \emph {et~al.}(2015)\citenamefont
  {{Giesen}}, \citenamefont {{Boudaud}}, \citenamefont {{G{\'e}nolini}},
  \citenamefont {{Poulin}}, \citenamefont {{Cirelli}}, \citenamefont
  {{Salati}},\ and\ \citenamefont {{Serpico}}}]{2015JCAP...09..023G}%
  \BibitemOpen
  \bibfield  {author} {\bibinfo {author} {\bibfnamefont {G.}~\bibnamefont
  {{Giesen}}}, \bibinfo {author} {\bibfnamefont {M.}~\bibnamefont {{Boudaud}}},
  \bibinfo {author} {\bibfnamefont {Y.}~\bibnamefont {{G{\'e}nolini}}},
  \bibinfo {author} {\bibfnamefont {V.}~\bibnamefont {{Poulin}}}, \bibinfo
  {author} {\bibfnamefont {M.}~\bibnamefont {{Cirelli}}}, \bibinfo {author}
  {\bibfnamefont {P.}~\bibnamefont {{Salati}}}, \ and\ \bibinfo {author}
  {\bibfnamefont {P.~D.}\ \bibnamefont {{Serpico}}},\ }\href {\doibase
  10.1088/1475-7516/2015/09/023} {\bibfield  {journal} {\bibinfo  {journal}
  {\jcap}\ }\textbf {\bibinfo {volume} {9}},\ \bibinfo {eid} {023} (\bibinfo
  {year} {2015})},\ \Eprint {http://arxiv.org/abs/1504.04276} {arXiv:1504.04276
  [astro-ph.HE]} \BibitemShut {NoStop}%
\bibitem [{\citenamefont {{Kappl}}\ \emph {et~al.}(2015)\citenamefont
  {{Kappl}}, \citenamefont {{Reinert}},\ and\ \citenamefont
  {{Winkler}}}]{2015JCAP...10..034K}%
  \BibitemOpen
  \bibfield  {author} {\bibinfo {author} {\bibfnamefont {R.}~\bibnamefont
  {{Kappl}}}, \bibinfo {author} {\bibfnamefont {A.}~\bibnamefont {{Reinert}}},
  \ and\ \bibinfo {author} {\bibfnamefont {M.~W.}\ \bibnamefont {{Winkler}}},\
  }\href {\doibase 10.1088/1475-7516/2015/10/034} {\bibfield  {journal}
  {\bibinfo  {journal} {\jcap}\ }\textbf {\bibinfo {volume} {10}},\ \bibinfo
  {eid} {034} (\bibinfo {year} {2015})},\ \Eprint
  {http://arxiv.org/abs/1506.04145} {arXiv:1506.04145 [astro-ph.HE]}
  \BibitemShut {NoStop}%
\bibitem [{\citenamefont {{Evoli}}\ \emph {et~al.}(2015)\citenamefont
  {{Evoli}}, \citenamefont {{Gaggero}},\ and\ \citenamefont
  {{Grasso}}}]{2015JCAP...12..039E}%
  \BibitemOpen
  \bibfield  {author} {\bibinfo {author} {\bibfnamefont {C.}~\bibnamefont
  {{Evoli}}}, \bibinfo {author} {\bibfnamefont {D.}~\bibnamefont {{Gaggero}}},
  \ and\ \bibinfo {author} {\bibfnamefont {D.}~\bibnamefont {{Grasso}}},\
  }\href {\doibase 10.1088/1475-7516/2015/12/039} {\bibfield  {journal}
  {\bibinfo  {journal} {\jcap}\ }\textbf {\bibinfo {volume} {12}},\ \bibinfo
  {eid} {039} (\bibinfo {year} {2015})},\ \Eprint
  {http://arxiv.org/abs/1504.05175} {arXiv:1504.05175 [astro-ph.HE]}
  \BibitemShut {NoStop}%
\bibitem [{\citenamefont {{Maurin}}\ \emph {et~al.}(2002)\citenamefont
  {{Maurin}}, \citenamefont {{Taillet}}, \citenamefont {{Donato}},
  \citenamefont {{Salati}}, \citenamefont {{Barrau}},\ and\ \citenamefont
  {{Boudoul}}}]{2002astro.ph.12111M}%
  \BibitemOpen
  \bibfield  {author} {\bibinfo {author} {\bibfnamefont {D.}~\bibnamefont
  {{Maurin}}}, \bibinfo {author} {\bibfnamefont {R.}~\bibnamefont {{Taillet}}},
  \bibinfo {author} {\bibfnamefont {F.}~\bibnamefont {{Donato}}}, \bibinfo
  {author} {\bibfnamefont {P.}~\bibnamefont {{Salati}}}, \bibinfo {author}
  {\bibfnamefont {A.}~\bibnamefont {{Barrau}}}, \ and\ \bibinfo {author}
  {\bibfnamefont {G.}~\bibnamefont {{Boudoul}}},\ }\href@noop {} {\bibfield
  {journal} {\bibinfo  {journal} {arXiv Astrophysics e-prints}\ } (\bibinfo
  {year} {2002})},\ \Eprint {http://arxiv.org/abs/astro-ph/0212111}
  {astro-ph/0212111} \BibitemShut {NoStop}%
\bibitem [{\citenamefont {{Lavalle}}\ and\ \citenamefont
  {{Salati}}(2012)}]{2012CRPhy..13..740L}%
  \BibitemOpen
  \bibfield  {author} {\bibinfo {author} {\bibfnamefont {J.}~\bibnamefont
  {{Lavalle}}}\ and\ \bibinfo {author} {\bibfnamefont {P.}~\bibnamefont
  {{Salati}}},\ }\href {\doibase 10.1016/j.crhy.2012.05.001} {\bibfield
  {journal} {\bibinfo  {journal} {\crp}\ }\textbf {\bibinfo {volume} {13}},\
  \bibinfo {pages} {740} (\bibinfo {year} {2012})},\ \Eprint
  {http://arxiv.org/abs/1205.1004} {arXiv:1205.1004 [astro-ph.HE]} \BibitemShut
  {NoStop}%
\bibitem [{\citenamefont {Boudaud}\ \emph {et~al.}(2015)\citenamefont
  {Boudaud}, \citenamefont {Cirelli}, \citenamefont {Giesen},\ and\
  \citenamefont {Salati}}]{Boudaud:2014qra}%
  \BibitemOpen
  \bibfield  {author} {\bibinfo {author} {\bibfnamefont {M.}~\bibnamefont
  {Boudaud}}, \bibinfo {author} {\bibfnamefont {M.}~\bibnamefont {Cirelli}},
  \bibinfo {author} {\bibfnamefont {G.}~\bibnamefont {Giesen}}, \ and\ \bibinfo
  {author} {\bibfnamefont {P.}~\bibnamefont {Salati}},\ }\href {\doibase
  10.1088/1475-7516/2015/05/013} {\bibfield  {journal} {\bibinfo  {journal}
  {JCAP}\ }\textbf {\bibinfo {volume} {1505}},\ \bibinfo {pages} {013}
  (\bibinfo {year} {2015})},\ \Eprint {http://arxiv.org/abs/1412.5696}
  {arXiv:1412.5696 [astro-ph.HE]} \BibitemShut {NoStop}%
\bibitem [{\citenamefont {{Conrad}}\ and\ \citenamefont
  {{Reimer}}(2017)}]{2017NatPh..13..224C}%
  \BibitemOpen
  \bibfield  {author} {\bibinfo {author} {\bibfnamefont {J.}~\bibnamefont
  {{Conrad}}}\ and\ \bibinfo {author} {\bibfnamefont {O.}~\bibnamefont
  {{Reimer}}},\ }\href {\doibase 10.1038/nphys4049} {\bibfield  {journal}
  {\bibinfo  {journal} {Nature Physics}\ }\textbf {\bibinfo {volume} {13}},\
  \bibinfo {pages} {224} (\bibinfo {year} {2017})},\ \Eprint
  {http://arxiv.org/abs/1705.11165} {arXiv:1705.11165 [astro-ph.HE]}
  \BibitemShut {NoStop}%
\bibitem [{\citenamefont {{Clark}}\ \emph {et~al.}(2018)\citenamefont
  {{Clark}}, \citenamefont {{Dutta}},\ and\ \citenamefont
  {{Strigari}}}]{2018PhRvD..97b3003C}%
  \BibitemOpen
  \bibfield  {author} {\bibinfo {author} {\bibfnamefont {S.~J.}\ \bibnamefont
  {{Clark}}}, \bibinfo {author} {\bibfnamefont {B.}~\bibnamefont {{Dutta}}}, \
  and\ \bibinfo {author} {\bibfnamefont {L.~E.}\ \bibnamefont {{Strigari}}},\
  }\href {\doibase 10.1103/PhysRevD.97.023003} {\bibfield  {journal} {\bibinfo
  {journal} {\prd}\ }\textbf {\bibinfo {volume} {97}},\ \bibinfo {eid} {023003}
  (\bibinfo {year} {2018})},\ \Eprint {http://arxiv.org/abs/1709.07410}
  {arXiv:1709.07410 [astro-ph.HE]} \BibitemShut {NoStop}%
\bibitem [{\citenamefont {{Cui}}\ \emph {et~al.}(2017)\citenamefont {{Cui}},
  \citenamefont {{Yuan}}, \citenamefont {{Tsai}},\ and\ \citenamefont
  {{Fan}}}]{2017PhRvL.118s1101C}%
  \BibitemOpen
  \bibfield  {author} {\bibinfo {author} {\bibfnamefont {M.-Y.}\ \bibnamefont
  {{Cui}}}, \bibinfo {author} {\bibfnamefont {Q.}~\bibnamefont {{Yuan}}},
  \bibinfo {author} {\bibfnamefont {Y.-L.~S.}\ \bibnamefont {{Tsai}}}, \ and\
  \bibinfo {author} {\bibfnamefont {Y.-Z.}\ \bibnamefont {{Fan}}},\ }\href
  {\doibase 10.1103/PhysRevLett.118.191101} {\bibfield  {journal} {\bibinfo
  {journal} {Physical Review Letters}\ }\textbf {\bibinfo {volume} {118}},\
  \bibinfo {eid} {191101} (\bibinfo {year} {2017})},\ \Eprint
  {http://arxiv.org/abs/1610.03840} {arXiv:1610.03840 [astro-ph.HE]}
  \BibitemShut {NoStop}%
\bibitem [{\citenamefont {{Cuoco}}\ \emph {et~al.}(2017)\citenamefont
  {{Cuoco}}, \citenamefont {{Kr{\"a}mer}},\ and\ \citenamefont
  {{Korsmeier}}}]{2017PhRvL.118s1102C}%
  \BibitemOpen
  \bibfield  {author} {\bibinfo {author} {\bibfnamefont {A.}~\bibnamefont
  {{Cuoco}}}, \bibinfo {author} {\bibfnamefont {M.}~\bibnamefont
  {{Kr{\"a}mer}}}, \ and\ \bibinfo {author} {\bibfnamefont {M.}~\bibnamefont
  {{Korsmeier}}},\ }\href {\doibase 10.1103/PhysRevLett.118.191102} {\bibfield
  {journal} {\bibinfo  {journal} {Physical Review Letters}\ }\textbf {\bibinfo
  {volume} {118}},\ \bibinfo {eid} {191102} (\bibinfo {year} {2017})},\ \Eprint
  {http://arxiv.org/abs/1610.03071} {arXiv:1610.03071 [astro-ph.HE]}
  \BibitemShut {NoStop}%
\bibitem [{\citenamefont {Cholis}\ \emph {et~al.}(2019)\citenamefont {Cholis},
  \citenamefont {Linden},\ and\ \citenamefont {Hooper}}]{Cholis:2019ejx}%
  \BibitemOpen
  \bibfield  {author} {\bibinfo {author} {\bibfnamefont {I.}~\bibnamefont
  {Cholis}}, \bibinfo {author} {\bibfnamefont {T.}~\bibnamefont {Linden}}, \
  and\ \bibinfo {author} {\bibfnamefont {D.}~\bibnamefont {Hooper}},\ }\href
  {\doibase 10.1103/PhysRevD.99.103026} {\bibfield  {journal} {\bibinfo
  {journal} {Phys. Rev.}\ }\textbf {\bibinfo {volume} {D99}},\ \bibinfo {pages}
  {103026} (\bibinfo {year} {2019})},\ \Eprint
  {http://arxiv.org/abs/1903.02549} {arXiv:1903.02549 [astro-ph.HE]}
  \BibitemShut {NoStop}%
\bibitem [{\citenamefont {Cuoco}\ \emph {et~al.}(2019)\citenamefont {Cuoco},
  \citenamefont {Heisig}, \citenamefont {Klamt}, \citenamefont {Korsmeier},\
  and\ \citenamefont {Kramer}}]{Cuoco:2019kuu}%
  \BibitemOpen
  \bibfield  {author} {\bibinfo {author} {\bibfnamefont {A.}~\bibnamefont
  {Cuoco}}, \bibinfo {author} {\bibfnamefont {J.}~\bibnamefont {Heisig}},
  \bibinfo {author} {\bibfnamefont {L.}~\bibnamefont {Klamt}}, \bibinfo
  {author} {\bibfnamefont {M.}~\bibnamefont {Korsmeier}}, \ and\ \bibinfo
  {author} {\bibfnamefont {M.}~\bibnamefont {Kramer}},\ }\href {\doibase
  10.1103/PhysRevD.99.103014} {\bibfield  {journal} {\bibinfo  {journal} {Phys.
  Rev.}\ }\textbf {\bibinfo {volume} {D99}},\ \bibinfo {pages} {103014}
  (\bibinfo {year} {2019})},\ \Eprint {http://arxiv.org/abs/1903.01472}
  {arXiv:1903.01472 [astro-ph.HE]} \BibitemShut {NoStop}%
\bibitem [{\citenamefont {{Lin}}\ \emph {et~al.}(2019)\citenamefont {{Lin}},
  \citenamefont {{Bi}},\ and\ \citenamefont {{Yin}}}]{2019arXiv190309545L}%
  \BibitemOpen
  \bibfield  {author} {\bibinfo {author} {\bibfnamefont {S.-J.}\ \bibnamefont
  {{Lin}}}, \bibinfo {author} {\bibfnamefont {X.-J.}\ \bibnamefont {{Bi}}}, \
  and\ \bibinfo {author} {\bibfnamefont {P.-F.}\ \bibnamefont {{Yin}}},\
  }\href@noop {} {\bibfield  {journal} {\bibinfo  {journal} {arXiv e-prints}\ }
  (\bibinfo {year} {2019})},\ \Eprint {http://arxiv.org/abs/1903.09545}
  {arXiv:1903.09545 [astro-ph.HE]} \BibitemShut {NoStop}%
\bibitem [{\citenamefont {{Anticic}}\ \emph {et~al.}(2010)\citenamefont
  {{Anticic}}, \citenamefont {{Baatar}}, \citenamefont {{Bartke}},
  \citenamefont {{Betev}}, \citenamefont {{Bia{\l}kowska}}, \citenamefont
  {{Blume}}, \citenamefont {{Boimska}}, \citenamefont {{Bracinik}},
  \citenamefont {{Cerny}}, \citenamefont {{Chvala}},\ and\ \citenamefont {{NA49
  Collaboration}}}]{2010EPJC...65....9A}%
  \BibitemOpen
  \bibfield  {author} {\bibinfo {author} {\bibfnamefont {T.}~\bibnamefont
  {{Anticic}}}, \bibinfo {author} {\bibfnamefont {B.}~\bibnamefont {{Baatar}}},
  \bibinfo {author} {\bibfnamefont {J.}~\bibnamefont {{Bartke}}}, \bibinfo
  {author} {\bibfnamefont {L.}~\bibnamefont {{Betev}}}, \bibinfo {author}
  {\bibfnamefont {H.}~\bibnamefont {{Bia{\l}kowska}}}, \bibinfo {author}
  {\bibfnamefont {C.}~\bibnamefont {{Blume}}}, \bibinfo {author} {\bibfnamefont
  {B.}~\bibnamefont {{Boimska}}}, \bibinfo {author} {\bibfnamefont
  {J.}~\bibnamefont {{Bracinik}}}, \bibinfo {author} {\bibfnamefont
  {V.}~\bibnamefont {{Cerny}}}, \bibinfo {author} {\bibfnamefont
  {O.}~\bibnamefont {{Chvala}}}, \ and\ \bibinfo {author} {\bibnamefont {{NA49
  Collaboration}}},\ }\href {\doibase 10.1140/epjc/s10052-009-1172-2}
  {\bibfield  {journal} {\bibinfo  {journal} {European Physical Journal C}\
  }\textbf {\bibinfo {volume} {65}},\ \bibinfo {pages} {9} (\bibinfo {year}
  {2010})},\ \Eprint {http://arxiv.org/abs/0904.2708} {arXiv:0904.2708
  [hep-ex]} \BibitemShut {NoStop}%
\bibitem [{\citenamefont {{Aduszkiewicz}}\ \emph {et~al.}(2017)\citenamefont
  {{Aduszkiewicz}}, \citenamefont {{Ali}}, \citenamefont {{Andronov}},
  \citenamefont {{Anti{\'c}i{\'c}}}, \citenamefont {{Baatar}}, \citenamefont
  {{Baszczyk}}, \citenamefont {{Bhosale}}, \citenamefont {{Blondel}},
  \citenamefont {{Bogomilov}}, \citenamefont {{Brandin}}, \citenamefont
  {{Bravar}} \emph {et~al.}}]{2017EPJC...77..671A}%
  \BibitemOpen
  \bibfield  {author} {\bibinfo {author} {\bibfnamefont {A.}~\bibnamefont
  {{Aduszkiewicz}}}, \bibinfo {author} {\bibfnamefont {Y.}~\bibnamefont
  {{Ali}}}, \bibinfo {author} {\bibfnamefont {E.}~\bibnamefont {{Andronov}}},
  \bibinfo {author} {\bibfnamefont {T.}~\bibnamefont {{Anti{\'c}i{\'c}}}},
  \bibinfo {author} {\bibfnamefont {B.}~\bibnamefont {{Baatar}}}, \bibinfo
  {author} {\bibfnamefont {M.}~\bibnamefont {{Baszczyk}}}, \bibinfo {author}
  {\bibfnamefont {S.}~\bibnamefont {{Bhosale}}}, \bibinfo {author}
  {\bibfnamefont {A.}~\bibnamefont {{Blondel}}}, \bibinfo {author}
  {\bibfnamefont {M.}~\bibnamefont {{Bogomilov}}}, \bibinfo {author}
  {\bibfnamefont {A.}~\bibnamefont {{Brandin}}}, \bibinfo {author}
  {\bibfnamefont {A.}~\bibnamefont {{Bravar}}},  \emph {et~al.},\ }\href
  {\doibase 10.1140/epjc/s10052-017-5260-4} {\bibfield  {journal} {\bibinfo
  {journal} {European Physical Journal C}\ }\textbf {\bibinfo {volume} {77}},\
  \bibinfo {eid} {671} (\bibinfo {year} {2017})},\ \Eprint
  {http://arxiv.org/abs/1705.02467} {arXiv:1705.02467 [nucl-ex]} \BibitemShut
  {NoStop}%
\bibitem [{\citenamefont {Aaij}\ \emph {et~al.}(2018)\citenamefont {Aaij} \emph
  {et~al.}}]{Aaij:2018svt}%
  \BibitemOpen
  \bibfield  {author} {\bibinfo {author} {\bibfnamefont {R.}~\bibnamefont
  {Aaij}} \emph {et~al.} (\bibinfo {collaboration} {LHCb}),\ }\href {\doibase
  10.1103/PhysRevLett.121.222001} {\bibfield  {journal} {\bibinfo  {journal}
  {Phys. Rev. Lett.}\ }\textbf {\bibinfo {volume} {121}},\ \bibinfo {pages}
  {222001} (\bibinfo {year} {2018})},\ \Eprint
  {http://arxiv.org/abs/1808.06127} {arXiv:1808.06127 [hep-ex]} \BibitemShut
  {NoStop}%
\bibitem [{\citenamefont {{di Mauro}}\ \emph {et~al.}(2014)\citenamefont {{di
  Mauro}}, \citenamefont {{Donato}}, \citenamefont {{Goudelis}},\ and\
  \citenamefont {{Serpico}}}]{2014PhRvD..90h5017D}%
  \BibitemOpen
  \bibfield  {author} {\bibinfo {author} {\bibfnamefont {M.}~\bibnamefont {{di
  Mauro}}}, \bibinfo {author} {\bibfnamefont {F.}~\bibnamefont {{Donato}}},
  \bibinfo {author} {\bibfnamefont {A.}~\bibnamefont {{Goudelis}}}, \ and\
  \bibinfo {author} {\bibfnamefont {P.~D.}\ \bibnamefont {{Serpico}}},\ }\href
  {\doibase 10.1103/PhysRevD.90.085017} {\bibfield  {journal} {\bibinfo
  {journal} {\prd}\ }\textbf {\bibinfo {volume} {90}},\ \bibinfo {eid} {085017}
  (\bibinfo {year} {2014})},\ \Eprint {http://arxiv.org/abs/1408.0288}
  {arXiv:1408.0288 [hep-ph]} \BibitemShut {NoStop}%
\bibitem [{\citenamefont {{Lin}}\ \emph {et~al.}(2017)\citenamefont {{Lin}},
  \citenamefont {{Bi}}, \citenamefont {{Feng}}, \citenamefont {{Yin}},\ and\
  \citenamefont {{Yu}}}]{2017PhRvD..96l3010L}%
  \BibitemOpen
  \bibfield  {author} {\bibinfo {author} {\bibfnamefont {S.-J.}\ \bibnamefont
  {{Lin}}}, \bibinfo {author} {\bibfnamefont {X.-J.}\ \bibnamefont {{Bi}}},
  \bibinfo {author} {\bibfnamefont {J.}~\bibnamefont {{Feng}}}, \bibinfo
  {author} {\bibfnamefont {P.-F.}\ \bibnamefont {{Yin}}}, \ and\ \bibinfo
  {author} {\bibfnamefont {Z.-H.}\ \bibnamefont {{Yu}}},\ }\href {\doibase
  10.1103/PhysRevD.96.123010} {\bibfield  {journal} {\bibinfo  {journal}
  {\prd}\ }\textbf {\bibinfo {volume} {96}},\ \bibinfo {eid} {123010} (\bibinfo
  {year} {2017})},\ \Eprint {http://arxiv.org/abs/1612.04001} {arXiv:1612.04001
  [astro-ph.HE]} \BibitemShut {NoStop}%
\bibitem [{\citenamefont {{Donato}}\ \emph {et~al.}(2017)\citenamefont
  {{Donato}}, \citenamefont {{Korsmeier}},\ and\ \citenamefont {{Di
  Mauro}}}]{2017PhRvD..96d3007D}%
  \BibitemOpen
  \bibfield  {author} {\bibinfo {author} {\bibfnamefont {F.}~\bibnamefont
  {{Donato}}}, \bibinfo {author} {\bibfnamefont {M.}~\bibnamefont
  {{Korsmeier}}}, \ and\ \bibinfo {author} {\bibfnamefont {M.}~\bibnamefont
  {{Di Mauro}}},\ }\href {\doibase 10.1103/PhysRevD.96.043007} {\bibfield
  {journal} {\bibinfo  {journal} {\prd}\ }\textbf {\bibinfo {volume} {96}},\
  \bibinfo {eid} {043007} (\bibinfo {year} {2017})},\ \Eprint
  {http://arxiv.org/abs/1704.03663} {arXiv:1704.03663 [astro-ph.HE]}
  \BibitemShut {NoStop}%
\bibitem [{\citenamefont
  {{Winkler}}(2017{\natexlab{a}})}]{2017JCAP...02..048W}%
  \BibitemOpen
  \bibfield  {author} {\bibinfo {author} {\bibfnamefont {M.~W.}\ \bibnamefont
  {{Winkler}}},\ }\href {\doibase 10.1088/1475-7516/2017/02/048} {\bibfield
  {journal} {\bibinfo  {journal} {\jcap}\ }\textbf {\bibinfo {volume} {2}},\
  \bibinfo {eid} {048} (\bibinfo {year} {2017}{\natexlab{a}})},\ \Eprint
  {http://arxiv.org/abs/1701.04866} {arXiv:1701.04866 [hep-ph]} \BibitemShut
  {NoStop}%
\bibitem [{\citenamefont {{Korsmeier}}\ \emph {et~al.}(2018)\citenamefont
  {{Korsmeier}}, \citenamefont {{Donato}},\ and\ \citenamefont {{Di
  Mauro}}}]{2018PhRvD..97j3019K}%
  \BibitemOpen
  \bibfield  {author} {\bibinfo {author} {\bibfnamefont {M.}~\bibnamefont
  {{Korsmeier}}}, \bibinfo {author} {\bibfnamefont {F.}~\bibnamefont
  {{Donato}}}, \ and\ \bibinfo {author} {\bibfnamefont {M.}~\bibnamefont {{Di
  Mauro}}},\ }\href {\doibase 10.1103/PhysRevD.97.103019} {\bibfield  {journal}
  {\bibinfo  {journal} {\prd}\ }\textbf {\bibinfo {volume} {97}},\ \bibinfo
  {eid} {103019} (\bibinfo {year} {2018})},\ \Eprint
  {http://arxiv.org/abs/1802.03030} {arXiv:1802.03030 [astro-ph.HE]}
  \BibitemShut {NoStop}%
\bibitem [{\citenamefont {{Maurin}}(2018)}]{2018arXiv180702968M}%
  \BibitemOpen
  \bibfield  {author} {\bibinfo {author} {\bibfnamefont {D.}~\bibnamefont
  {{Maurin}}},\ }\href@noop {} {\bibfield  {journal} {\bibinfo  {journal}
  {ArXiv e-prints}\ } (\bibinfo {year} {2018})},\ \Eprint
  {http://arxiv.org/abs/1807.02968} {arXiv:1807.02968 [astro-ph.IM]}
  \BibitemShut {NoStop}%
\bibitem [{\citenamefont {{Korsmeier}}\ and\ \citenamefont
  {{Cuoco}}(2016)}]{2016PhRvD..94l3019K}%
  \BibitemOpen
  \bibfield  {author} {\bibinfo {author} {\bibfnamefont {M.}~\bibnamefont
  {{Korsmeier}}}\ and\ \bibinfo {author} {\bibfnamefont {A.}~\bibnamefont
  {{Cuoco}}},\ }\href {\doibase 10.1103/PhysRevD.94.123019} {\bibfield
  {journal} {\bibinfo  {journal} {\prd}\ }\textbf {\bibinfo {volume} {94}},\
  \bibinfo {eid} {123019} (\bibinfo {year} {2016})},\ \Eprint
  {http://arxiv.org/abs/1607.06093} {arXiv:1607.06093 [astro-ph.HE]}
  \BibitemShut {NoStop}%
\bibitem [{\citenamefont {{Feng}}\ \emph {et~al.}(2016)\citenamefont {{Feng}},
  \citenamefont {{Tomassetti}},\ and\ \citenamefont
  {{Oliva}}}]{2016PhRvD..94l3007F}%
  \BibitemOpen
  \bibfield  {author} {\bibinfo {author} {\bibfnamefont {J.}~\bibnamefont
  {{Feng}}}, \bibinfo {author} {\bibfnamefont {N.}~\bibnamefont
  {{Tomassetti}}}, \ and\ \bibinfo {author} {\bibfnamefont {A.}~\bibnamefont
  {{Oliva}}},\ }\href {\doibase 10.1103/PhysRevD.94.123007} {\bibfield
  {journal} {\bibinfo  {journal} {\prd}\ }\textbf {\bibinfo {volume} {94}},\
  \bibinfo {eid} {123007} (\bibinfo {year} {2016})},\ \Eprint
  {http://arxiv.org/abs/1610.06182} {arXiv:1610.06182 [astro-ph.HE]}
  \BibitemShut {NoStop}%
\bibitem [{\citenamefont {{Kachelrie{\ss}}}\ \emph {et~al.}(2015)\citenamefont
  {{Kachelrie{\ss}}}, \citenamefont {{Neronov}},\ and\ \citenamefont
  {{Semikoz}}}]{2015PhRvL.115r1103K}%
  \BibitemOpen
  \bibfield  {author} {\bibinfo {author} {\bibfnamefont {M.}~\bibnamefont
  {{Kachelrie{\ss}}}}, \bibinfo {author} {\bibfnamefont {A.}~\bibnamefont
  {{Neronov}}}, \ and\ \bibinfo {author} {\bibfnamefont {D.~V.}\ \bibnamefont
  {{Semikoz}}},\ }\href {\doibase 10.1103/PhysRevLett.115.181103} {\bibfield
  {journal} {\bibinfo  {journal} {Physical Review Letters}\ }\textbf {\bibinfo
  {volume} {115}},\ \bibinfo {eid} {181103} (\bibinfo {year} {2015})},\ \Eprint
  {http://arxiv.org/abs/1504.06472} {arXiv:1504.06472 [astro-ph.HE]}
  \BibitemShut {NoStop}%
\bibitem [{\citenamefont {G\'enolini}\ \emph {et~al.}(2019)\citenamefont
  {G\'enolini}, \citenamefont {Boudaud} \emph {et~al.}}]{Genolini:2019ewc}%
  \BibitemOpen
  \bibfield  {author} {\bibinfo {author} {\bibfnamefont {Y.}~\bibnamefont
  {G\'enolini}}, \bibinfo {author} {\bibfnamefont {M.}~\bibnamefont {Boudaud}},
   \emph {et~al.},\ }\href {\doibase 10.1103/PhysRevD.99.123028} {\bibfield
  {journal} {\bibinfo  {journal} {Phys. Rev.}\ }\textbf {\bibinfo {volume}
  {D99}},\ \bibinfo {pages} {123028} (\bibinfo {year} {2019})},\ \Eprint
  {http://arxiv.org/abs/1904.08917} {arXiv:1904.08917 [astro-ph.HE]}
  \BibitemShut {NoStop}%
\bibitem [{\citenamefont {Derome}\ \emph {et~al.}(2019)\citenamefont {Derome},
  \citenamefont {Maurin}, \citenamefont {Salati}, \citenamefont {Boudaud},
  \citenamefont {G\'enolini},\ and\ \citenamefont {Kunz\'e}}]{Derome:2019jfs}%
  \BibitemOpen
  \bibfield  {author} {\bibinfo {author} {\bibfnamefont {L.}~\bibnamefont
  {Derome}}, \bibinfo {author} {\bibfnamefont {D.}~\bibnamefont {Maurin}},
  \bibinfo {author} {\bibfnamefont {P.}~\bibnamefont {Salati}}, \bibinfo
  {author} {\bibfnamefont {M.}~\bibnamefont {Boudaud}}, \bibinfo {author}
  {\bibfnamefont {Y.}~\bibnamefont {G\'enolini}}, \ and\ \bibinfo {author}
  {\bibfnamefont {P.}~\bibnamefont {Kunz\'e}},\ }\href {\doibase
  10.1051/0004-6361/201935717} {\bibfield  {journal} {\bibinfo  {journal}
  {Astron. Astrophys.}\ }\textbf {\bibinfo {volume} {627}},\ \bibinfo {pages}
  {A158} (\bibinfo {year} {2019})},\ \Eprint {http://arxiv.org/abs/1904.08210}
  {arXiv:1904.08210 [astro-ph.HE]} \BibitemShut {NoStop}%
\bibitem [{\citenamefont {{James}}\ and\ \citenamefont
  {{Roos}}(1975)}]{1975CoPhC..10..343J}%
  \BibitemOpen
  \bibfield  {author} {\bibinfo {author} {\bibfnamefont {F.}~\bibnamefont
  {{James}}}\ and\ \bibinfo {author} {\bibfnamefont {M.}~\bibnamefont
  {{Roos}}},\ }\href {\doibase 10.1016/0010-4655(75)90039-9} {\bibfield
  {journal} {\bibinfo  {journal} {Computer Physics Communications}\ }\textbf
  {\bibinfo {volume} {10}},\ \bibinfo {pages} {343} (\bibinfo {year}
  {1975})}\BibitemShut {NoStop}%
\bibitem [{\citenamefont {{Di Bernardo}}\ \emph {et~al.}(2010)\citenamefont
  {{Di Bernardo}}, \citenamefont {{Evoli}}, \citenamefont {{Gaggero}},
  \citenamefont {{Grasso}},\ and\ \citenamefont
  {{Maccione}}}]{2010APh....34..274D}%
  \BibitemOpen
  \bibfield  {author} {\bibinfo {author} {\bibfnamefont {G.}~\bibnamefont {{Di
  Bernardo}}}, \bibinfo {author} {\bibfnamefont {C.}~\bibnamefont {{Evoli}}},
  \bibinfo {author} {\bibfnamefont {D.}~\bibnamefont {{Gaggero}}}, \bibinfo
  {author} {\bibfnamefont {D.}~\bibnamefont {{Grasso}}}, \ and\ \bibinfo
  {author} {\bibfnamefont {L.}~\bibnamefont {{Maccione}}},\ }\href {\doibase
  10.1016/j.astropartphys.2010.08.006} {\bibfield  {journal} {\bibinfo
  {journal} {Astroparticle Physics}\ }\textbf {\bibinfo {volume} {34}},\
  \bibinfo {pages} {274} (\bibinfo {year} {2010})},\ \Eprint
  {http://arxiv.org/abs/0909.4548} {arXiv:0909.4548 [astro-ph.HE]} \BibitemShut
  {NoStop}%
\bibitem [{\citenamefont {{Aguilar}}\ \emph {et~al.}(015a)\citenamefont
  {{Aguilar}}, \citenamefont {{Aisa}}, \citenamefont {{Alpat}}, \citenamefont
  {{Alvino}}, \citenamefont {{Ambrosi}}, \citenamefont {{Andeen}},
  \citenamefont {{Arruda}}, \citenamefont {{Attig}}, \citenamefont
  {{Azzarello}}, \citenamefont {{Bachlechner}},\ and\ \citenamefont
  {et~al.}}]{2015PhRvL.114q1103A}%
  \BibitemOpen
  \bibfield  {author} {\bibinfo {author} {\bibfnamefont {M.}~\bibnamefont
  {{Aguilar}}}, \bibinfo {author} {\bibfnamefont {D.}~\bibnamefont {{Aisa}}},
  \bibinfo {author} {\bibfnamefont {B.}~\bibnamefont {{Alpat}}}, \bibinfo
  {author} {\bibfnamefont {A.}~\bibnamefont {{Alvino}}}, \bibinfo {author}
  {\bibfnamefont {G.}~\bibnamefont {{Ambrosi}}}, \bibinfo {author}
  {\bibfnamefont {K.}~\bibnamefont {{Andeen}}}, \bibinfo {author}
  {\bibfnamefont {L.}~\bibnamefont {{Arruda}}}, \bibinfo {author}
  {\bibfnamefont {N.}~\bibnamefont {{Attig}}}, \bibinfo {author} {\bibfnamefont
  {P.}~\bibnamefont {{Azzarello}}}, \bibinfo {author} {\bibfnamefont
  {A.}~\bibnamefont {{Bachlechner}}}, \ and\ \bibinfo {author} {\bibnamefont
  {et~al.}},\ }\href {\doibase 10.1103/PhysRevLett.114.171103} {\bibfield
  {journal} {\bibinfo  {journal} {\prl}\ }\textbf {\bibinfo {volume} {114}},\
  \bibinfo {eid} {171103} (\bibinfo {year} {2015a})}\BibitemShut {NoStop}%
\bibitem [{\citenamefont {{Aguilar}}\ \emph {et~al.}(2017)\citenamefont
  {{Aguilar}}, \citenamefont {{Ali Cavasonza}}, \citenamefont {{Alpat}},
  \citenamefont {{Ambrosi}}, \citenamefont {{Arruda}}, \citenamefont {{Attig}},
  \citenamefont {{Aupetit}}, \citenamefont {{Azzarello}}, \citenamefont
  {{Bachlechner}}, \citenamefont {{Barao}},\ and\ \citenamefont
  {et~al.}}]{2017PhRvL.119y1101A}%
  \BibitemOpen
  \bibfield  {author} {\bibinfo {author} {\bibfnamefont {M.}~\bibnamefont
  {{Aguilar}}}, \bibinfo {author} {\bibfnamefont {L.}~\bibnamefont {{Ali
  Cavasonza}}}, \bibinfo {author} {\bibfnamefont {B.}~\bibnamefont {{Alpat}}},
  \bibinfo {author} {\bibfnamefont {G.}~\bibnamefont {{Ambrosi}}}, \bibinfo
  {author} {\bibfnamefont {L.}~\bibnamefont {{Arruda}}}, \bibinfo {author}
  {\bibfnamefont {N.}~\bibnamefont {{Attig}}}, \bibinfo {author} {\bibfnamefont
  {S.}~\bibnamefont {{Aupetit}}}, \bibinfo {author} {\bibfnamefont
  {P.}~\bibnamefont {{Azzarello}}}, \bibinfo {author} {\bibfnamefont
  {A.}~\bibnamefont {{Bachlechner}}}, \bibinfo {author} {\bibfnamefont
  {F.}~\bibnamefont {{Barao}}}, \ and\ \bibinfo {author} {\bibnamefont
  {et~al.}},\ }\href {\doibase 10.1103/PhysRevLett.119.251101} {\bibfield
  {journal} {\bibinfo  {journal} {Physical Review Letters}\ }\textbf {\bibinfo
  {volume} {119}},\ \bibinfo {eid} {251101} (\bibinfo {year}
  {2017})}\BibitemShut {NoStop}%
\bibitem [{\citenamefont {{Lodders}}(2003)}]{2003ApJ...591.1220L}%
  \BibitemOpen
  \bibfield  {author} {\bibinfo {author} {\bibfnamefont {K.}~\bibnamefont
  {{Lodders}}},\ }\href {\doibase 10.1086/375492} {\bibfield  {journal}
  {\bibinfo  {journal} {\apj}\ }\textbf {\bibinfo {volume} {591}},\ \bibinfo
  {pages} {1220} (\bibinfo {year} {2003})}\BibitemShut {NoStop}%
\bibitem [{\citenamefont {{Engelmann}}\ \emph {et~al.}(1990)\citenamefont
  {{Engelmann}}, \citenamefont {{Ferrando}}, \citenamefont {{Soutoul}},
  \citenamefont {{Goret}},\ and\ \citenamefont
  {{Juliusson}}}]{1990A&A...233...96E}%
  \BibitemOpen
  \bibfield  {author} {\bibinfo {author} {\bibfnamefont {J.~J.}\ \bibnamefont
  {{Engelmann}}}, \bibinfo {author} {\bibfnamefont {P.}~\bibnamefont
  {{Ferrando}}}, \bibinfo {author} {\bibfnamefont {A.}~\bibnamefont
  {{Soutoul}}}, \bibinfo {author} {\bibfnamefont {P.}~\bibnamefont {{Goret}}},
  \ and\ \bibinfo {author} {\bibfnamefont {E.}~\bibnamefont {{Juliusson}}},\
  }\href@noop {} {\bibfield  {journal} {\bibinfo  {journal} {\aap}\ }\textbf
  {\bibinfo {volume} {233}},\ \bibinfo {pages} {96} (\bibinfo {year}
  {1990})}\BibitemShut {NoStop}%
\bibitem [{\citenamefont {{Garcia-Munoz}}\ \emph {et~al.}(1975)\citenamefont
  {{Garcia-Munoz}}, \citenamefont {{Mason}},\ and\ \citenamefont
  {{Simpson}}}]{1975ApJ...202..265G}%
  \BibitemOpen
  \bibfield  {author} {\bibinfo {author} {\bibfnamefont {M.}~\bibnamefont
  {{Garcia-Munoz}}}, \bibinfo {author} {\bibfnamefont {G.~M.}\ \bibnamefont
  {{Mason}}}, \ and\ \bibinfo {author} {\bibfnamefont {J.~A.}\ \bibnamefont
  {{Simpson}}},\ }\href {\doibase 10.1086/153973} {\bibfield  {journal}
  {\bibinfo  {journal} {\apj}\ }\textbf {\bibinfo {volume} {202}},\ \bibinfo
  {pages} {265} (\bibinfo {year} {1975})}\BibitemShut {NoStop}%
\bibitem [{\citenamefont {{Casadei}}\ and\ \citenamefont
  {{Bindi}}(2004)}]{2004ApJ...612..262C}%
  \BibitemOpen
  \bibfield  {author} {\bibinfo {author} {\bibfnamefont {D.}~\bibnamefont
  {{Casadei}}}\ and\ \bibinfo {author} {\bibfnamefont {V.}~\bibnamefont
  {{Bindi}}},\ }\href {\doibase 10.1086/422514} {\bibfield  {journal} {\bibinfo
   {journal} {\apj}\ }\textbf {\bibinfo {volume} {612}},\ \bibinfo {pages}
  {262} (\bibinfo {year} {2004})},\ \Eprint
  {http://arxiv.org/abs/astro-ph/0302307} {astro-ph/0302307} \BibitemShut
  {NoStop}%
\bibitem [{\citenamefont {{O'Neill}}(2006)}]{2006AdSpR..37.1727O}%
  \BibitemOpen
  \bibfield  {author} {\bibinfo {author} {\bibfnamefont {P.~M.}\ \bibnamefont
  {{O'Neill}}},\ }\href {\doibase 10.1016/j.asr.2005.02.001} {\bibfield
  {journal} {\bibinfo  {journal} {Advances in Space Research}\ }\textbf
  {\bibinfo {volume} {37}},\ \bibinfo {pages} {1727} (\bibinfo {year}
  {2006})}\BibitemShut {NoStop}%
\bibitem [{\citenamefont {{Shikaze}}\ \emph {et~al.}(2007)\citenamefont
  {{Shikaze}}, \citenamefont {{Haino}}, \citenamefont {{Abe}}, \citenamefont
  {{Fuke}}, \citenamefont {{Hams}}, \citenamefont {{Kim}}, \citenamefont
  {{Makida}}, \citenamefont {{Matsuda}}, \citenamefont {{Mitchell}},
  \citenamefont {{Moiseev}} \emph {et~al.}}]{2007APh....28..154S}%
  \BibitemOpen
  \bibfield  {author} {\bibinfo {author} {\bibfnamefont {Y.}~\bibnamefont
  {{Shikaze}}}, \bibinfo {author} {\bibfnamefont {S.}~\bibnamefont {{Haino}}},
  \bibinfo {author} {\bibfnamefont {K.}~\bibnamefont {{Abe}}}, \bibinfo
  {author} {\bibfnamefont {H.}~\bibnamefont {{Fuke}}}, \bibinfo {author}
  {\bibfnamefont {T.}~\bibnamefont {{Hams}}}, \bibinfo {author} {\bibfnamefont
  {K.~C.}\ \bibnamefont {{Kim}}}, \bibinfo {author} {\bibfnamefont
  {Y.}~\bibnamefont {{Makida}}}, \bibinfo {author} {\bibfnamefont
  {S.}~\bibnamefont {{Matsuda}}}, \bibinfo {author} {\bibfnamefont {J.~W.}\
  \bibnamefont {{Mitchell}}}, \bibinfo {author} {\bibfnamefont {A.~A.}\
  \bibnamefont {{Moiseev}}},  \emph {et~al.},\ }\href {\doibase
  10.1016/j.astropartphys.2007.05.001} {\bibfield  {journal} {\bibinfo
  {journal} {Astroparticle Physics}\ }\textbf {\bibinfo {volume} {28}},\
  \bibinfo {pages} {154} (\bibinfo {year} {2007})},\ \Eprint
  {http://arxiv.org/abs/arXiv:astro-ph/0611388} {arXiv:astro-ph/0611388}
  \BibitemShut {NoStop}%
\bibitem [{\citenamefont {{Ghelfi}}\ \emph {et~al.}(2017)\citenamefont
  {{Ghelfi}}, \citenamefont {{Barao}}, \citenamefont {{Derome}},\ and\
  \citenamefont {{Maurin}}}]{2017A&A...605C...2G}%
  \BibitemOpen
  \bibfield  {author} {\bibinfo {author} {\bibfnamefont {A.}~\bibnamefont
  {{Ghelfi}}}, \bibinfo {author} {\bibfnamefont {F.}~\bibnamefont {{Barao}}},
  \bibinfo {author} {\bibfnamefont {L.}~\bibnamefont {{Derome}}}, \ and\
  \bibinfo {author} {\bibfnamefont {D.}~\bibnamefont {{Maurin}}},\ }\href
  {\doibase 10.1051/0004-6361/201527852e} {\bibfield  {journal} {\bibinfo
  {journal} {\aap}\ }\textbf {\bibinfo {volume} {605}},\ \bibinfo {eid} {C2}
  (\bibinfo {year} {2017})}\BibitemShut {NoStop}%
\bibitem [{\citenamefont {{Potgieter}}(2013)}]{2013LRSP...10....3P}%
  \BibitemOpen
  \bibfield  {author} {\bibinfo {author} {\bibfnamefont {M.}~\bibnamefont
  {{Potgieter}}},\ }\href {\doibase 10.12942/lrsp-2013-3} {\bibfield  {journal}
  {\bibinfo  {journal} {Living Reviews in Solar Physics}\ }\textbf {\bibinfo
  {volume} {10}},\ \bibinfo {pages} {3} (\bibinfo {year} {2013})},\ \Eprint
  {http://arxiv.org/abs/1306.4421} {arXiv:1306.4421 [physics.space-ph]}
  \BibitemShut {NoStop}%
\bibitem [{\citenamefont {{Perko}}(1987)}]{1987A&A...184..119P}%
  \BibitemOpen
  \bibfield  {author} {\bibinfo {author} {\bibfnamefont {J.~S.}\ \bibnamefont
  {{Perko}}},\ }\href@noop {} {\bibfield  {journal} {\bibinfo  {journal}
  {\aap}\ }\textbf {\bibinfo {volume} {184}},\ \bibinfo {pages} {119} (\bibinfo
  {year} {1987})}\BibitemShut {NoStop}%
\bibitem [{\citenamefont {Vittino}\ \emph {et~al.}(2018)\citenamefont
  {Vittino}, \citenamefont {Evoli},\ and\ \citenamefont
  {Gaggero}}]{Vittino:2017fuh}%
  \BibitemOpen
  \bibfield  {author} {\bibinfo {author} {\bibfnamefont {A.}~\bibnamefont
  {Vittino}}, \bibinfo {author} {\bibfnamefont {C.}~\bibnamefont {Evoli}}, \
  and\ \bibinfo {author} {\bibfnamefont {D.}~\bibnamefont {Gaggero}},\
  }\bibfield  {booktitle} {\emph {\bibinfo {booktitle} {{The Fluorescence
  detector Array of Single-pixel Telescopes: Contributions to the 35th
  International Cosmic Ray Conference (ICRC 2017)}}},\ }\href {\doibase
  10.22323/1.301.0024} {\bibfield  {journal} {\bibinfo  {journal} {PoS}\
  }\textbf {\bibinfo {volume} {ICRC2017}},\ \bibinfo {pages} {024} (\bibinfo
  {year} {2018})},\ \bibinfo {note} {[35,24(2017)]},\ \Eprint
  {http://arxiv.org/abs/1707.09003} {arXiv:1707.09003 [astro-ph.HE]}
  \BibitemShut {NoStop}%
\bibitem [{\citenamefont {{Cirelli}}\ \emph {et~al.}(2014)\citenamefont
  {{Cirelli}}, \citenamefont {{Gaggero}}, \citenamefont {{Giesen}},
  \citenamefont {{Taoso}},\ and\ \citenamefont
  {{Urbano}}}]{2014JCAP...12..045C}%
  \BibitemOpen
  \bibfield  {author} {\bibinfo {author} {\bibfnamefont {M.}~\bibnamefont
  {{Cirelli}}}, \bibinfo {author} {\bibfnamefont {D.}~\bibnamefont
  {{Gaggero}}}, \bibinfo {author} {\bibfnamefont {G.}~\bibnamefont {{Giesen}}},
  \bibinfo {author} {\bibfnamefont {M.}~\bibnamefont {{Taoso}}}, \ and\
  \bibinfo {author} {\bibfnamefont {A.}~\bibnamefont {{Urbano}}},\ }\href
  {\doibase 10.1088/1475-7516/2014/12/045} {\bibfield  {journal} {\bibinfo
  {journal} {\jcap}\ }\textbf {\bibinfo {volume} {12}},\ \bibinfo {eid} {045}
  (\bibinfo {year} {2014})},\ \Eprint {http://arxiv.org/abs/1407.2173}
  {arXiv:1407.2173 [hep-ph]} \BibitemShut {NoStop}%
\bibitem [{\citenamefont {{Aguilar}}\ \emph {et~al.}(2018)\citenamefont
  {{Aguilar}}, \citenamefont {{Ali Cavasonza}}, \citenamefont {{Ambrosi}},
  \citenamefont {{Arruda}}, \citenamefont {{Attig}}, \citenamefont {{Aupetit}},
  \citenamefont {{Azzarello}}, \citenamefont {{Bachlechner}}, \citenamefont
  {{Barao}}, \citenamefont {{Barrau}},\ and\ \citenamefont
  {et~al.}}]{2018PhRvL.120b1101A}%
  \BibitemOpen
  \bibfield  {author} {\bibinfo {author} {\bibfnamefont {M.}~\bibnamefont
  {{Aguilar}}}, \bibinfo {author} {\bibfnamefont {L.}~\bibnamefont {{Ali
  Cavasonza}}}, \bibinfo {author} {\bibfnamefont {G.}~\bibnamefont
  {{Ambrosi}}}, \bibinfo {author} {\bibfnamefont {L.}~\bibnamefont {{Arruda}}},
  \bibinfo {author} {\bibfnamefont {N.}~\bibnamefont {{Attig}}}, \bibinfo
  {author} {\bibfnamefont {S.}~\bibnamefont {{Aupetit}}}, \bibinfo {author}
  {\bibfnamefont {P.}~\bibnamefont {{Azzarello}}}, \bibinfo {author}
  {\bibfnamefont {A.}~\bibnamefont {{Bachlechner}}}, \bibinfo {author}
  {\bibfnamefont {F.}~\bibnamefont {{Barao}}}, \bibinfo {author} {\bibfnamefont
  {A.}~\bibnamefont {{Barrau}}}, \ and\ \bibinfo {author} {\bibnamefont
  {et~al.}},\ }\href {\doibase 10.1103/PhysRevLett.120.021101} {\bibfield
  {journal} {\bibinfo  {journal} {\prl}\ }\textbf {\bibinfo {volume} {120}},\
  \bibinfo {eid} {021101} (\bibinfo {year} {2018})}\BibitemShut {NoStop}%
\bibitem [{\citenamefont {{Andrae}}\ \emph {et~al.}(2010)\citenamefont
  {{Andrae}}, \citenamefont {{Schulze-Hartung}},\ and\ \citenamefont
  {{Melchior}}}]{2010arXiv1012.3754A}%
  \BibitemOpen
  \bibfield  {author} {\bibinfo {author} {\bibfnamefont {R.}~\bibnamefont
  {{Andrae}}}, \bibinfo {author} {\bibfnamefont {T.}~\bibnamefont
  {{Schulze-Hartung}}}, \ and\ \bibinfo {author} {\bibfnamefont
  {P.}~\bibnamefont {{Melchior}}},\ }\href@noop {} {\bibfield  {journal}
  {\bibinfo  {journal} {arXiv e-prints}\ ,\ \bibinfo {eid} {arXiv:1012.3754}}
  (\bibinfo {year} {2010})},\ \Eprint {http://arxiv.org/abs/1012.3754}
  {arXiv:1012.3754 [astro-ph.IM]} \BibitemShut {NoStop}%
\bibitem [{\citenamefont {Gaggero}\ \emph {et~al.}(2015)\citenamefont
  {Gaggero}, \citenamefont {Urbano}, \citenamefont {Valli},\ and\ \citenamefont
  {Ullio}}]{Gaggero:2014xla}%
  \BibitemOpen
  \bibfield  {author} {\bibinfo {author} {\bibfnamefont {D.}~\bibnamefont
  {Gaggero}}, \bibinfo {author} {\bibfnamefont {A.}~\bibnamefont {Urbano}},
  \bibinfo {author} {\bibfnamefont {M.}~\bibnamefont {Valli}}, \ and\ \bibinfo
  {author} {\bibfnamefont {P.}~\bibnamefont {Ullio}},\ }\href {\doibase
  10.1103/PhysRevD.91.083012} {\bibfield  {journal} {\bibinfo  {journal} {Phys.
  Rev.}\ }\textbf {\bibinfo {volume} {D91}},\ \bibinfo {pages} {083012}
  (\bibinfo {year} {2015})},\ \Eprint {http://arxiv.org/abs/1411.7623}
  {arXiv:1411.7623 [astro-ph.HE]} \BibitemShut {NoStop}%
\bibitem [{\citenamefont {{Putze}}\ \emph {et~al.}(2009)\citenamefont
  {{Putze}}, \citenamefont {{Derome}}, \citenamefont {{Maurin}}, \citenamefont
  {{Perotto}},\ and\ \citenamefont {{Taillet}}}]{2009A&A...497..991P}%
  \BibitemOpen
  \bibfield  {author} {\bibinfo {author} {\bibfnamefont {A.}~\bibnamefont
  {{Putze}}}, \bibinfo {author} {\bibfnamefont {L.}~\bibnamefont {{Derome}}},
  \bibinfo {author} {\bibfnamefont {D.}~\bibnamefont {{Maurin}}}, \bibinfo
  {author} {\bibfnamefont {L.}~\bibnamefont {{Perotto}}}, \ and\ \bibinfo
  {author} {\bibfnamefont {R.}~\bibnamefont {{Taillet}}},\ }\href {\doibase
  10.1051/0004-6361/200810824} {\bibfield  {journal} {\bibinfo  {journal}
  {\aap}\ }\textbf {\bibinfo {volume} {497}},\ \bibinfo {pages} {991} (\bibinfo
  {year} {2009})},\ \Eprint {http://arxiv.org/abs/0808.2437} {arXiv:0808.2437}
  \BibitemShut {NoStop}%
\bibitem [{\citenamefont {{Adriani}}\ \emph {et~al.}(2019)\citenamefont
  {{Adriani}}, \citenamefont {{Akaike}}, \citenamefont {{Asano}}, \citenamefont
  {{Asaoka}}, \citenamefont {{Bagliesi}}, \citenamefont {{Berti}},
  \citenamefont {{Bigongiari}}, \citenamefont {{Binns}}, \citenamefont
  {{Bonechi}}, \citenamefont {{Bongi}},\ and\ \citenamefont {{Calet
  Collaboration}}}]{2019PhRvL.122r1102A}%
  \BibitemOpen
  \bibfield  {author} {\bibinfo {author} {\bibfnamefont {O.}~\bibnamefont
  {{Adriani}}}, \bibinfo {author} {\bibfnamefont {Y.}~\bibnamefont {{Akaike}}},
  \bibinfo {author} {\bibfnamefont {K.}~\bibnamefont {{Asano}}}, \bibinfo
  {author} {\bibfnamefont {Y.}~\bibnamefont {{Asaoka}}}, \bibinfo {author}
  {\bibfnamefont {M.~G.}\ \bibnamefont {{Bagliesi}}}, \bibinfo {author}
  {\bibfnamefont {E.}~\bibnamefont {{Berti}}}, \bibinfo {author} {\bibfnamefont
  {G.}~\bibnamefont {{Bigongiari}}}, \bibinfo {author} {\bibfnamefont {W.~R.}\
  \bibnamefont {{Binns}}}, \bibinfo {author} {\bibfnamefont {S.}~\bibnamefont
  {{Bonechi}}}, \bibinfo {author} {\bibfnamefont {M.}~\bibnamefont {{Bongi}}},
  \ and\ \bibinfo {author} {\bibnamefont {{Calet Collaboration}}},\ }\href
  {\doibase 10.1103/PhysRevLett.122.181102} {\bibfield  {journal} {\bibinfo
  {journal} {Physical Review Letters}\ }\textbf {\bibinfo {volume} {122}},\
  \bibinfo {eid} {181102} (\bibinfo {year} {2019})},\ \Eprint
  {http://arxiv.org/abs/1905.04229} {arXiv:1905.04229 [astro-ph.HE]}
  \BibitemShut {NoStop}%
\bibitem [{\citenamefont {{Winkler}}(2017{\natexlab{b}})}]{Winkler2017}%
  \BibitemOpen
  \bibfield  {author} {\bibinfo {author} {\bibfnamefont {M.~W.}\ \bibnamefont
  {{Winkler}}},\ }\href {\doibase 10.1088/1475-7516/2017/02/048} {\bibfield
  {journal} {\bibinfo  {journal} {\jcap}\ }\textbf {\bibinfo {volume} {2}},\
  \bibinfo {eid} {048} (\bibinfo {year} {2017}{\natexlab{b}})},\ \Eprint
  {http://arxiv.org/abs/1701.04866} {arXiv:1701.04866 [hep-ph]} \BibitemShut
  {NoStop}%
\bibitem [{\citenamefont {{Adriani}}\ \emph {et~al.}(2011)\citenamefont
  {{Adriani}}, \citenamefont {{Barbarino}}, \citenamefont {{Bazilevskaya}},
  \citenamefont {{Bellotti}}, \citenamefont {{Boezio}}, \citenamefont
  {{Bogomolov}}, \citenamefont {{Bonechi}}, \citenamefont {{Bongi}},
  \citenamefont {{Bonvicini}}, \citenamefont {{Borisov}}, \citenamefont
  {{Bottai}} \emph {et~al.}}]{2011Sci...332...69A}%
  \BibitemOpen
  \bibfield  {author} {\bibinfo {author} {\bibfnamefont {O.}~\bibnamefont
  {{Adriani}}}, \bibinfo {author} {\bibfnamefont {G.~C.}\ \bibnamefont
  {{Barbarino}}}, \bibinfo {author} {\bibfnamefont {G.~A.}\ \bibnamefont
  {{Bazilevskaya}}}, \bibinfo {author} {\bibfnamefont {R.}~\bibnamefont
  {{Bellotti}}}, \bibinfo {author} {\bibfnamefont {M.}~\bibnamefont
  {{Boezio}}}, \bibinfo {author} {\bibfnamefont {E.~A.}\ \bibnamefont
  {{Bogomolov}}}, \bibinfo {author} {\bibfnamefont {L.}~\bibnamefont
  {{Bonechi}}}, \bibinfo {author} {\bibfnamefont {M.}~\bibnamefont {{Bongi}}},
  \bibinfo {author} {\bibfnamefont {V.}~\bibnamefont {{Bonvicini}}}, \bibinfo
  {author} {\bibfnamefont {S.}~\bibnamefont {{Borisov}}}, \bibinfo {author}
  {\bibfnamefont {S.}~\bibnamefont {{Bottai}}},  \emph {et~al.},\ }\href
  {\doibase 10.1126/science.1199172} {\bibfield  {journal} {\bibinfo  {journal}
  {Science}\ }\textbf {\bibinfo {volume} {332}},\ \bibinfo {pages} {69}
  (\bibinfo {year} {2011})},\ \Eprint {http://arxiv.org/abs/1103.4055}
  {arXiv:1103.4055 [astro-ph.HE]} \BibitemShut {NoStop}%
\bibitem [{\citenamefont {{Duperray}}\ \emph {et~al.}(2005)\citenamefont
  {{Duperray}}, \citenamefont {{Baret}}, \citenamefont {{Maurin}},
  \citenamefont {{Boudoul}}, \citenamefont {{Barrau}}, \citenamefont
  {{Derome}}, \citenamefont {{Protasov}},\ and\ \citenamefont
  {{Bu{\'e}nerd}}}]{2005PhRvD..71h3013D}%
  \BibitemOpen
  \bibfield  {author} {\bibinfo {author} {\bibfnamefont {R.}~\bibnamefont
  {{Duperray}}}, \bibinfo {author} {\bibfnamefont {B.}~\bibnamefont {{Baret}}},
  \bibinfo {author} {\bibfnamefont {D.}~\bibnamefont {{Maurin}}}, \bibinfo
  {author} {\bibfnamefont {G.}~\bibnamefont {{Boudoul}}}, \bibinfo {author}
  {\bibfnamefont {A.}~\bibnamefont {{Barrau}}}, \bibinfo {author}
  {\bibfnamefont {L.}~\bibnamefont {{Derome}}}, \bibinfo {author}
  {\bibfnamefont {K.}~\bibnamefont {{Protasov}}}, \ and\ \bibinfo {author}
  {\bibfnamefont {M.}~\bibnamefont {{Bu{\'e}nerd}}},\ }\href {\doibase
  10.1103/PhysRevD.71.083013} {\bibfield  {journal} {\bibinfo  {journal}
  {\prd}\ }\textbf {\bibinfo {volume} {71}},\ \bibinfo {eid} {083013} (\bibinfo
  {year} {2005})},\ \Eprint {http://arxiv.org/abs/astro-ph/0503544}
  {astro-ph/0503544} \BibitemShut {NoStop}%
\bibitem [{\citenamefont {{Asplund}}\ \emph {et~al.}(2009)\citenamefont
  {{Asplund}}, \citenamefont {{Grevesse}}, \citenamefont {{Sauval}},\ and\
  \citenamefont {{Scott}}}]{2009ARA&A..47..481A}%
  \BibitemOpen
  \bibfield  {author} {\bibinfo {author} {\bibfnamefont {M.}~\bibnamefont
  {{Asplund}}}, \bibinfo {author} {\bibfnamefont {N.}~\bibnamefont
  {{Grevesse}}}, \bibinfo {author} {\bibfnamefont {A.~J.}\ \bibnamefont
  {{Sauval}}}, \ and\ \bibinfo {author} {\bibfnamefont {P.}~\bibnamefont
  {{Scott}}},\ }\href {\doibase 10.1146/annurev.astro.46.060407.145222}
  {\bibfield  {journal} {\bibinfo  {journal} {\araa}\ }\textbf {\bibinfo
  {volume} {47}},\ \bibinfo {pages} {481} (\bibinfo {year} {2009})},\ \Eprint
  {http://arxiv.org/abs/0909.0948} {arXiv:0909.0948 [astro-ph.SR]} \BibitemShut
  {NoStop}%
\bibitem [{\citenamefont {{Davis}}\ \emph {et~al.}(1995)\citenamefont
  {{Davis}}, \citenamefont {{Menn}}, \citenamefont {{Barbier}}, \citenamefont
  {{Christian}}, \citenamefont {{Golden}}, \citenamefont {{Hof}}, \citenamefont
  {{Krombel}}, \citenamefont {{Labrador}}, \citenamefont {{Mewaldt}},
  \citenamefont {{Mitchell}} \emph {et~al.}}]{1995ICRC....2..622D}%
  \BibitemOpen
  \bibfield  {author} {\bibinfo {author} {\bibfnamefont {A.~J.}\ \bibnamefont
  {{Davis}}}, \bibinfo {author} {\bibfnamefont {W.}~\bibnamefont {{Menn}}},
  \bibinfo {author} {\bibfnamefont {L.~M.}\ \bibnamefont {{Barbier}}}, \bibinfo
  {author} {\bibfnamefont {E.~R.}\ \bibnamefont {{Christian}}}, \bibinfo
  {author} {\bibfnamefont {R.~L.}\ \bibnamefont {{Golden}}}, \bibinfo {author}
  {\bibfnamefont {M.}~\bibnamefont {{Hof}}}, \bibinfo {author} {\bibfnamefont
  {K.~E.}\ \bibnamefont {{Krombel}}}, \bibinfo {author} {\bibfnamefont {A.~W.}\
  \bibnamefont {{Labrador}}}, \bibinfo {author} {\bibfnamefont {R.~A.}\
  \bibnamefont {{Mewaldt}}}, \bibinfo {author} {\bibnamefont {{Mitchell}}},
  \emph {et~al.},\ }\href@noop {} {\bibfield  {journal} {\bibinfo  {journal}
  {International Cosmic Ray Conference}\ }\textbf {\bibinfo {volume} {2}},\
  \bibinfo {pages} {622} (\bibinfo {year} {1995})}\BibitemShut {NoStop}%
\end{thebibliography}%
\end{document}